\documentclass[12pt]{article}

\usepackage{amssymb,amsmath,graphicx}
\usepackage{simplewick}
\usepackage{epsf}
\usepackage{graphicx,epsfig}
\usepackage{amsfonts}
\usepackage{amssymb}
\usepackage{cite}

\def\bk{{\bf k}}
\def\bp{{\bf p}}
\def\bq{{\bf q}}
\def\bx{{\bf x}}

\def\CA{{\cal A}}

\def\CH{{\cal H}}
\def\CL{{\cal L}}
\def\CO{{\cal O}}

\def\CT{{\cal T}}

\def\mpl{M_{\rm Pl}}
\def\half{\frac{1}{2}}

\makeatletter
\renewcommand\section{\@startsection {section}{1}{\z@}%
                                 {-3.5ex \@plus -1ex \@minus -.2ex}
                                   {2.3ex \@plus.2ex}%
                                   {\normalfont\large\bfseries}}
\renewcommand\subsection{\@startsection{subsection}{2}{\z@}%
                                   {-3.25ex\@plus -1ex \@minus -.2ex}%
                                     {1.5ex \@plus .2ex}%
                                     {\normalfont\bfseries}}
\renewcommand\subsubsection{\@startsection{subsubsection}{3}{\z@}%
                                   {-3.25ex\@plus -1ex \@minus -.2ex}%
                                     {1.5ex \@plus .2ex}%
                                     {\normalfont\itshape}}
\makeatother

\newcommand{\Letter}{
\setlength{\textwidth}{16.5cm}
   \setlength{\textheight}{22.6cm}
    \hoffset=-0.5in
\voffset=-2.1cm }

\Letter

\begin{document}
\newcommand{\be}{\begin{equation}}
\newcommand{\ee}{\end{equation}}
\newcommand{\bea}{\begin{eqnarray}}
\newcommand{\eea}{\end{eqnarray}}
\newcommand{\barr}{\begin{array}}
\newcommand{\earr}{\end{array}}

\thispagestyle{empty}
\begin{flushright}
\parbox[t]{1.8in}{MIT-CTP-4039\\
CAS-KITPC/ITP-111\\
CERN-PH-TH/2009-064\\ MAD-TH-09-04}
\end{flushright}

\vspace*{0.3in}

\begin{center}
{\large \bf Large Primordial Trispectra in General Single Field
Inflation}

\vspace*{0.5in} {Xingang Chen$^1$, Bin Hu$^2$, Min-xin Huang$^3$,
Gary Shiu$^{4,5}$, and Yi Wang$^2$}
\\[.3in]
{\em
$^1$ Center for Theoretical Physics\\
Massachusetts Institute of Technology, Cambridge, MA 02139, USA
\\[.1in]
$^2$ Kavli Institute for Theoretical Physics China, \\Key Laboratory
of Frontiers in Theoretical Physics, \\Institute of Theoretical
Physics,
Chinese Academy of Sciences, \\Beijing 100190, P.R.China \\[.1in]
$^3$ Theory Division, Department of Physics, CERN,\\ CH-1211 Geneva,
Switzerland\\[.1in]
$^4$ Department of Physics, University of Wisconsin, \\Madison, WI
53706, USA \\[.1in]
$^5$School of Natural Sciences, Institute for Advanced Study, \\
Princeton, NJ 08540, USA
\\[0.3in]}
\end{center}

\begin{center}
{\bf Abstract}
\end{center}
\noindent We compute the large scalar four-point correlation functions
in general single field inflation models, where the inflaton
Lagrangian is an
arbitrary function of the inflaton and its first derivative.
We find that the
leading order trispectra have four different shapes determined by
three parameters. We study features in these shapes that can be used to
distinguish among themselves, and between them and the
 trispectra of the local form. For the
purpose of data analyses, we give two simple representative forms
for these ``equilateral trispectra''. We
also study the effects on the trispectra if the initial state of inflation
deviates from the standard Bunch-Davies
vacuum.

\vfill

\newpage
\setcounter{page}{1}

\tableofcontents

\newpage

\section{Introduction}
\setcounter{equation}{0}

Primordial non-Gaussianity is potentially one of the most promising
probes of the inflationary universe \cite{Komatsu:2009kd}. Like the role
colliders play in particle physics, measurements of primordial
non-Gaussian features
provide microscopic information on the
interactions of the inflatons and/or the curvatons. Constraining and detecting
primordial non-Gaussianities has become one of the major efforts in
modern cosmology. Theoretical predictions of non-Gaussianities, especially their
explicit forms, play an important role in this program. On the one
hand, they are needed as inputs of data analyses
\cite{Komatsu:2003iq,Creminelli:2006gc,Smith:2006ud,Fergusson:2008ra}
which eventually constrain the parameters defining the non-Gaussian features;
on the other hand, different forms of non-Gaussianities are
associated with different inflaton or curvaton interactions, and so if
detected can help us  understand the nature of inflation.

A variety of potentially detectable forms of non-Gaussian features from inflation
models have been
proposed and classified, in terms of their
shapes and running. The scalar three-point functions,
i.e.~the scalar bispectra,
are by far the most well-studied. For single field
inflation, a brief summary of the status is as follows. Minimal slow-roll
inflation gives undetectable amount of primordial non-Gaussianities
\cite{Maldacena:2002vr,Acquaviva:2002ud,Seery:2005wm};
non-canonical kinetic terms can generate large
bispectra of the equilateral shapes
\cite{Chen:2006nt,Cheung:2007st};
non-Bunch-Davies vacuum can boost
the folded shape \cite{Chen:2006nt,Holman:2007na,Meerburg:2009ys}; and
features in the Lagrangian (sharp or periodic)
 can give rise to large bispectra with oscillatory
running \cite{Chen:2006xjb,Chen:2008wn}. Multifield inflation models
provide many other possibilities due to various kinds of
isocurvature modes, such as curvatons \cite{Lyth:2002my}, turning
\cite{Vernizzi:2006ve,Huang:2007hh,Gao:2008dt,Langlois:2008qf,Arroja:2008yy,Byrnes:2008zy}
or bifurcating \cite{Naruko:2008sq,Li:2009sp} trajectories, thermal
effects \cite{Moss:2007cv,Chen:2007gd} and etc. These models give
many additional forms of large bispectra, notably ones with a large
local shape.

\begin{figure}
\begin{center}
\epsfig{file=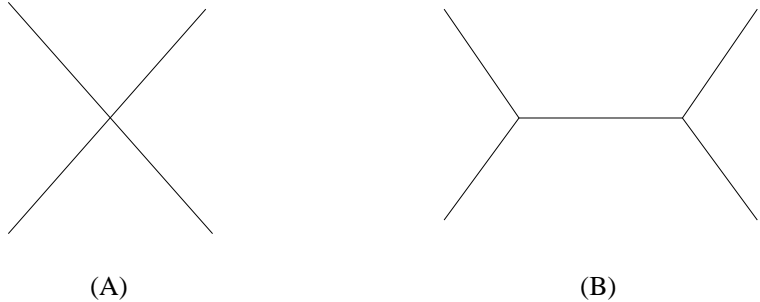, width=10cm}
\end{center}
\medskip
\caption{Two diagrams that contribute to the large trispectra.}
\label{Fdiagrams}
\end{figure}

We will be getting much more data in the near future from new
generations of experiments, ranging from cosmic microwave
background, large scale structure and possibly even 21-cm hydrogen
line. Compared with the current WMAP, these experiments will be
measuring signals from shorter scales and/or in three dimensions.
Therefore a significant larger number of modes will become
available. This makes the study of four- or higher point functions
interesting, as they provide information on new interaction terms
and refined distinctions among models. In this paper we extend the
work of Ref.~\cite{Chen:2006nt} and classify the forms of large
scalar trispectra (i.e.~the scalar four-point function) in general
single field inflation models. There have been some preliminary
works in this direction \cite{Huang:2006eh,Arroja:2008ga},
calculating contributions from the contact interaction diagram
(Fig.~\ref{Fdiagrams} (A)). For models with a large trispectrum,
there is yet another set of diagrams involving the exchange of a
scalar (Fig.~\ref{Fdiagrams} (B)) that contributes at the same order
of magnitude.\footnote{Note that for
  slow-roll inflation, the contribution from the scalar-exchange
  diagram Fig.~\ref{Fdiagrams} (B)
  is subleading, while the graviton-exchange contribution belongs to
  the leading order \cite{Seery:2008ax}.} In this paper, we
complete this program and classify all possible shapes arising in this
framework.

For the bispectra in general single field inflation,
the leading large non-Gaussianities have two different
shapes controlled by two parameters \cite{Chen:2006nt}.
As we will see here, for
trispectra, we have four different shapes controlled by three
parameters. Some of them have complicated momentum-dependence.
For the purpose of data analyses, we give simple
representative shapes that can capture the main features of these
functions. We point out
the features in the shapes that can be used to distinguish among
themselves, as well as to distinguish them
from the trispectra of the local form. We also study the effects of a
non-Bunch-Davies initial state of inflation on these trispectra.

This paper is organized as follows. In Section 2, we review the
basic formalisms and main results for the power spectrum and
bispectra in general single field inflation. In Section 3, we
calculate the leading order trispectra, and summarize the final
results. At leading order, the trispectra can be classified into
four shapes, controlled by three parameters. In Section 4, we
investigate the shapes of the trispectra, including consistency
relations, figures in various limits, and also give two simple
representative forms of these equilateral trispectra to facilitate future data
analyses. In Section 5, we discuss DBI and K-inflation as two
examples to illustrate our results. In Section 6, we study the trispectra
when the initial state of inflation is in a non-Bunch-Davies vacuum. We conclude in Section 7.

\section{Formalism and review}
\label{SecReview}
\setcounter{equation}{0}

In this section, we review the formalisms and main results of
Ref.~\cite{Chen:2006nt}.
As in Ref.~\cite{Garriga:1999vw}, we consider the following
general Lagrangian
for the inflaton field
$\phi$,
\bea
S= \half \int d^4x \sqrt{-g} \left[ \mpl^2 R + 2 P(X,\phi) \right] ~,
\eea
where $X \equiv - \half g^{\mu\nu} \partial_\mu \phi \partial_\nu
\phi$ and
the signature of the metric is $(-1,1,1,1)$.

Irrespective of the specific mechanism that is responsible for the
inflation, once it is achieved we require the following set of
slow-variation parameters to be small, \bea \epsilon = -\frac{\dot
H}{H^2} ~, ~~~~\eta = \frac{\dot \epsilon}{\epsilon H} ~, ~~~~s =
\frac{\dot c_s}{c_s H} ~, \eea where $H$ is Hubble parameter and
\bea c_s^2 \equiv \frac{P_{,X}}{P_{,X} + 2X P_{,XX}} \eea is the
sound speed. The slow-variation parameters can be large temporarily
or quickly oscillating \cite{Chen:2006xjb,Chen:2008wn,Bean:2008na},
but we do not consider such cases here.

The power spectrum $P_\zeta$ is defined from the two-point function
of the curvature perturbation $\zeta$, \bea \langle \zeta(\bk_1)
\zeta(\bk_2) \rangle = (2\pi)^5 \delta^3 (\bk_1 +\bk_2)
\frac{1}{2k_1^3} P_\zeta ~. \eea For the class of inflation models
that we consider, \bea P_\zeta =
\frac{1}{8\pi^2\mpl^2}\frac{H^2}{c_s \epsilon} ~. \eea

In order to parametrize the three-point function, we need to define
a parameter $\lambda/\Sigma$ related to the third derivative of
the inflaton Lagrangian $P$ with respect to $X$,
\bea
\lambda &=& X^2 P_{,XX} + \frac{2}{3} X^3 P_{,XXX} ~,
\\
\Sigma &=& X P_{,X} + 2 X^2 P_{,XX} = \frac{H^2\epsilon}{c_s^2} ~.
\eea
The bispectrum form factor $\CA(k_1,k_2,k_3)$ is defined as
\bea
\langle \zeta(\bk_1) \zeta(\bk_2) \zeta(\bk_3) \rangle
= (2\pi)^7 \delta^3 (\bk_1 +\bk_2 +\bk_3)
P_\zeta^2 \prod_{i=1}^{3} \frac{1}{k_i^3} ~\CA ~.
\eea
Up to $\CO(\epsilon)$, this bispectrum is determined by five
parameters, $c_s$, $\lambda/\Sigma$, $\epsilon$, $\eta$ and $s$.
For the most interesting cases $c_s \ll 1$ or $\lambda/\Sigma \gg 1$
where the non-Gaussianities are large, the leading bispectrum is given
by
\bea
\CA &=& \left( \frac{1}{c_s^2} - 1 - \frac{2\lambda}{\Sigma} \right)
\frac{3k_1^2k_2^2k_3^2}{2K^3}
\cr
&+& \left( \frac{1}{c_s^2} - 1 \right)
\left( - \frac{1}{K} \sum_{i>j} k_i^2 k_j^2 + \frac{1}{2K^2}
\sum_{i\ne j} k_i^2 k_j^3 + \frac{1}{8} \sum_i k_i^3 \right) ~.
\label{Aform}
\eea
So we have two different forms determined by two parameters.
In such cases, the effect of the non-canonical kinetic terms of
the inflaton has to
become large enough so that the inflationary mechanism is no longer
slow-roll (in slow-roll the canonical kinetic term dominates over
the non-canonical terms).
Since
inflation gives approximately scale-invariant spectrum, ignoring the
mild running of the non-Gaussianity \cite{Chen:2005fe}, the bispectrum
is approximately a function of two variables in terms of the momentum
ratios, $k_2/k_1$ and $k_3/k_1$ \cite{Babich:2004gb}.
The two forms in (\ref{Aform}) have
very similar shapes and
they are usually referred to as the equilateral shapes. Because the two
shapes do have a small difference, for fine-tuned model parameters
$c_s$ and $\lambda/\Sigma$, they can cancel each other to a
large extent
and leave an approximately orthogonal component. One can use this
component and the one of the originals to form a new bases of the
shapes.\footnote{We would like to thank Eiichiro Komatsu for
  discussions on this point \cite{SenatoreUnpub}.}

\section{Large trispectra}
\label{SecTri}
\setcounter{equation}{0}

As in the bispectrum case, we are most interested in cases where the
trispectra are large. In general single field inflationary models,
this is achieved by the non-canonical kinetic terms.
The origin of large non-Gaussianities
come from terms with derivatives of the
inflaton
Lagrangian $P$ with respect to $X$.
The contribution from the gravity sector is negligibly
small. The derivative of $P$ with respective to $\phi$ is also
small due to the approximate shift symmetry associated with the
inflaton. Another equivalent way to see this is to work in the
comoving gauge \cite{Maldacena:2002vr}
where the scalar perturbation $\zeta$ only appears in the metric.
So $P_{,\phi}$ explicitly does not appear in the
expansion. Using the leading order relation
\bea
\zeta \approx - \frac{H}{\dot \phi} \alpha
\label{zetaalpha}
\eea
to convert $\zeta$ into
$\alpha \equiv \delta \phi$, again we see that $P_{,\phi}$ does not
appear.

Therefore for our purpose, it is convenient to choose the inflaton
gauge where the scalar perturbation only appears in the inflaton
\cite{Maldacena:2002vr},
\bea
\phi = \phi_0(t) + \alpha(t,\bx) ~;
\eea
and when we expand the inflaton Lagrangian $P$,
we only concentrate on terms that have derivatives with respect to
$X$. Such a method has also been used in
Ref.~\cite{Creminelli:2003iq,Gruzinov:2004jx}.

\subsection{Scalar-exchange diagram}
\label{SecSE}

In this subsection, we compute the scalar-exchange diagram,
Fig.~\ref{Fdiagrams} (B).
Using the inflaton gauge, we get the cubic terms of the Lagrangian
in the small $c_s$ or large $\lambda/\Sigma$ limit,
\bea
\CL_3 = \left(\half P_{,XX} \dot \phi + \frac{1}{6} P_{,XXX} \dot
\phi^3 \right) a^3 \dot \alpha^3
-\half P_{,XX} \dot \phi ~a \dot \alpha (\nabla \alpha)^2 ~.
\eea
Written in terms of $\zeta$ using (\ref{zetaalpha}), we get
\bea
\CL_3 =
-2a^3 \frac{\lambda}{H^3} \dot \zeta^3 + a \frac{\Sigma}{H^3}
(1-c_s^2) \dot
\zeta (\partial_i \zeta)^2 ~.
\label{CL3}
\eea
Despite of its
different appearance from the three leading cubic terms in
\cite{Chen:2006nt}, one can show, using the linear equation of
motion and integration by part, that the difference is a total
derivative.

In terms of the interaction Hamiltonian, $\CH^I_3 = - \CL_3$,
we denote the two terms in (\ref{CL3}) as
\bea H^I_3 =  - \int d^3x \CL_3 = H_a + H_b ~, \eea
where
\bea H_a(t) &=& 2a^3 \frac{\lambda}{H^3}
\int \prod_{i=1}^3
\frac{d^3\bp_i}{(2\pi)^3} \dot\zeta_I(\bp_1,t) \dot\zeta_I(\bp_2,t)
\dot\zeta_I(\bp_3,t)
(2\pi)^3
\delta^3(\sum_{i=1}^3 \bp_i) ~,
\label{H_a}
\\
H_b(t) &=& a \frac{\Sigma}{H^3} (1-c_s^2)
\int \prod_{i=1}^3
\frac{d^3\bp_i}{(2\pi)^3} (\bp_2 \cdot \bp_3)
\dot\zeta_I(\bp_1,t) \zeta_I(\bp_2,t)
\zeta_I(\bp_3,t)
(2\pi)^3 \delta^3(\sum_{i=1}^3 \bp_i) ~.
\label{H_b}
\eea
The $\zeta_I$ is in the interaction picture and satisfies the equation
of motion followed from the kinematic Hamiltonian.

The scalar trispectrum is the expectation value of the curvature
perturbation $\zeta_I^4$ in the interaction vacuum. According to the
in-in formalism \cite{Weinberg:2005vy}, there are three terms
contributing to the diagram Fig.~\ref{Fdiagrams} (B),\footnote{The
in-in
  formalism is often used in the literature in terms of a
  commutator form which is equivalent to the form presented here.
  However, for a subset of terms,
  the algebra in the
  commutator form is more complicated than the one we use here. We
  discuss this equivalence in Appendix \ref{AppCom}.}
\bea
\langle \zeta^4
\rangle &=& \langle 0| \left[{\bar T} e^{i\int_{t_0}^t dt'
  H_I(t')} \right]
  \zeta_I(\bk_1,t) \zeta_I(\bk_2,t) \zeta_I(\bk_3,t)
  \zeta_I(\bk_4,t)
  \left[ T e^{-i\int_{t_0}^t dt' H_I(t')} \right] |0 \rangle
\label{Def}
\\
&\supset& \int_{t_0}^t dt' \int_{t_0}^t dt'' ~\langle 0| ~H_I(t')
~\zeta_I^4 ~H_I(t'') ~|0 \rangle
\nonumber \\
&-& \int_{t_0}^t dt' \int_{t_0}^{t'} dt'' ~\langle 0| ~H_I(t'')
~H_I(t') ~\zeta_I^4 ~|0\rangle
\nonumber \\
&-& \int_{t_0}^t dt' \int_{t_0}^{t'} dt'' ~\langle 0| ~\zeta_I^4
~H_I(t') ~H_I(t'') ~|0\rangle ~.
\label{4pt3terms}
\eea
Here $t$ is
a time several efolds after the modes exit the horizon and $t_0$ is
a time when modes are all well within the horizon. In terms of the
conformal time $\tau$, $dt=a(\tau) d\tau$,
we take $\tau=0$ and $\tau_0=-\infty$.

We evaluate (\ref{4pt3terms}) using the standard technique of normal
ordering. We decompose (omitting the subscript ``I'' for $\zeta$ in
the following)
\bea \zeta (\bk,\tau) = \zeta^+ + \zeta^- = u(\bk,\tau) a_\bk +
u^*(-\bk,\tau) a^\dagger_{-\bk} ~, \eea where \bea u(\bk,\tau) =
\frac{H}{\sqrt{4\epsilon c_s k^3}} (1+i k c_s \tau)e^{-i k c_s\tau}
~, \eea
and
\bea [a_\bk, a^\dagger_\bp] = (2\pi)^3 \delta^3(\bk-\bp)
~. \eea
After normal ordering, the only terms that are non-vanishing
are those with all terms contracted. A contraction between the two
terms, $\zeta(\bk,\tau')$ (on the left) and $\zeta(\bp,\tau'')$ (on
the right), gives
\bea
[\zeta^+(\bk,\tau'), \zeta^-(\bp,\tau'')] = u(\bk,\tau')
u^*(-\bp,\tau'') (2\pi)^3 \delta^3(\bk+\bp) ~.
\eea
We sum over all possible
contractions that represent the Feynman
diagram Fig.~\ref{Fdiagrams} (B), where the four external
legs are connected to $\zeta(\bk_i,t)$'s.

To give an example,
we look at the 1st term of (\ref{4pt3terms}) with the
component (\ref{H_a}). One example of such contractions is
\bea
&&
\contraction{\dot\zeta(} {\bp_1 }{ ,t') \dot\zeta(\bp_2,t')
\dot\zeta(\bp_3,t') \zeta( } { \bk_1 }
\contraction[2ex]{\dot\zeta(\bp_1,t') \dot\zeta(} {\bp_2} {,t')
\dot\zeta(\bp_3,t') \zeta(\bk_1,t) \zeta(} {\bk_2}
\contraction[1ex]{\dot\zeta(\bp_1,t') \dot\zeta(\bp_2,t')
\dot\zeta(\bp_3,t') \zeta(\bk_1,t) \zeta(\bk_2,t) \zeta(} {\bk_3}
{,t) \zeta(\bk_4,t) \dot\zeta(} {\bq_1}
\contraction[2ex]{\dot\zeta(\bp_1,t') \dot\zeta(\bp_2,t')
\dot\zeta(\bp_3,t') \zeta(\bk_1,t) \zeta(\bk_2,t) \zeta(\bk_3,t)
\zeta(} {\bk_4} {,t) \dot\zeta(\bq_1,t'') \dot\zeta(} {\bq_2}
\bcontraction[1ex]{\dot\zeta(\bp_1,t') \dot\zeta(\bp_2,t')
\dot\zeta(} {\bp_3} {,t') \zeta(\bk_1,t) \zeta(\bk_2,t) \zeta(\bk_3,t)
\zeta(\bk_4,t) \dot\zeta(\bq_1,t'') \dot\zeta(\bq_2,t'')
\dot\zeta(} {\bq_3}
\dot\zeta(\bp_1,t') \dot\zeta(\bp_2,t')
\dot\zeta(\bp_3,t') \zeta(\bk_1,t) \zeta(\bk_2,t) \zeta(\bk_3,t)
\zeta(\bk_4,t) \dot\zeta(\bq_1,t'') \dot\zeta(\bq_2,t'')
\dot\zeta(\bq_3,t'')
\nonumber\\
&=&
[\dot\zeta^+(\bp_1,t'), \zeta^-(\bk_1,t)]
[\dot\zeta^+(\bp_2,t'), \zeta^-(\bk_2,t)]
[\zeta^+(\bk_3,t), \dot\zeta^-(\bq_1,t'')]
[\zeta^+(\bk_4,t), \dot\zeta^-(\bq_2,t'')]
\nonumber \\
&&[\dot\zeta^+(\bp_3,t'), \dot\zeta^-(\bq_3,t'')] ~.
\eea
There are three ways of picking two of the three $\bp_i$'s ($\bq_i$'s),
so we have a symmetry factor 9. Also, there are 24 permutations of the
$\bk_i$'s.
The overall contribution
to the correlation function is
\footnote{The integrations
  are conveniently done in terms of the conformal time $\tau$.
Integrals such as
$\int_{-\infty}^0 dx~ x^2 e^{\pm ix} = \pm 2i$ are constantly used
in the evaluation in this paper. As in \cite{Maldacena:2002vr}, the
convergence at $x\to -\infty$ is achieved by $x\to x(1\mp
i\varepsilon)$.}
\bea
&& 9\cdot 4\frac{\lambda^2}{H^6} ~u_{k_1}^*(t)
u_{k_2}^*(t) u_{k_3}(t) u_{k_4}(t)
\nonumber \\
&\times& \left[ \int_{t_0}^t dt' \int_{t_0}^t dt'' a^3(t') a^3(t'')
\dot u_{k_1}(t') \dot u_{k_2}(t') \dot u_{k_3}^*(t'') \dot
u_{k_4}^*(t'') \dot u_{k_{12}}(t') \dot u_{k_{12}}^*(t'') \right]
\nonumber \\
&\times& (2\pi)^3 \delta^3(\sum_{i=1}^4 \bk_i) + {\rm 23~ perm.}
\nonumber \\
&=& \frac{9}{8} \left( \frac{\lambda}{\Sigma} \right)^2
\frac{k_{12}}{k_1k_2k_3k_4} \frac{1}{(k_1+k_2+k_{12})^3}
\frac{1}{(k_3+k_4+k_{12})^3}
\nonumber \\
&\times& (2\pi)^9 P_\zeta^3 \delta^3(\sum_{i=1}^4 \bk_i) + {\rm 23~
perm.} ~,
\label{aa1}
\eea
where
\bea
\bk_{12} = \bk_1 + \bk_2 ~.
\eea

The 2nd and 3rd term in (\ref{4pt3terms}) has a time-ordered double
integration, and so is more complicated. Their integrands are complex
conjugate to each other, and we get
\bea
&& 2\cdot \frac{9}{8}
\left( \frac{\lambda}{\Sigma} \right)^2 \frac{k_{12}}{k_1k_2k_3k_4}
\frac{1}{(k_1+k_2+k_{12})^3}
\nonumber \\
&\times& \left[ 6 \frac{(k_1+k_2+k_{12})^2}{K^5} +
  3\frac{k_1+k_2+k_{12}}{K^4} +
  \frac{1}{K^3} \right]
\nonumber \\
&\times& (2\pi)^9 P_\zeta^3 \delta^3(\sum_{i=1}^4 \bk_i) + {\rm 23~
perm.} ~,
\label{aa2}
\eea
where
\bea K=k_1+k_2+k_3+k_4 ~. \eea

The other terms are similarly computed. We leave the details to
Appendix \ref{AppSEDetail}.

\subsection{Contact-interaction diagram}
\label{SecCI}

In this subsection, we compute the contact-interaction diagram,
Fig.~\ref{Fdiagrams} (A).
We define
\bea
\mu \equiv \half X^2 P_{,XX} + 2X^3 P_{,XXX} + \frac{2}{3} X^4
P_{,XXXX} ~.
\eea
The fourth order expansion is \cite{Huang:2006eh}
\bea
\CL_4 = a^3 \frac{\mu}{H^4} \dot \zeta^4 - \frac{a}{H^4}
\left(3\lambda-\Sigma(1-c_s^2)\right) (\partial \zeta)^2 \dot
\zeta^2
+ \frac{1}{4aH^4} \Sigma (1-c_s^2) (\partial \zeta)^4 ~.
\label{CL4}
\eea

Generally speaking, the Lagrangian of the form
\bea
\CL_2 &=& f_0 \dot \zeta^2 + j_2 ~,
\\
\CL_3 &=& g_0 \dot \zeta^3 + g_1 \dot \zeta^2 + g_2 \dot \zeta + j_3
~,
\\
\CL_4 &=& h_0 \dot \zeta^4 + h_1 \dot \zeta^3 + h_2 \dot \zeta^2 + h_3
\dot \zeta + j_4 ~
\eea
gives the following interaction Hamiltonian at the fourth order in
$\dot \zeta_I$ \cite{Huang:2006eh},
\bea
\CH_4^{I} &=& \left( \frac{9g_0^2}{4f_0} - h_0 \right) \dot\zeta_I^4
+ \left( \frac{3g_0g_1}{f_0} - h_1 \right) \dot \zeta_I^3
\cr
&+& \left( \frac{3g_0g_2}{2f_0} + \frac{g_1^2}{f_0} -
h_2 \right) \dot\zeta_I^2
+ \left(\frac{g_1g_2}{f_0} - h_3 \right) \dot \zeta_I
+ \frac{g_2^2}{4f_0} - j_4 ~,
\eea
where $f$, $g$, $h$ and $j$'s are functions of $\zeta$, $\partial_i
\zeta$ and $t$, and the subscripts denote the orders of $\zeta$.
So for (\ref{CL4}) we have
\bea
\CH_4^{I} &=& \frac{a^3}{H^4} (-\mu + 9\frac{\lambda^2}{\Sigma})
\dot\zeta_I^4 + \frac{a}{H^4} \left( 3\lambda c_s^2 - \Sigma (1-c_s^2)
\right) (\partial \zeta_I)^2 \dot\zeta_I^2
\cr
&+& \frac{1}{4aH^4} \Sigma (-c_s^2 + c_s^4) (\partial \zeta_I)^4 ~.
\label{CH4}
\eea
Note that in the second term the order $\lambda$ term
cancelled, in the third term the order $\Sigma$ term
cancelled.

The following are the contributions to the form factor $\CT$ defined
as
\bea
\langle \zeta^4 \rangle =
(2\pi)^9 P_\zeta^3 \delta^3 (\sum_{i=1}^4 \bk_i)
~\prod_{i=1}^4 \frac{1}{k_i^3} ~\CT ~.
\eea
The contribution from the first term in (\ref{CH4}) is
\bea
36 \left( \frac{\mu}{\Sigma} - \frac{9 \lambda^2}{\Sigma^2}
\right)
\frac{\prod_{i=1}^4 k_i^2}{K^5} ~;
\label{CT_c1}
\eea
from the second term,
\bea
-\frac{1}{8} \left(\frac{3\lambda}{\Sigma} -
\frac{1}{c_s^2} +1 \right)
\frac{k_1^2k_2^2 (\bk_3 \cdot \bk_4)}{K^3}
\left[ 1+ \frac{3(k_3+k_4)}{K} + \frac{12k_3k_4}{K^2} \right]
+ {\rm 23~perm.} ~;
\eea
from the third term,
\bea
&&\frac{1}{32} \left( \frac{1}{c_s^2} -1 \right)
\frac{(\bk_1 \cdot \bk_2)(\bk_3 \cdot \bk_4)}{K}
\left[ 1+ \frac{\sum_{i<j} k_i k_j}{K^2}
+ \frac{3k_1k_2k_3k_4}{K^3} (\sum_{i=1}^4 \frac{1}{k_i} )
+ 12 \frac{k_1k_2k_3k_4}{K^4} \right]
\cr
&&+ ~{\rm 23 ~ perm.} ~.
\label{CT_c3}
\eea

\subsection{Summary of final results}

Here we summarize the final results from Sec.~\ref{SecSE}, \ref{SecCI}
and Appendix \ref{AppSEDetail}.
For the general single field inflation
$\CL({\phi,X})$, we define
\bea
c_s^2 &\equiv& \frac{P_{,X}}{P_{,X}+2X P_{,XX}} ~,
\cr
\Sigma &\equiv& X P_{,X} + 2X^2 P_{,XX} ~,
\cr
\lambda &\equiv& X^2 P_{,XX} + \frac{2}{3} X^3 P_{,XXX} ~,
\cr
\mu &\equiv& \half X^2 P_{,XX} + 2X^3 P_{,XXX} + \frac{2}{3} X^4
P_{,XXXX} ~.
\eea
If any of $\mu/\Sigma$, $\lambda^2/\Sigma^2$,
$1/c_s^4 \gtrsim 1$, the single field inflation generates
a large primordial trispectrum, whose
leading terms are given by
\bea
\langle \zeta^4 \rangle = (2\pi)^9
P_\zeta^3 \delta^3(\sum_{i=1}^4 \bk_i)
\prod_{i=1}^4 \frac{1}{k_i^3} ~ \CT(k_1,k_2,k_3,k_4,k_{12},k_{14}) ~,
\eea
where $\CT$ has the following six components:
\bea
\CT &=&
\left( \frac{\lambda}{\Sigma} \right)^2 T_{s1}
+ \frac{\lambda}{\Sigma} \left( \frac{1}{c_s^2}-1 \right) T_{s2}
+ \left( \frac{1}{c_s^2}-1 \right)^2 T_{s3}
+ \left( \frac{\mu}{\Sigma} - \frac{9\lambda^2}{\Sigma^2} \right)
T_{c1}
\cr
&+&
\left(\frac{3\lambda}{\Sigma} - \frac{1}{c_s^2} +1 \right) T_{c2}
+ \left( \frac{1}{c_s^2} -1 \right) T_{c3} ~.
\label{shapesum}
\eea
The $T_{s1,s2,s3}$ are contributions from the scalar-exchange
diagrams and are given in Appendix~\ref{AppSEDetail}, $T_{c1,c2,c3}$ are
contributions from the contact-interaction diagram and are given by
(\ref{CT_c1})-(\ref{CT_c3}).
For the most interesting cases, where any of
$\mu/\Sigma$, $\lambda^2/\Sigma^2$,
$1/c_s^4 \gg 1$, the first four terms in (\ref{shapesum})
are the leading contributions. So we have four shapes determined by
three parameters,\footnote{More generally,
we have six shapes controlled by three parameters. However,
the second line of (\ref{shapesum}) are negligible unless
$\mu/\Sigma$, $\lambda^2/\Sigma^2$,
$1/c_s^4$ are all $\sim 1$, in which case the trispectra is only
marginally large, $\sim \CO(1)$. Also note that, for $\mu/\Sigma$,
$\lambda^2/\Sigma^2$,
$1/c_s^4 \gg 1$, the second line of (\ref{shapesum}) does not capture all
the subleading contributions.} $\lambda/\Sigma$,
$1/c_s^2$ and $\mu/\Sigma$.
A large bispectrum necessarily implies a
large
trispectrum, because either $1/c_s^4$ or $(\lambda/\Sigma)^2$ is
large. But the reverse is not necessarily true. One can in principle
have a large $\mu/\Sigma$ but small $1/c_s^4$ and
$(\lambda/\Sigma)^2$.

To quantify the size (i.e., magnitude) of the non-Gaussianity for
each shape, we define the following estimator $t_{NL}$ for each
shape component,
\bea
\langle \zeta^4 \rangle_{\rm component}
\xrightarrow[\rm limit]{\rm RT} (2\pi)^9 P_\zeta^3 \delta^3 (\sum_i
\bk_i) \frac{1}{k^9} ~ t_{NL} ~, \label{tNLdef}
\eea
where the RT
limit stands for the regular tetrahedron limit
($k_1=k_2=k_3=k_4=k_{12}=k_{14}\equiv k$). The parameter $t_{NL}$ is
analogous to the $f_{NL}$ parameter for bispectra.
This definition applies to
both the cases of interest here, and the non-Gaussianities of the
local form that we will discuss shortly. Unlike the convention in
the bispectrum case where the normalization of $f_{NL}$ is chosen
according to the local form non-Gaussianity, here we conveniently
choose the normalization of $t_{NL}$ according to (\ref{tNLdef}).
This is because, for the trispectra, even the local form has two
different shapes.

The size of non-Gaussianity for
each shape in (\ref{shapesum}) is then given by
\bea
t_{NL}^{s1} = 0.250 \left(\frac{\lambda}{\Sigma} \right)^2 ,
~~
t_{NL}^{s2} = 0.420 \frac{\lambda}{\Sigma}
\left(\frac{1}{c_2^2}-1\right) ,
~~
t_{NL}^{s3} = 0.305 \left( \frac{1}{c_s^2}-1 \right)^2 ,
\cr
t_{NL}^{c1} = 0.0352 \left(
\frac{\mu}{\Sigma}-\frac{9\lambda^2}{\Sigma^2} \right) ,
~~
t_{NL}^{c2} = 0.0508 \left(
\frac{3\lambda}{\Sigma}-\frac{1}{c_s^2}+1\right) ,
~~
t_{NL}^{c3} = 0.0503 \left(\frac{1}{c_s^2}-1\right) ~.
\eea

For comparison, let us also look at the trispectrum of the local
form. This is obtained from the ansatz in real space
\cite{Okamoto:2002ik,Kogo:2006kh},
\bea
\zeta (\bx)
= \zeta_g + \frac{3}{5} f_{NL} \left( \zeta_g^2-\langle \zeta_g^2
\rangle \right) + \frac{9}{25} g_{NL} \left( \zeta_g^3 -3\langle
\zeta_g^2 \rangle \zeta_g \right) ~,
\eea
where $\zeta_g$ is Gaussian and the shifts in the 2nd and 3rd
terms are introduced to cancel the disconnected diagrams.
Such a form constantly arises in multi-field models, where
the large non-Gaussianities are converted from
isocurvature modes at super-horizon scales.
The resulting trispectrum is
\bea
\CT = f_{NL}^2 T_{loc1} + g_{NL} T_{loc2} ~.
\eea
The two shapes are
\bea
T_{loc1} &=& \frac{9}{50} \left( \frac{k_1^3 k_2^3}{k_{13}^3} + {\rm
  11~perm.} \right) ~,
\label{Tloc1}
\\
T_{loc2} &=&
 \frac{27}{100} \sum_{i=1}^4 k_i^3 ~,
\label{Tloc2}
\eea
where the 11 permutations includes $k_{13} \to k_{14}$ and 6 choices of
picking two momenta such as $k_1$ and $k_2$.
The size of the trispectrum for each shape is
\bea
t_{NL}^{loc1} = 2.16 f_{NL}^2 ~, ~~~
t_{NL}^{loc2} = 1.08 g_{NL} ~.
\eea
So again a large bispectrum implies a large trispectrum, but not
reversely.

\section{Shapes of trispectra}
\setcounter{equation}{0}
\label{shapesection}

In this section, we investigate the shape of the trispectra. We take
various limits of the shape functions $T_{s1}$, $T_{s2}$, $T_{s3}$ and
$T_{c1}$, and then compare among themselves, and with the local
shapes $T_{loc1}$ and $T_{loc2}$. We will summarize the main results
at the end of this section.

\begin{figure}
  \center
  \includegraphics[width=0.6\textwidth]{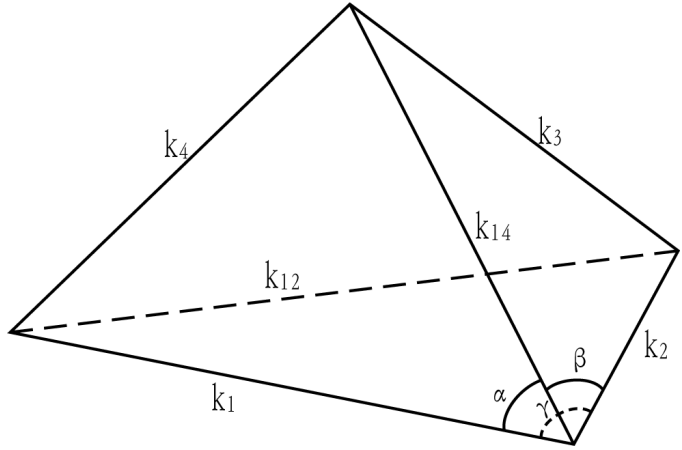}
    \caption{\label{tFig} This figure illustrates the tetrahedron we
      consider. }
\end{figure}

Before the discussion of the shape functions, we note that the
arguments of the shape functions are six momenta
$k_1,k_2,k_3,k_4,k_{12},k_{14}$. In order for these momenta to form a
tetrahedron (as in Fig. \ref{tFig}), the following two conditions are
required:

Firstly, we define three angles at one vertex:
\begin{align}
  \cos(\alpha) &= \frac{k_1^2+k_{14}^2-k_4^2}{2k_1k_{14}}~, \nonumber \\
\cos(\beta) &= \frac{k_2^2+k_{14}^2-k_3^2}{2k_2k_{14}}~, \nonumber \\
\cos(\gamma) &= \frac{k_1^2+k_{2}^2-k_{12}^2}{2k_1k_{2}}~.
\end{align}
These three angles should satisfy $\cos(\alpha-\beta)\geq \cos(\gamma)
\geq \cos(\alpha+\beta)$. This inequality is equivalent to
\begin{align}\label{cosineq}
  1-\cos^2(\alpha)-\cos^2(\beta)-\cos^2(\gamma)+2\cos(\alpha)\cos(\beta)\cos(\gamma) \geq 0~.
\end{align}

Secondly, the four momenta should satisfy all the triangle
inequalities. We need
\begin{align}
  \label{ineq}
  k_1+k_4>k_{14}~,\quad   k_1+k_2>k_{12}~,\quad
  k_2+k_3>k_{14}~,\nonumber\\
  k_1+k_{14}>k_4~,\quad   k_1+k_{12}>k_2~,\quad
  k_2+k_{14}>k_3~,\nonumber\\
  k_4+k_{14}>k_1~,\quad   k_2+k_{12}>k_1~,\quad
  k_3+k_{14}>k_2~.
\end{align}
The last triangle inequality involving $(k_3,k_4,k_{12})$ is always
satisfied given Eq. \eqref{cosineq} and Eqs. \eqref{ineq}.

We also would like to mention a symmetry in our trispectrum.
As $k_1,
k_2,k_3,k_4$ are symmetric in our model,
we have
\begin{align}\label{symeq}
  {\cal T}(k_1,k_2,k_3,k_4,k_{12},k_{14})=  {\cal
    T}(k_1,k_2,k_4,k_3,k_{12},k_{13}) =  {\cal
    T}(k_1,k_3,k_2,k_4,k_{13},k_{14}) ~,
\end{align}
and etc, where
\bea
k_{13}\equiv |{\bf k}_1+{\bf
  k}_3|=\sqrt{k_1^2+k_2^2+k_3^2+k_4^2-k_{12}^2-k_{14}^2} ~.
\label{k13Ex}
\eea

The first set of limits we would like to take is those involved in the
consistency relations. There are two known consistency relations for
the trispectra to satisfy.

Firstly, we discuss the consistency relation in the squeezed limit.
When one external momentum, say, $k_4$ goes to zero,
this mode can be treated as a classical background for the other
modes, and the
trispectrum should reduce to the product of a
power spectrum and a running of bispectrum
\cite{Seery:2006vu,Huang:2006eh,Li:2008gg}:
\begin{align}\label{consistency1}
  \langle \zeta_{{\bf k}_1}\zeta_{{\bf k}_2}\zeta_{{\bf
      k}_3}\zeta_{{\bf k}_4} \rangle
\sim -
{\mathbb P}(k_4)
\frac{d}{d\ln a}\langle \zeta_{{\bf k}_1}\zeta_{{\bf
      k}_2}\zeta_{{\bf k}_3} \rangle~,
\end{align}
where ${\mathbb P}(k)\equiv \frac{2\pi^2}{k^3} P_\zeta(k)$,
and $P_\zeta(k)$ is
the dimensionless power spectrum.

In our case, the leading order contribution to the trispectra scales
as $c_s^{-4}$, or $c_s^{-2}\lambda/\Sigma$, or $(\lambda/\Sigma)^2$,
or $\mu/\Sigma$.
However, RHS scales as $c_s^{-3}$, or $c_s^{-1}\lambda/\Sigma$. In
order that Eq. \eqref{consistency1} holds, $k_4^3\langle \zeta_{{\bf
    k}_1}\zeta_{{\bf k}_2}\zeta_{{\bf k}_3}\zeta_{{\bf k}_4} \rangle$
must vanish at the leading order in the $k_4\rightarrow 0$
limit. One can check that our result indeed vanish in this limit,
$T_{s1,2,3},T_{c1} \to \CO(k_4^2)$.

Secondly, we check the folded limit, say $k_{12} \to 0$.
For the s-channel (in which the exchanged scalar carries the momentum
$\bk_{12}$),
the four-point function can be regarded as
a pair of two-point functions modulated by the same classical
background generated by
the long wave mode $\bk_{12}$, and we have \cite{Seery:2008ax}
\begin{align}\label{consistency2}
  \langle \zeta_{{\bf k}_1}\zeta_{{\bf k}_2}\zeta_{{\bf
      k}_3}\zeta_{{\bf k}_4} \rangle
\sim (n_s-1)^2 {\mathbb P}(k_1){\mathbb P}(k_3)\langle \zeta_{-{\bf
    k}_{12}} \zeta_{{\bf k}_{12}}\rangle~.
\end{align}
Note that the RHS takes the same shape as (\ref{Tloc1}).
Again, RHS scales as $c_s^{-3}$ in our case.
So in the $k_{12}\rightarrow 0$
limit, we expect $k_{12}^3\langle \zeta_{{\bf
    k}_1}\zeta_{{\bf k}_2}\zeta_{{\bf k}_3}\zeta_{{\bf k}_4} \rangle$
to vanish for the s-channel.
For the t-, u- and the contact interaction channels, there are
neither
propagators that give rise to the pole behavior $1/k_{12}^3$ nor
inverse Laplacians in our Lagrangian. Therefore in our case
$k_{12}^3\langle \zeta_{{\bf
    k}_1}\zeta_{{\bf k}_2}\zeta_{{\bf k}_3}\zeta_{{\bf k}_4} \rangle
\to 0$
trivially for these channels.

One can check that our results indeed satisfy the condition.
In fact, we have
$\langle \zeta_{{\bf
    k}_1}\zeta_{{\bf k}_2}\zeta_{{\bf k}_3}\zeta_{{\bf k}_4} \rangle
\to \CO(k_{12})$ for the s-channel, so the pole behavior at $k_{12}=0$
is cancelled more than enough to satisfy the condition. (Note that
summing over all channels gives $\langle \zeta^4 \rangle \to {\rm
  constant}$.)

\begin{figure}
  \center
  \includegraphics[width=0.43\textwidth]{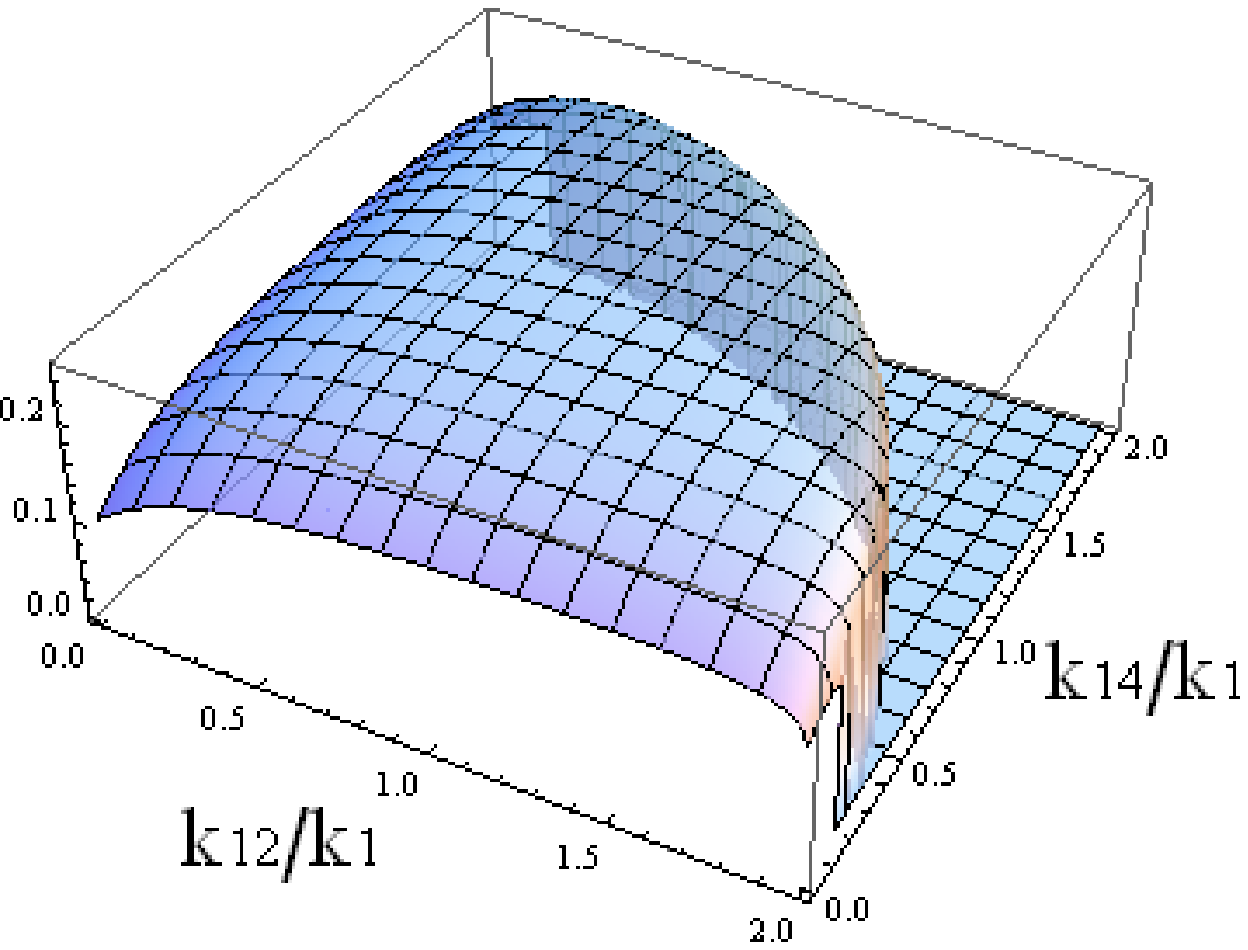}
\hspace{0.02\textwidth}
  \includegraphics[width=0.4\textwidth]{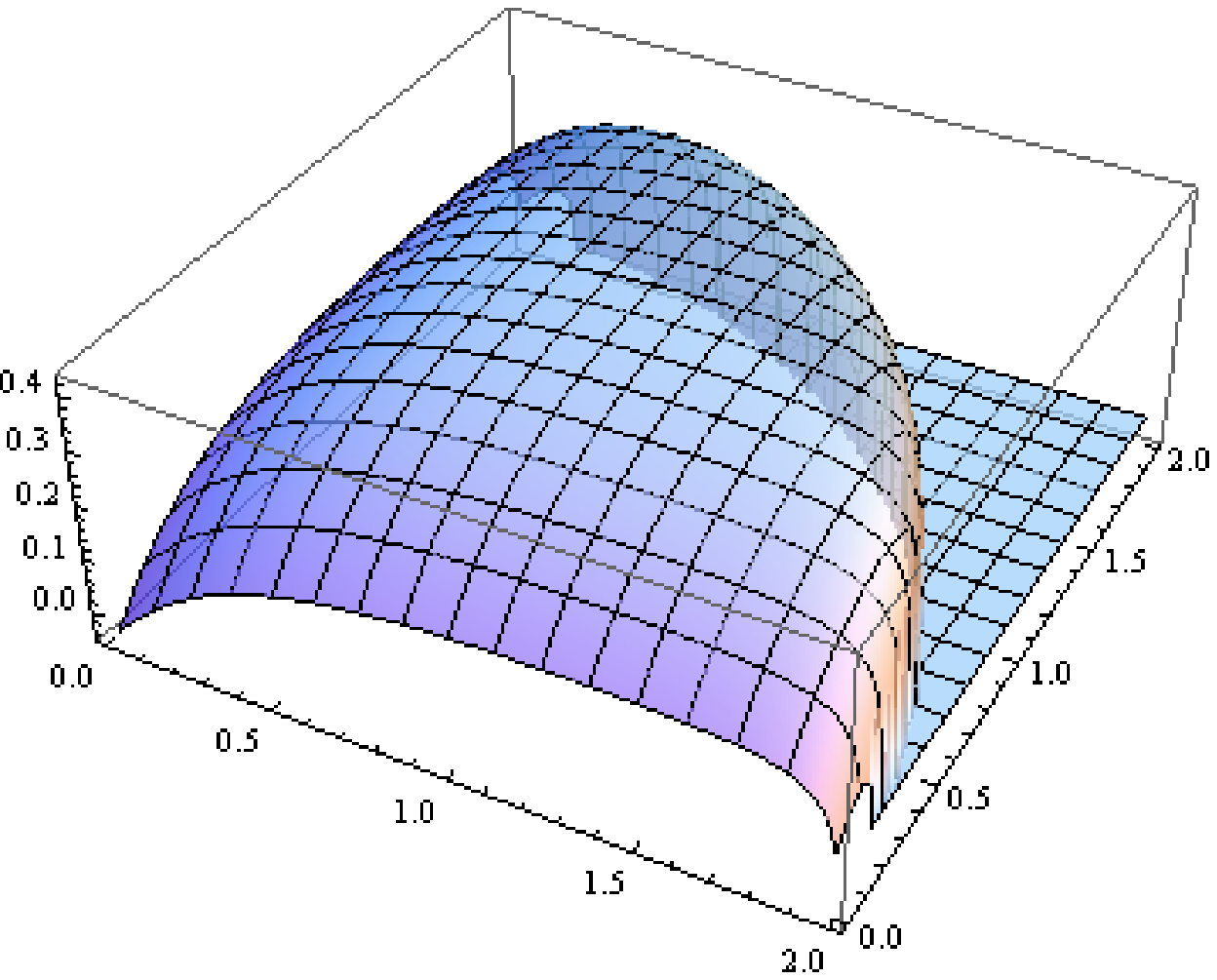}

  \includegraphics[width=0.4\textwidth]{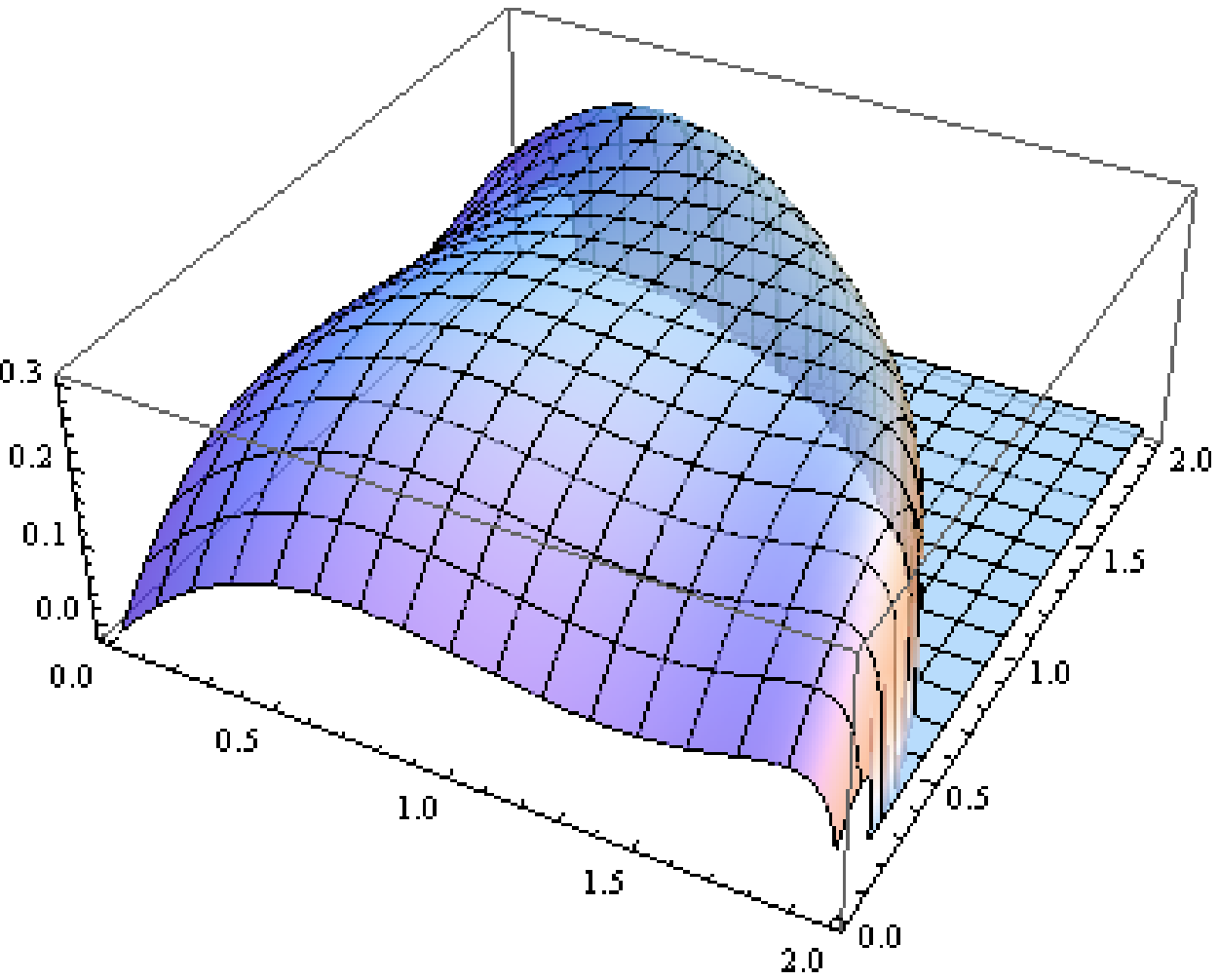}
\hspace{0.05\textwidth}
  \includegraphics[width=0.4\textwidth]{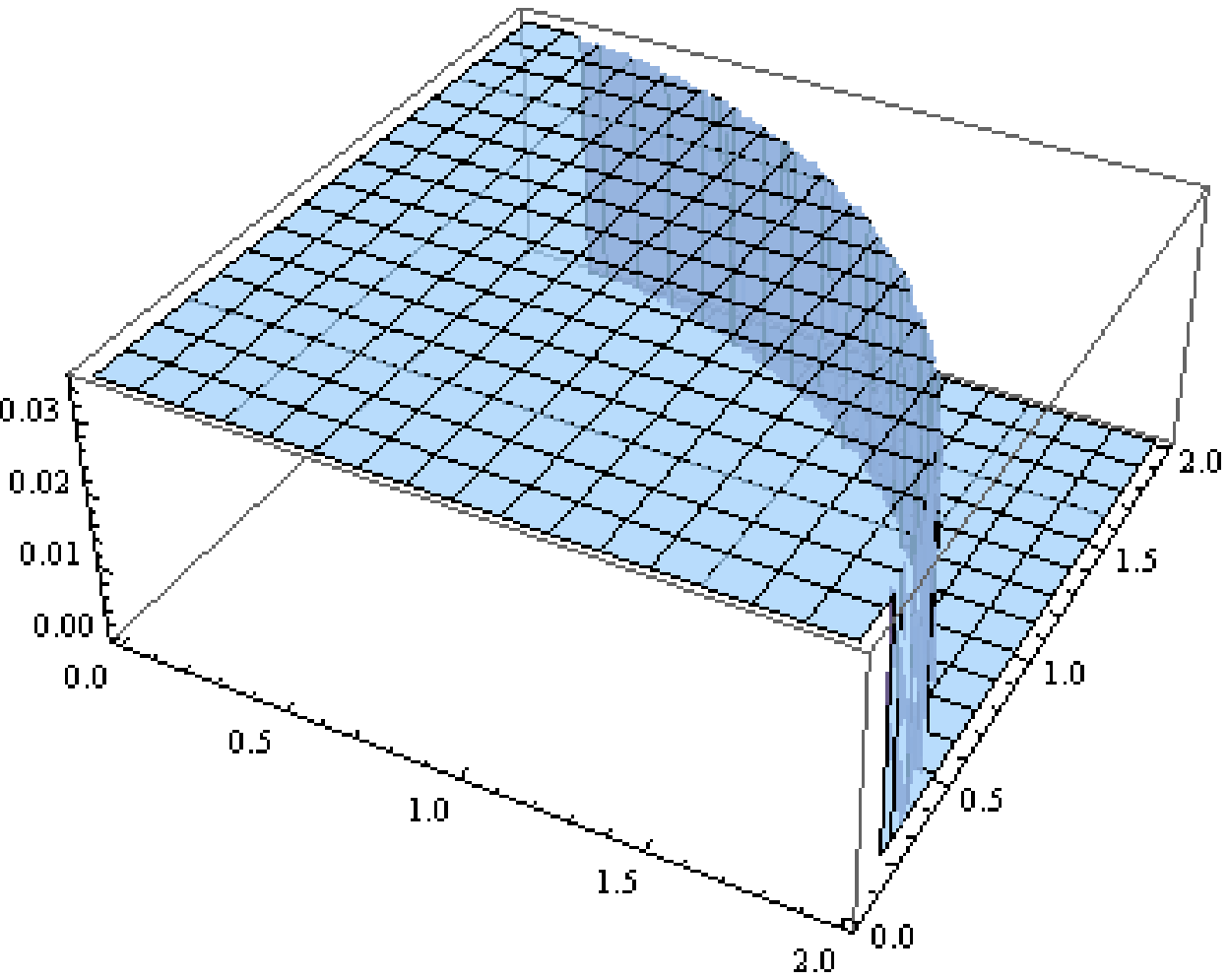}

  \includegraphics[width=0.4\textwidth]{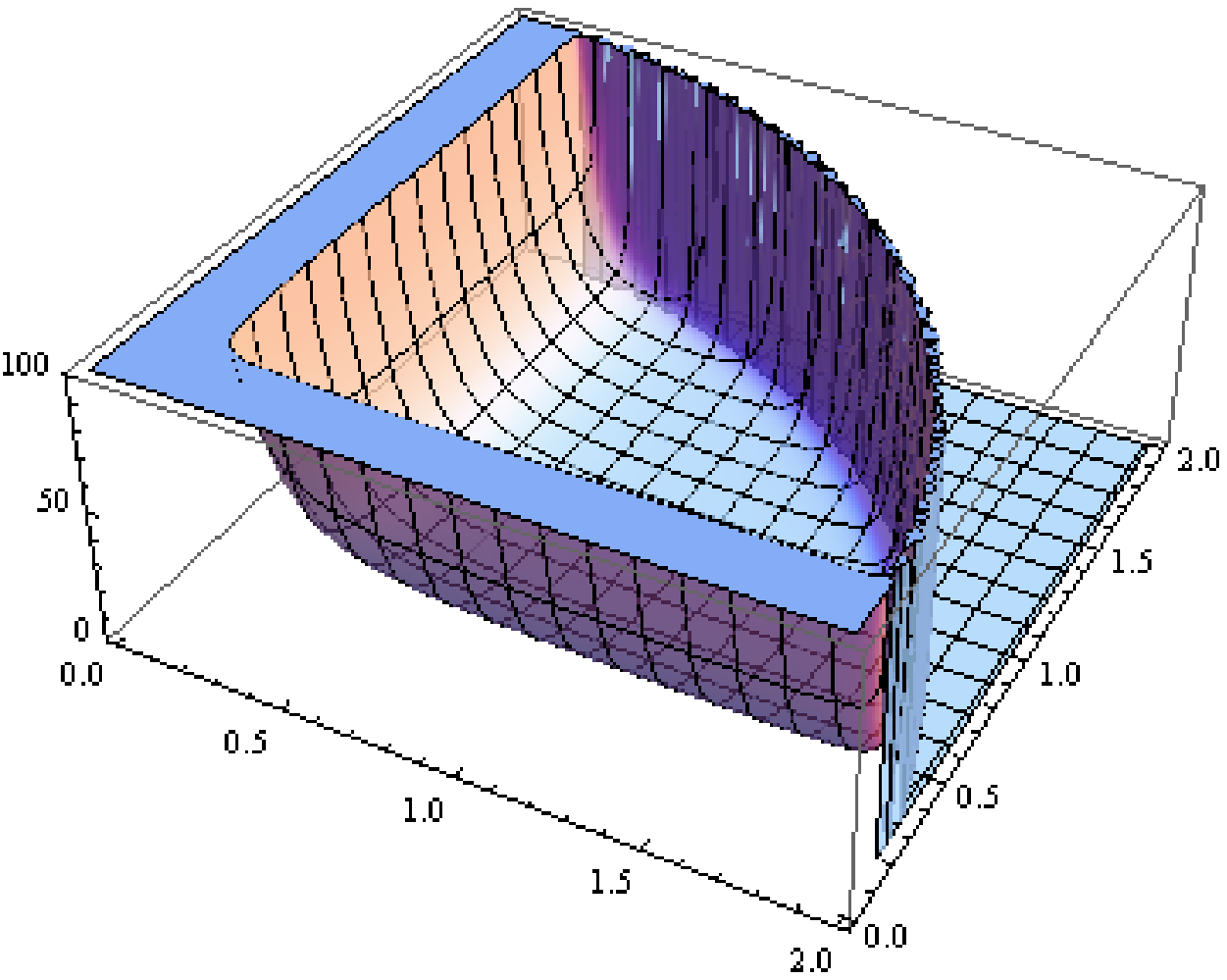}
\hspace{0.05\textwidth}
  \includegraphics[width=0.4\textwidth]{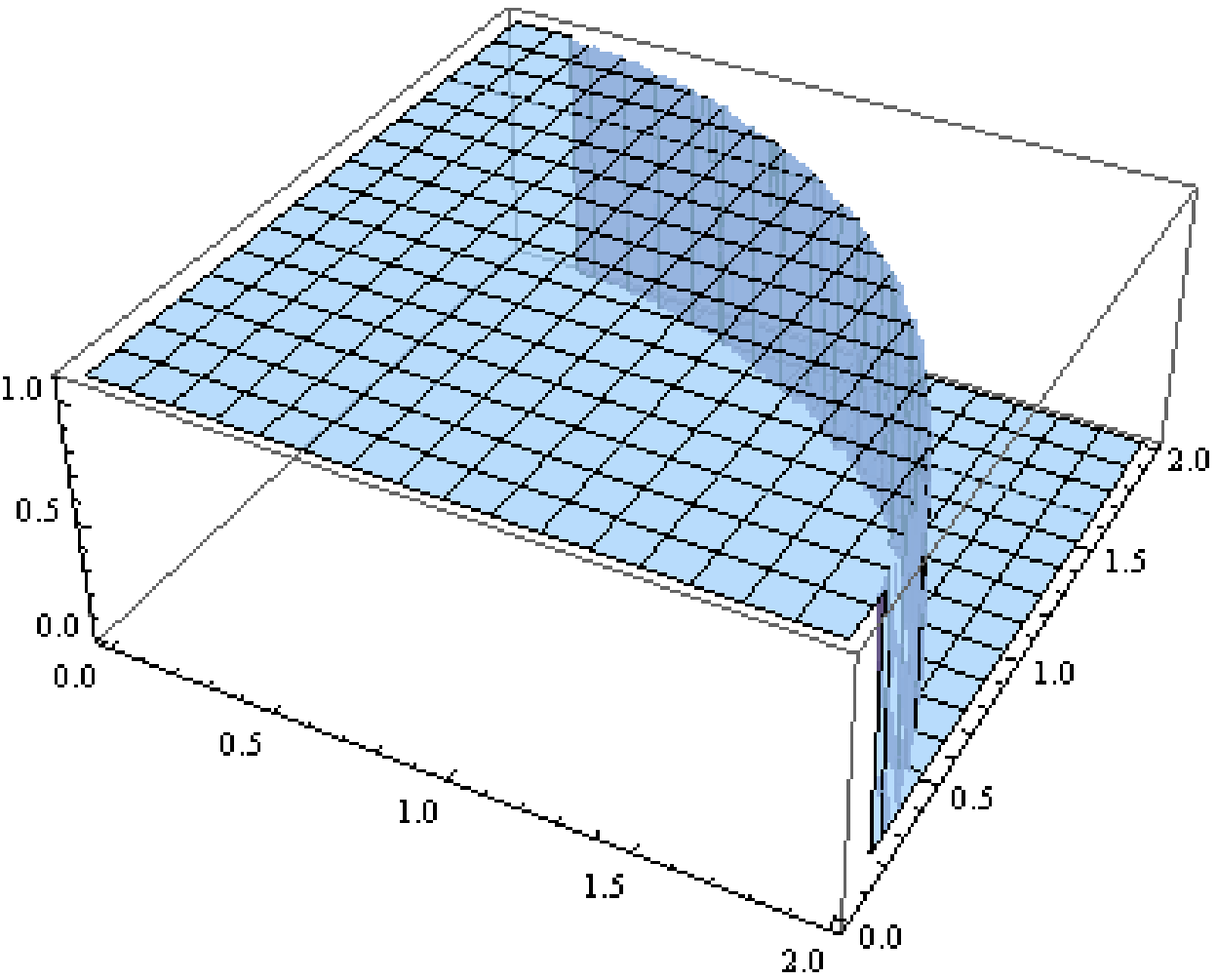}
    \caption{\label{equilateral} In this group of figures, we consider
      the equilateral limit $k_1=k_2=k_3=k_4$, and plot $T_{s1}$,
      $T_{s2}$, $T_{s3}$, $T_{c1}$, $T_{loc1}$ and $T_{loc2}$,
      respectively, as
      functions of $k_{12}/k_1$ and
      $k_{14}/k_1$. Note that
    $T_{loc1}$ blows up when $k_{12}\ll k_1$ and $k_{14} \ll
    k_1$. $T_{loc1}$ also blows up in the other boundary, because this
  boundary corresponds to $k_{13}\ll k_1$. So $T_{loc1}$ is
distinguishable from all other shapes in this limit. We also note
that $T_{c1}$ and $T_{loc2}$ are both independent of $k_{12}$ and $k_{14}$.}
\end{figure}

\begin{figure}
  \center
  \includegraphics[width=0.43\textwidth]{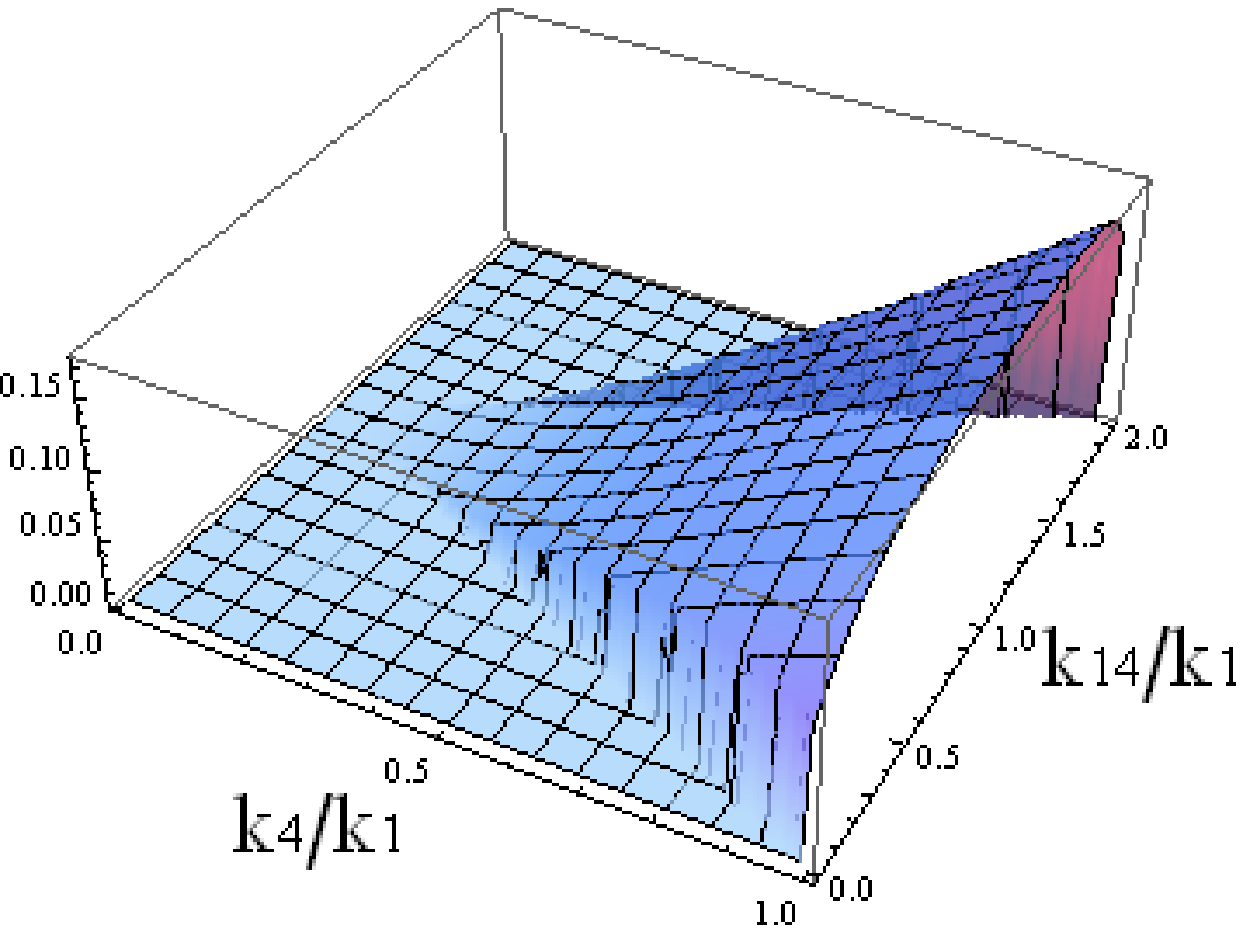}
\hspace{0.02\textwidth}
  \includegraphics[width=0.4\textwidth]{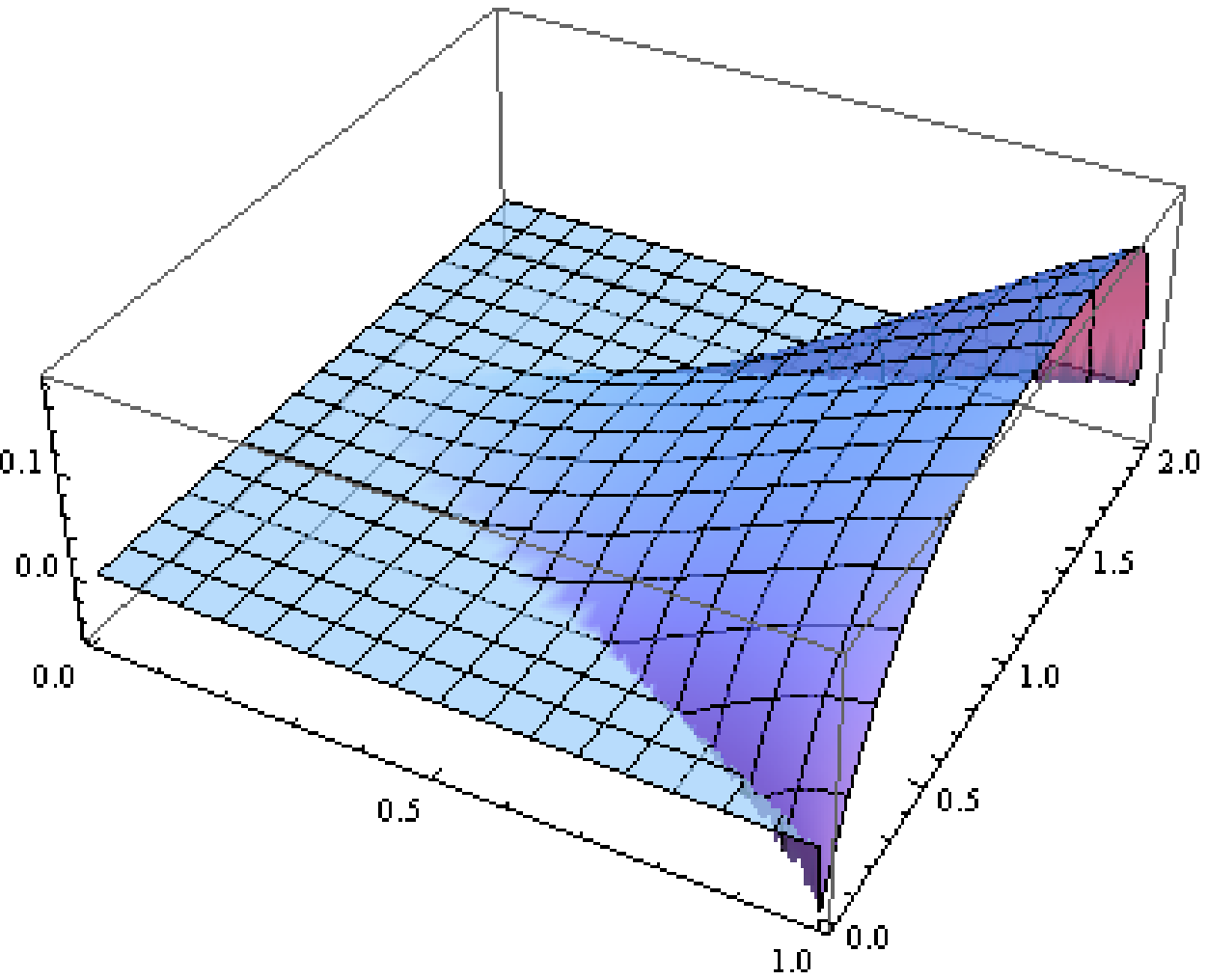}

  \includegraphics[width=0.4\textwidth]{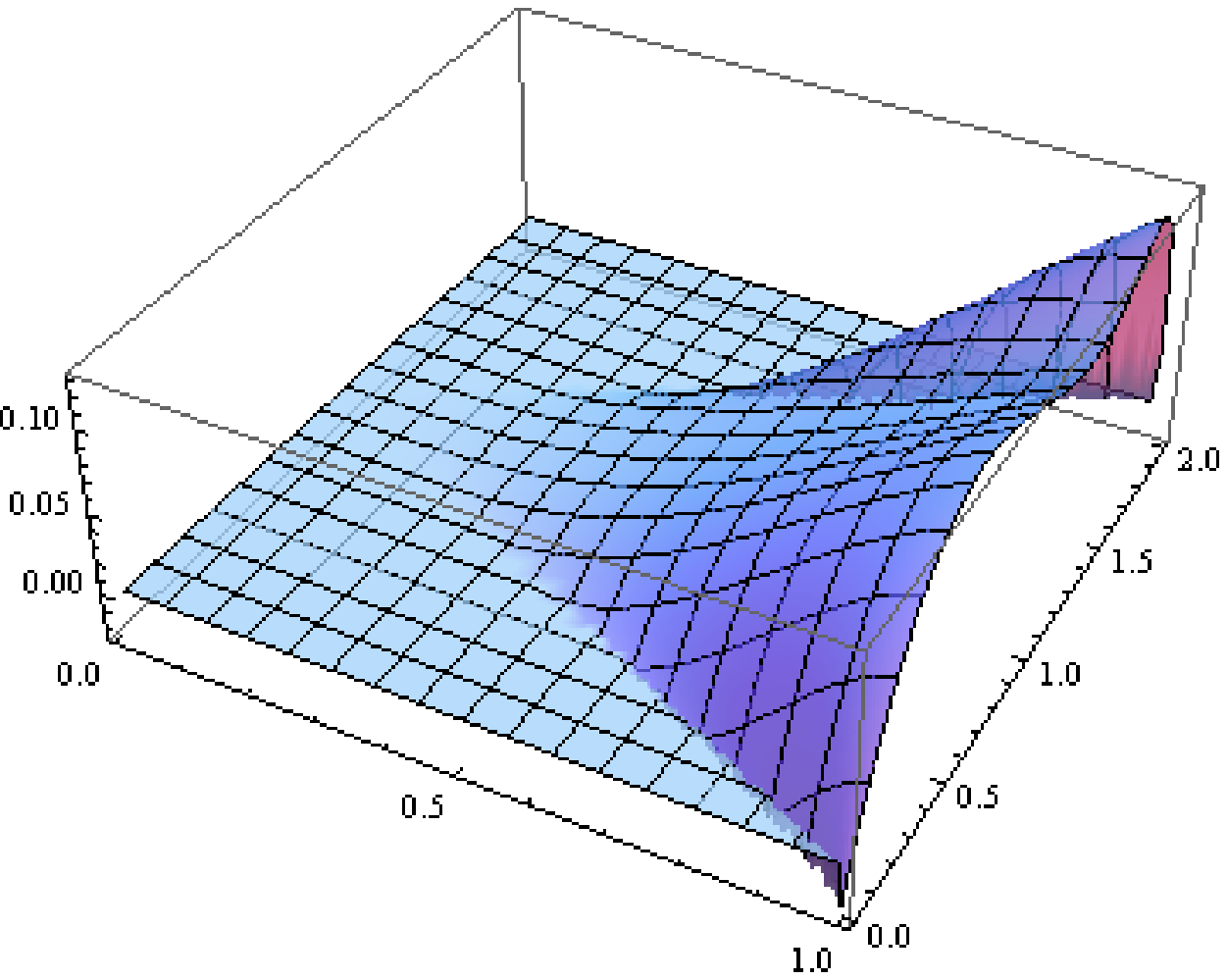}
\hspace{0.05\textwidth}
  \includegraphics[width=0.4\textwidth]{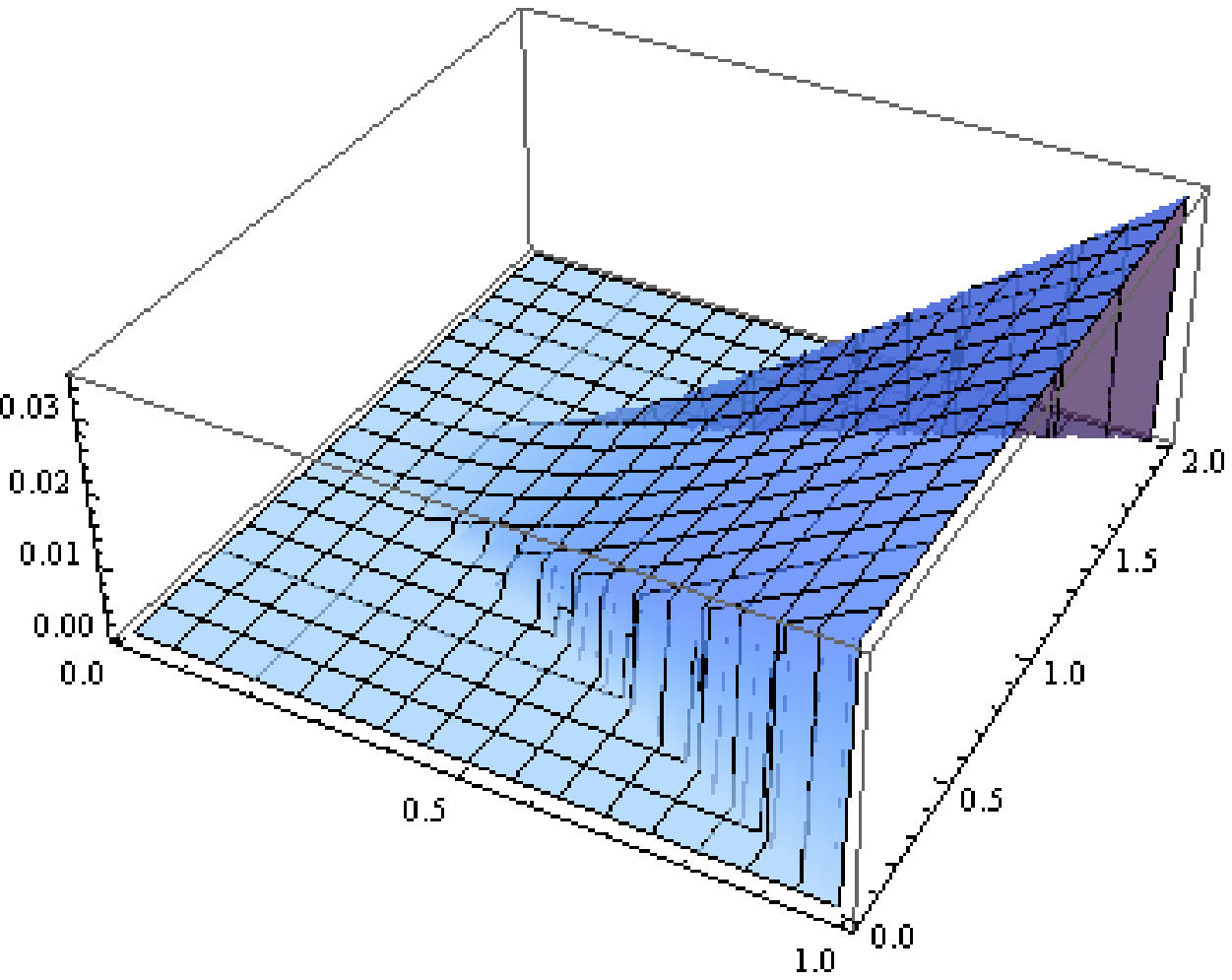}

  \includegraphics[width=0.4\textwidth]{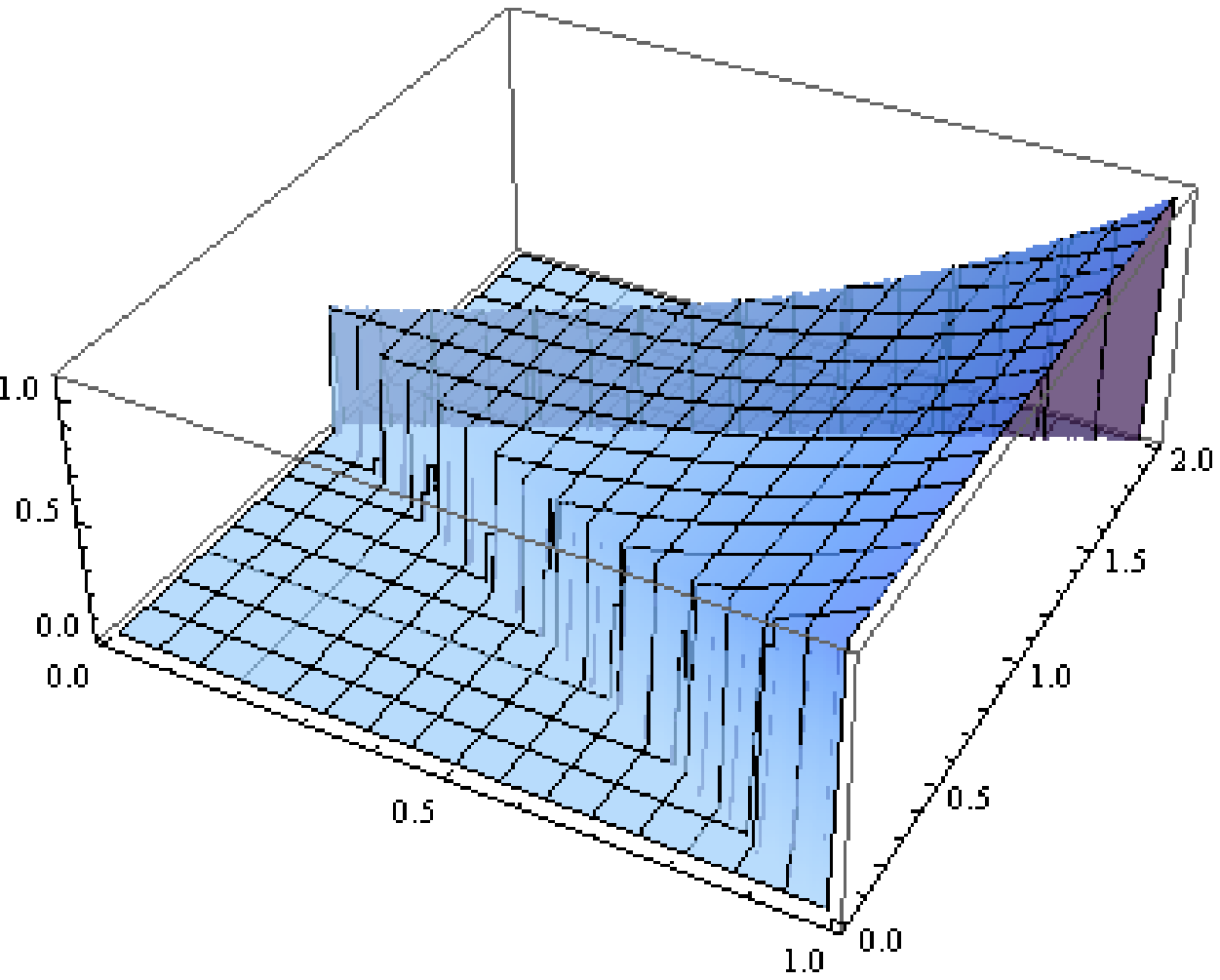}
    \caption{\label{folded} In this group of figures, we consider
      the folded limit $k_{12}=0$, and plot
      $T_{s1}$, $T_{s2}$, $T_{s3}$, $T_{c1}$ and $T_{loc2}$,
      respectively, as
      functions of $k_{14}/k_1$ and
      $k_{4}/k_1$. $T_{loc1}$
      blows up in this limit. Note that when $k_4\rightarrow 0$, all
      shape functions except $T_{loc1}$ and $T_{loc2}$ vanish.}
\end{figure}

\begin{figure}
  \center
  \includegraphics[width=0.41\textwidth]{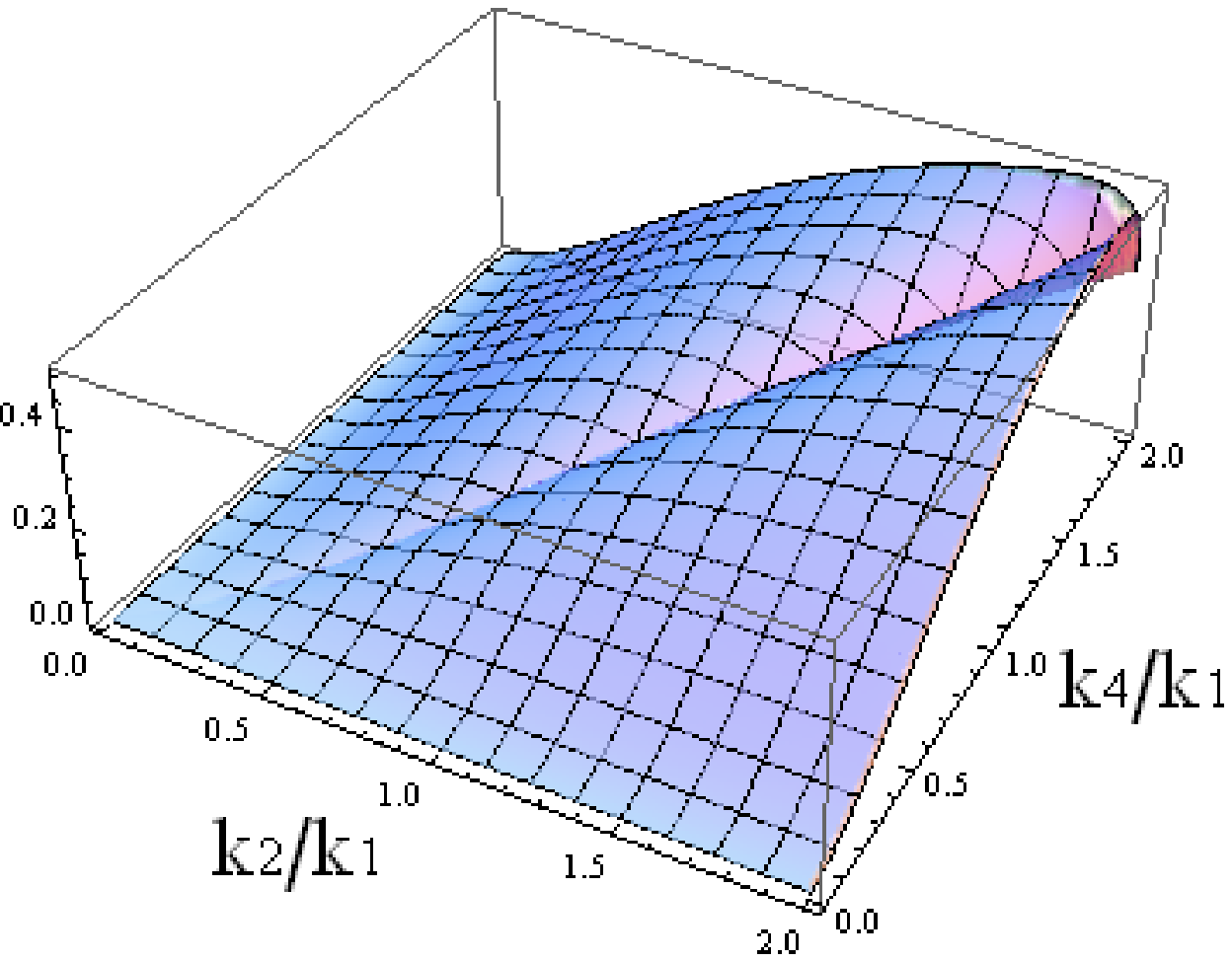}
\hspace{0.04\textwidth}
  \includegraphics[width=0.4\textwidth]{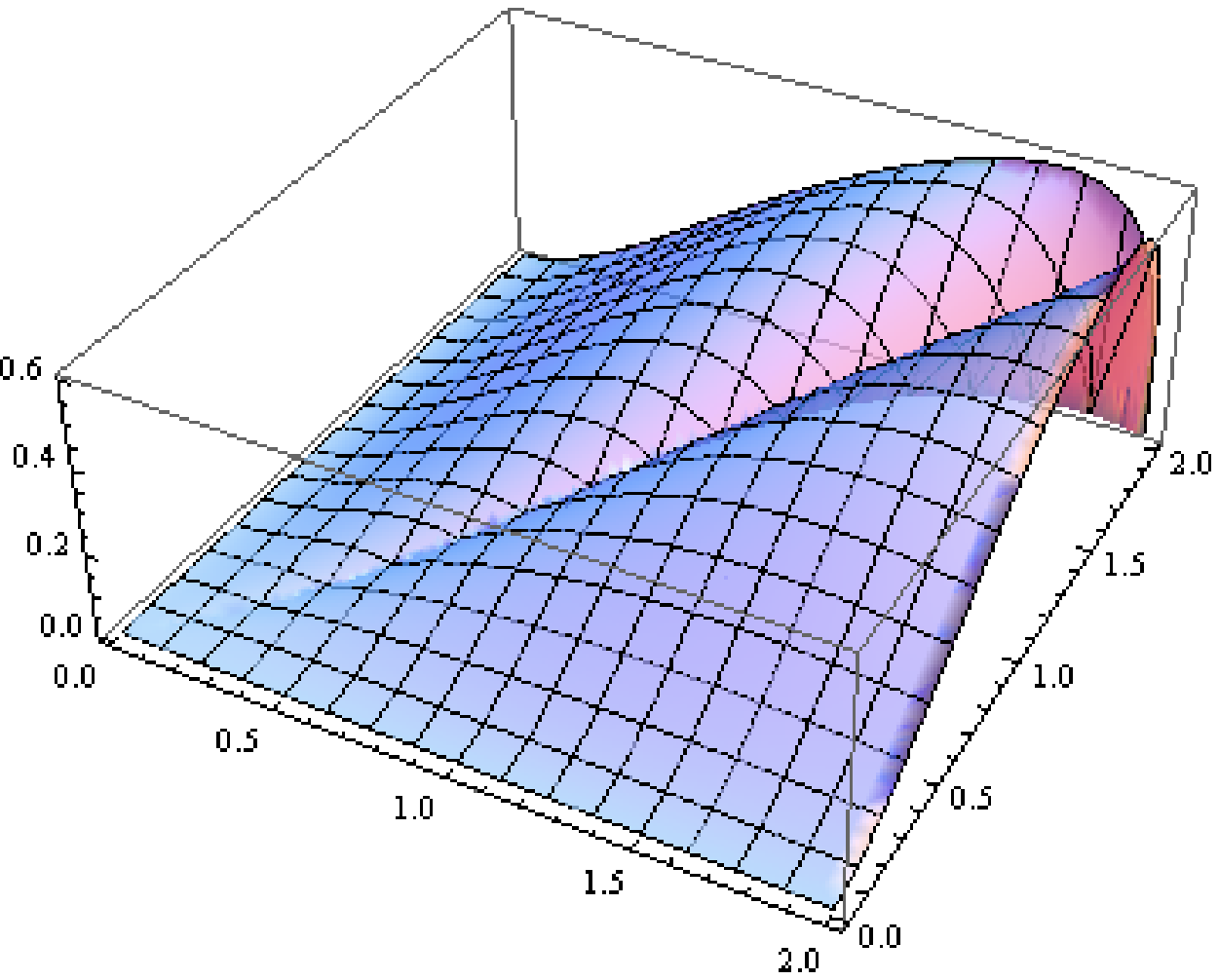}

  \includegraphics[width=0.4\textwidth]{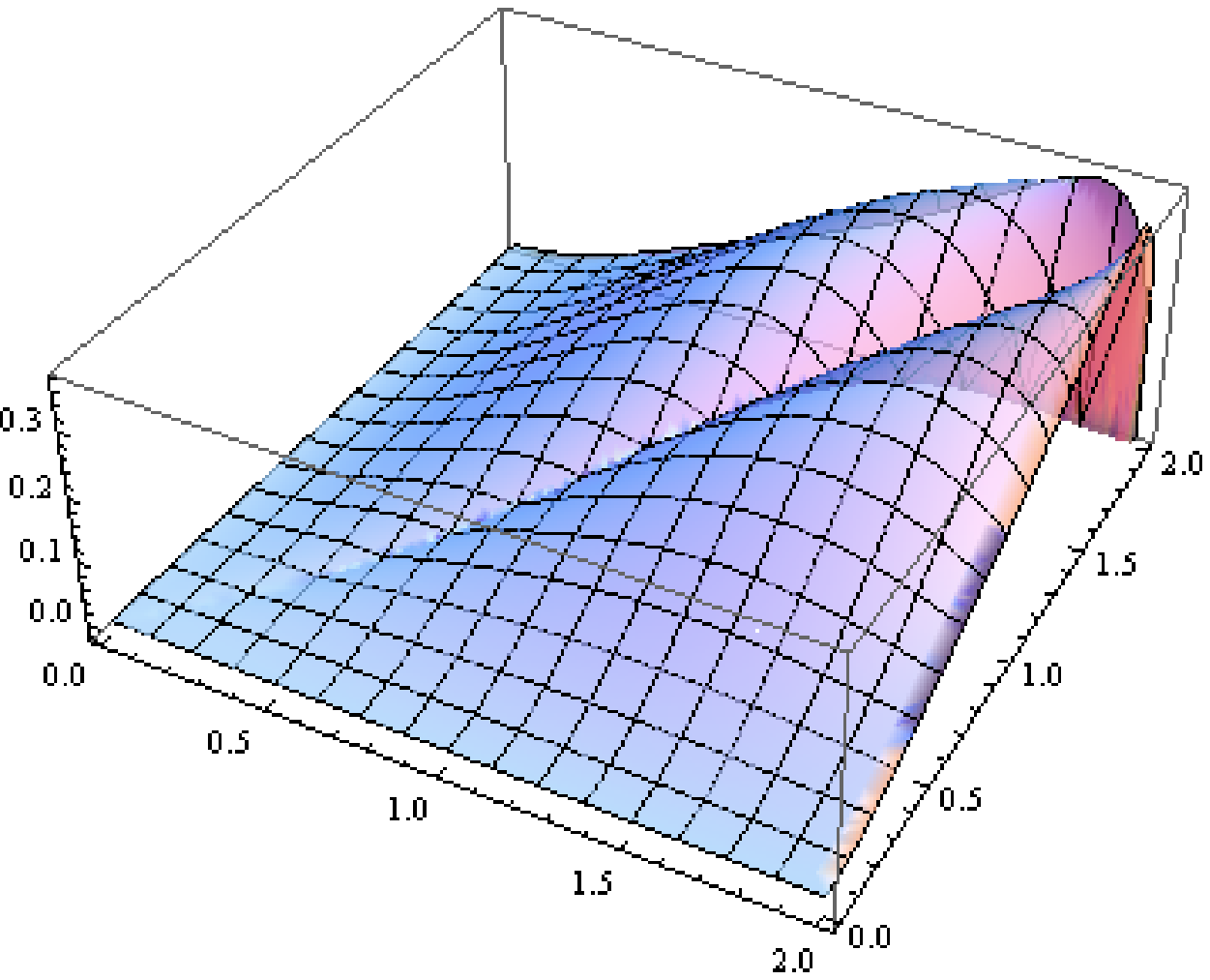}
\hspace{0.05\textwidth}
  \includegraphics[width=0.4\textwidth]{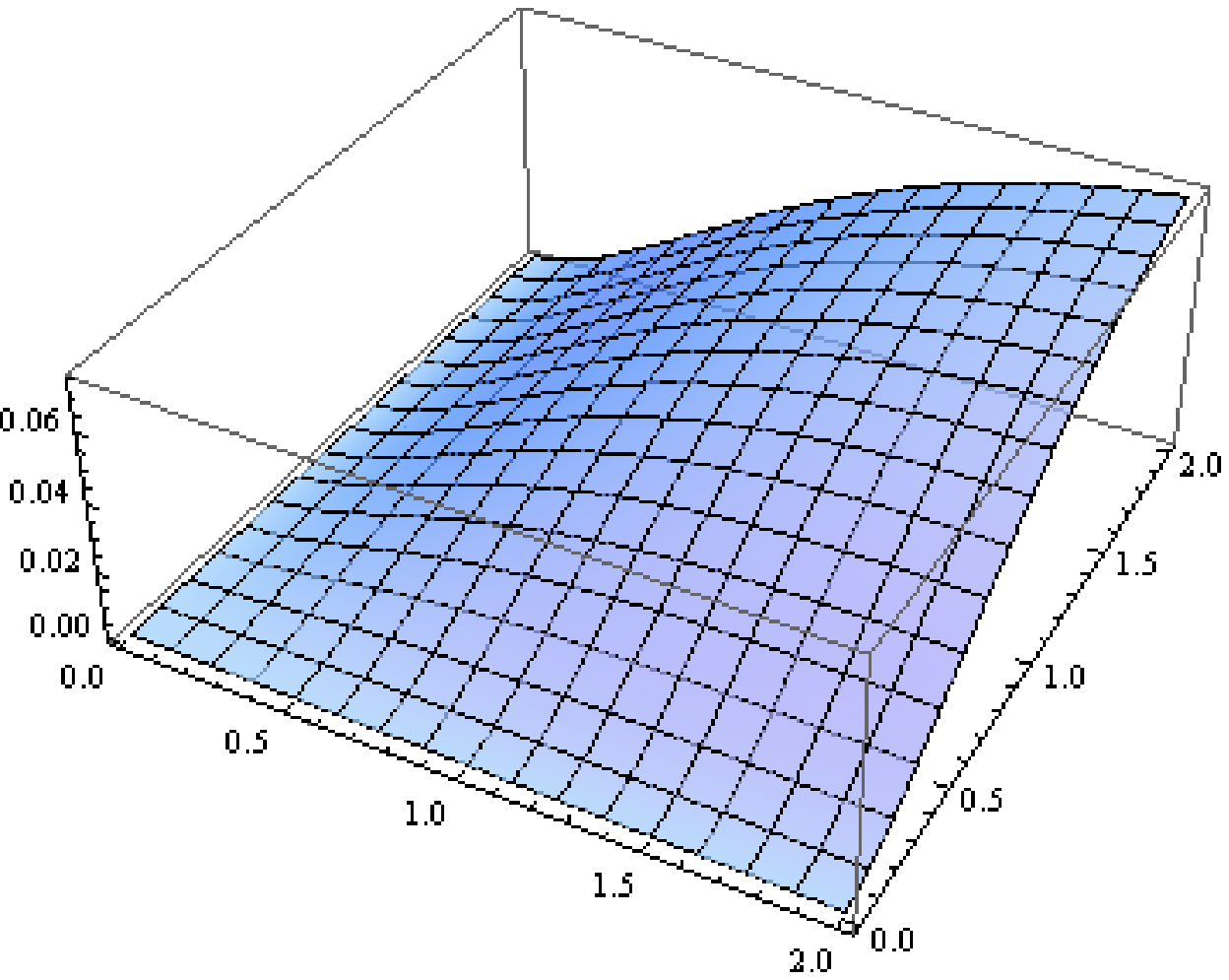}

  \includegraphics[width=0.4\textwidth]{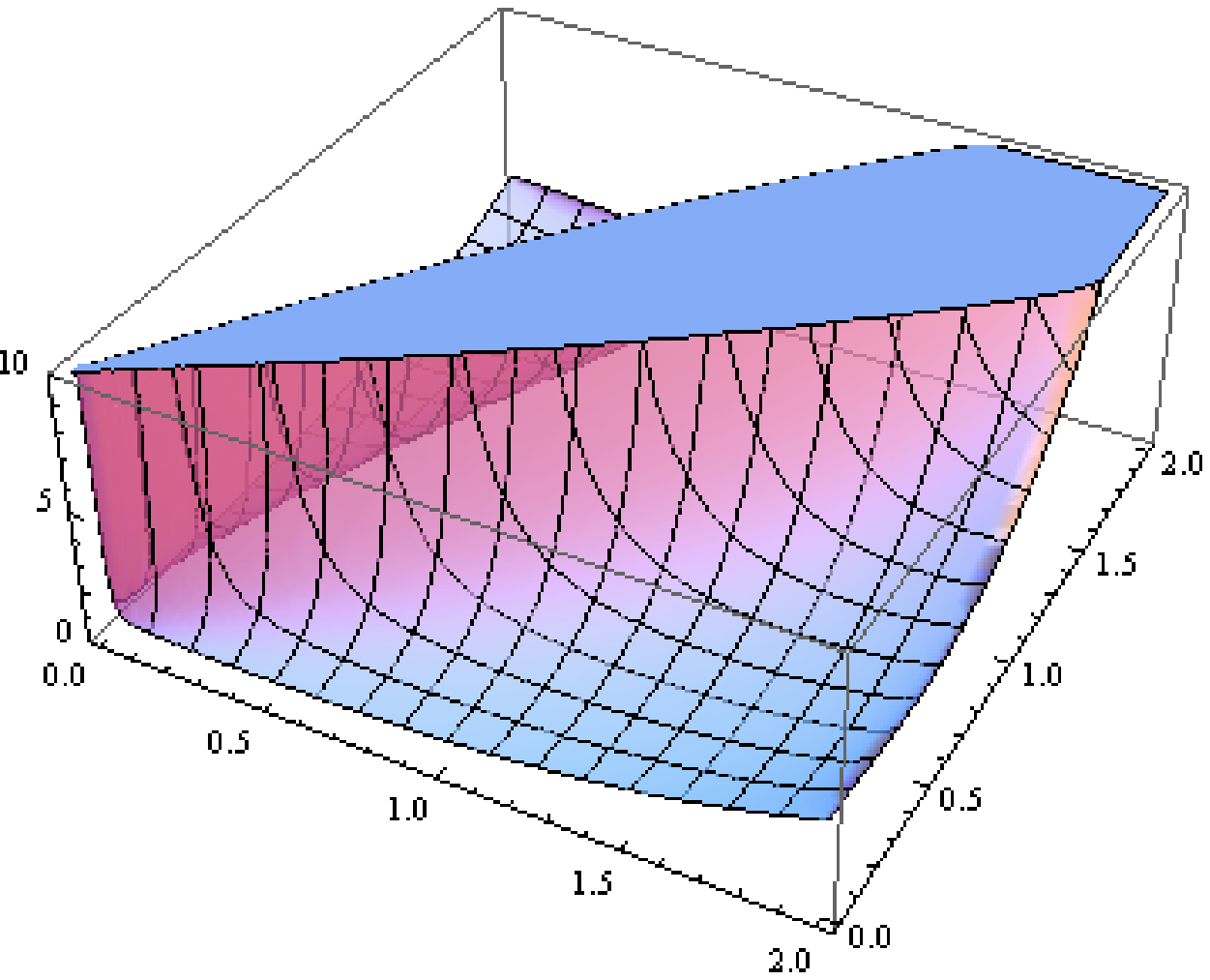}
\hspace{0.05\textwidth}
  \includegraphics[width=0.4\textwidth]{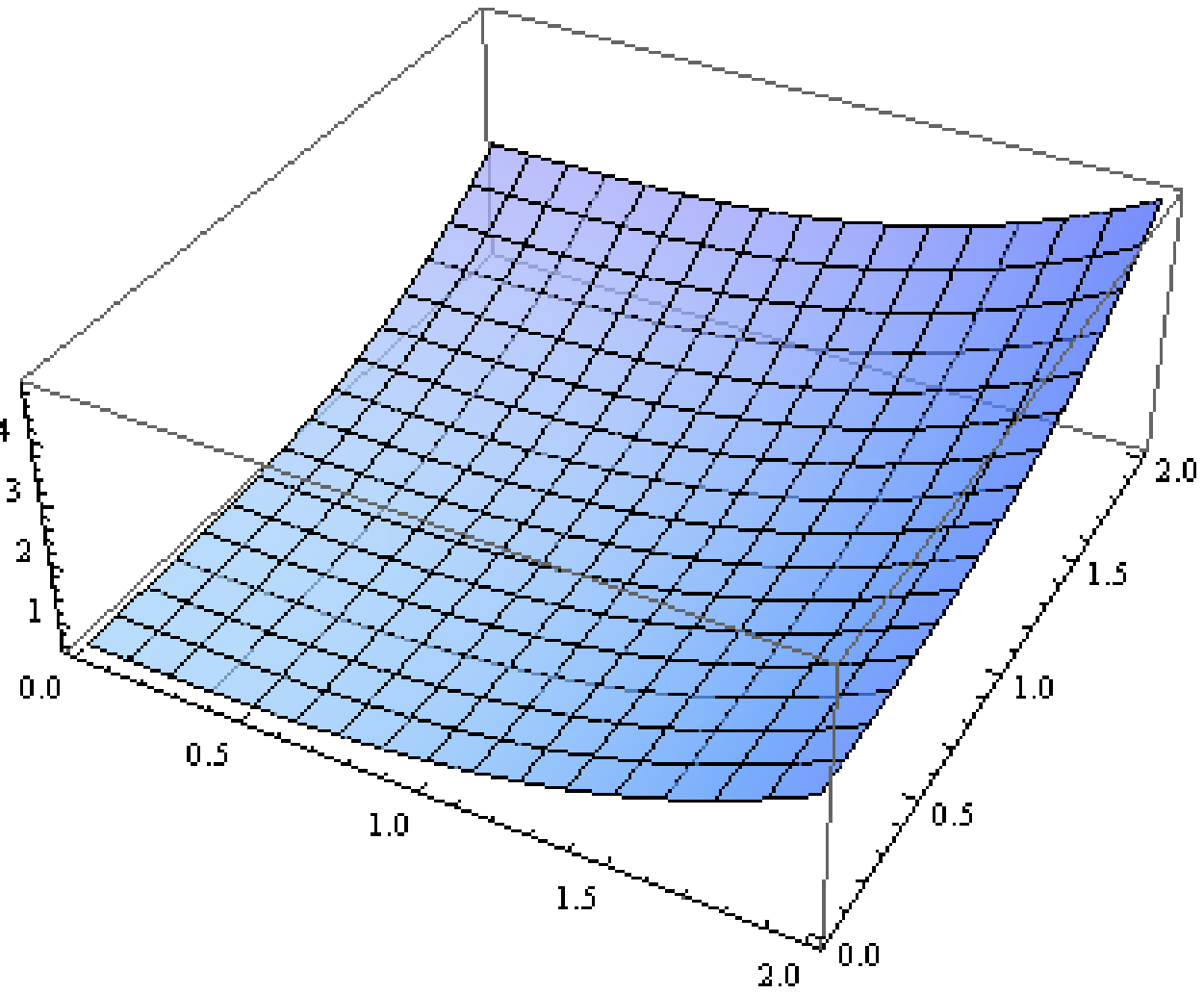}
    \caption{\label{specplanar} In this group of figures, we consider
      the specialized planar limit with $k_1=k_3=k_{14}$, and plot
      $T_{s1}$, $T_{s2}$, $T_{s3}$, $T_{c1}$, $T_{loc1}$ and
      $T_{loc2}$, respectively, as
      functions of $k_{2}/k_1$ and
      $k_{4}/k_1$. Again, in
      the $k_2\rightarrow 0$ or $k_4\rightarrow 0$ limit, our shape
      functions vanish as $\CO(k_2^2)$ and $\CO(k_4^2)$ respectively.
      This is different from that of the local
      shape.
$T_{loc1}$ blows up when $k_2 \rightarrow k_4$. This is because in
this limit, $k_{13}\rightarrow 0$. }
\end{figure}

\begin{figure}
  \center
  \includegraphics[width=0.48\textwidth]{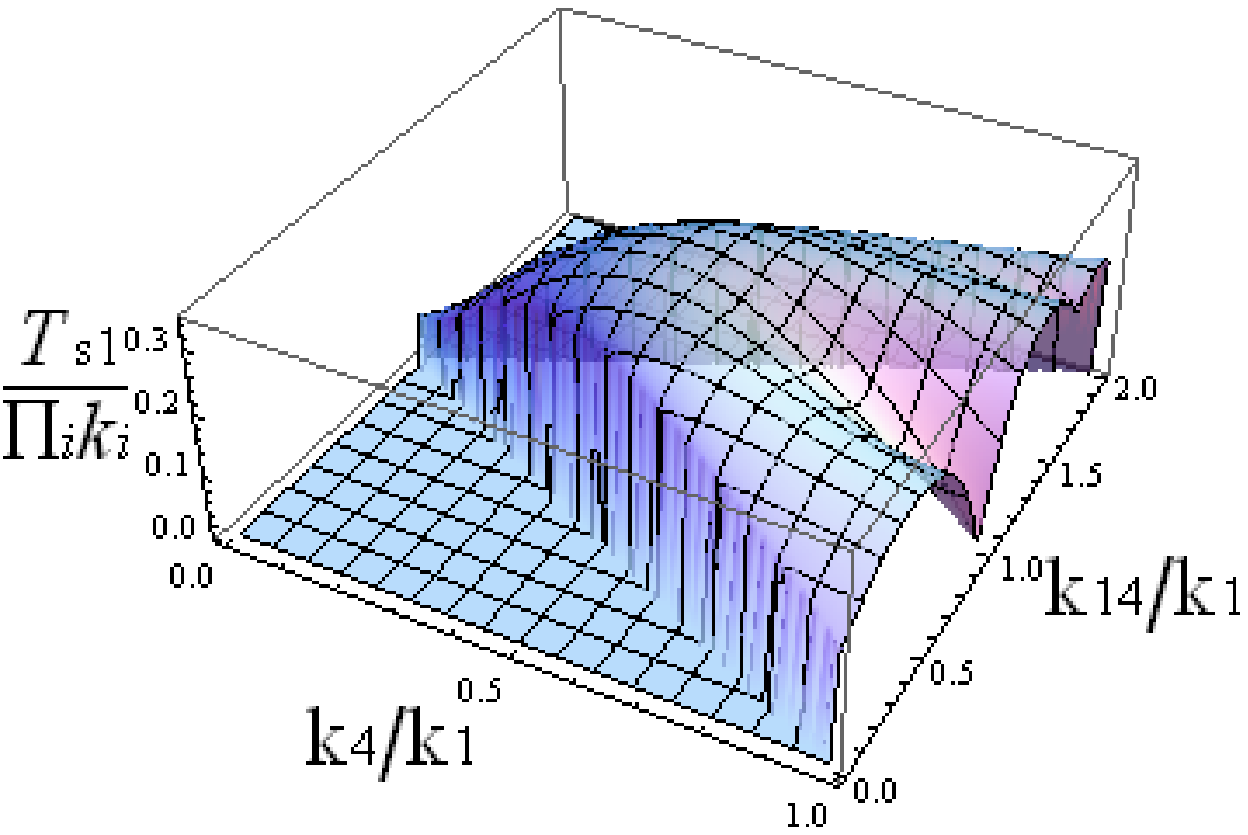}
\hspace{0.\textwidth}
  \includegraphics[width=0.4\textwidth]{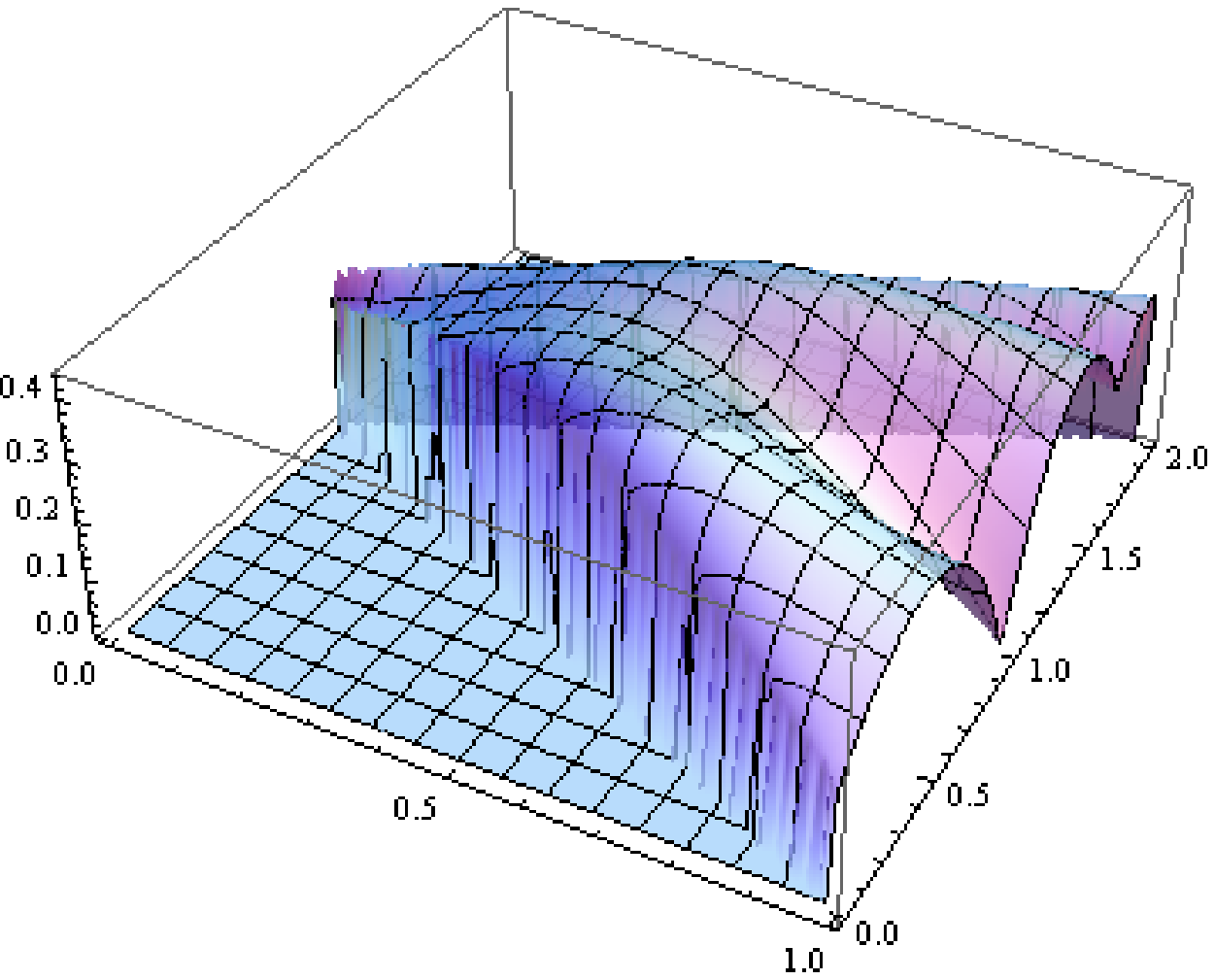}

  \includegraphics[width=0.4\textwidth]{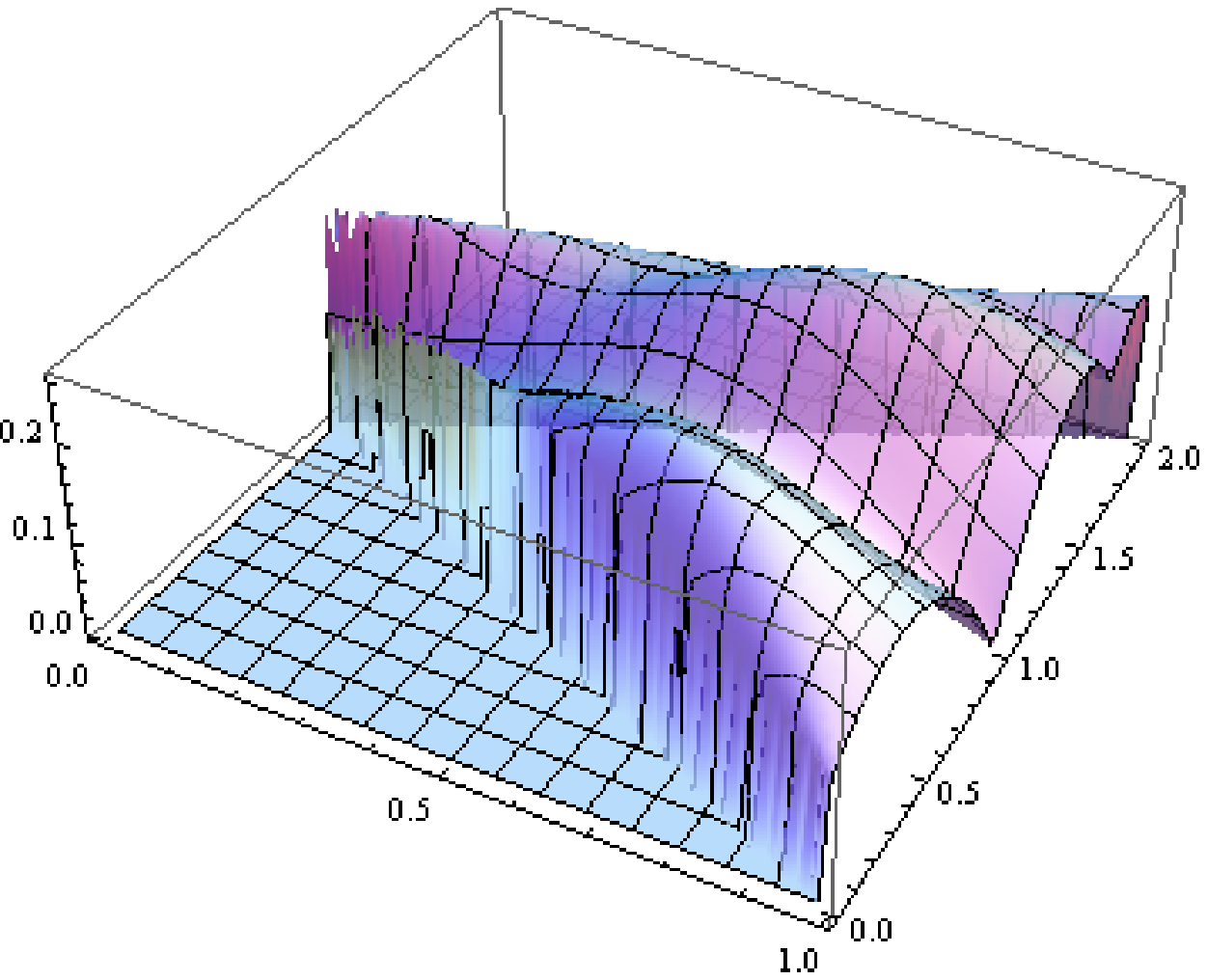}
\hspace{0.05\textwidth}
  \includegraphics[width=0.4\textwidth]{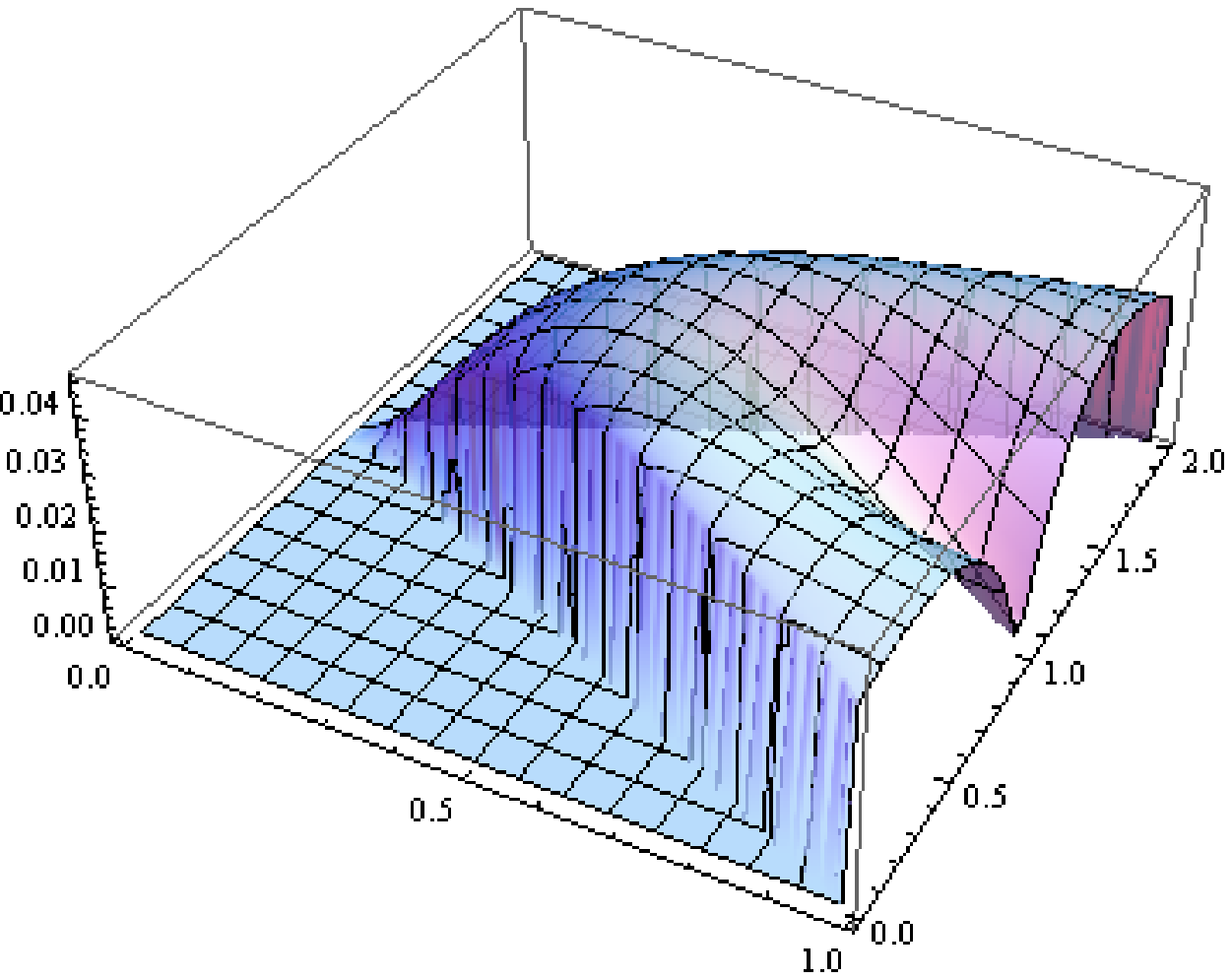}

  \includegraphics[width=0.4\textwidth]{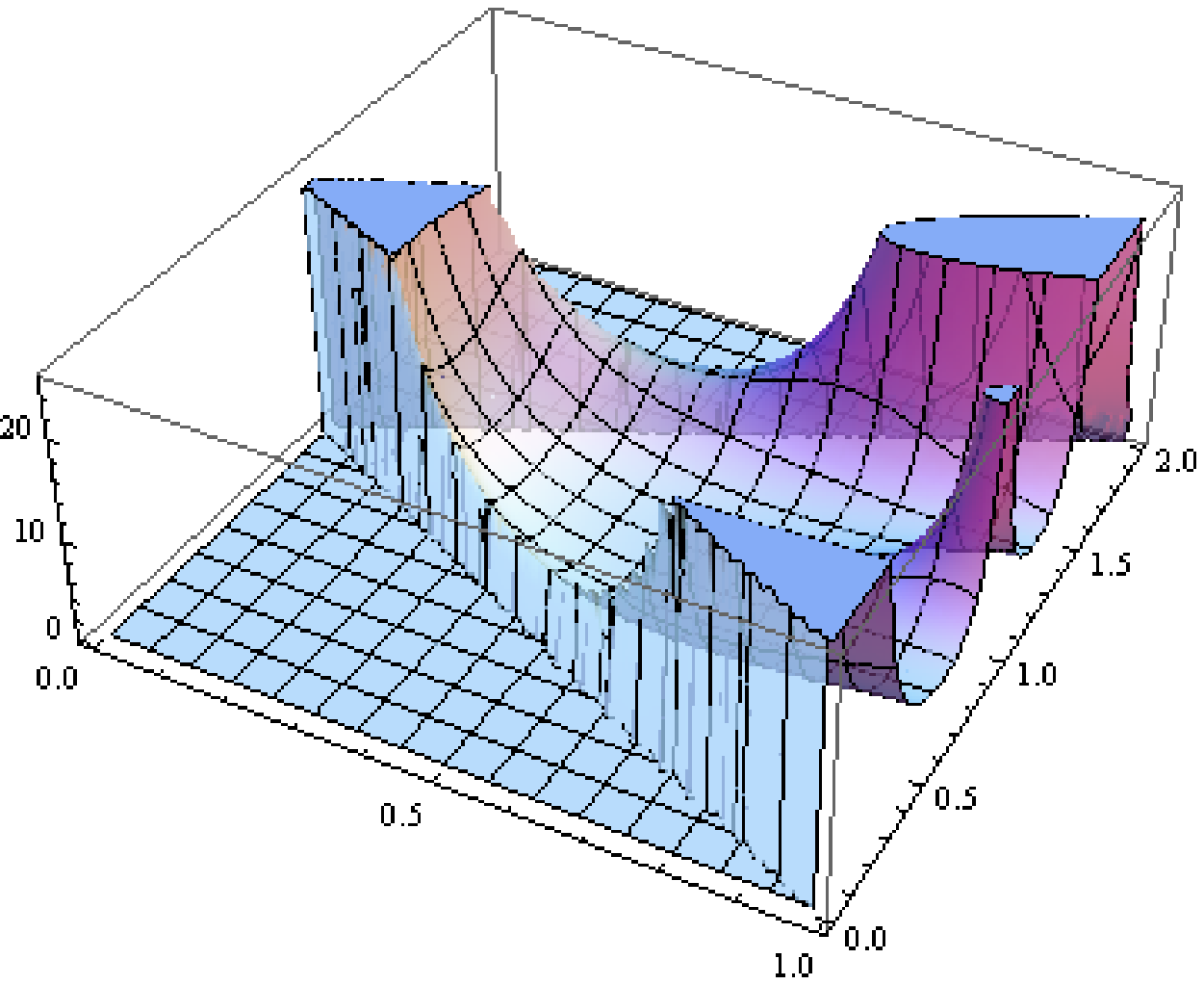}
\hspace{0.05\textwidth}
  \includegraphics[width=0.4\textwidth]{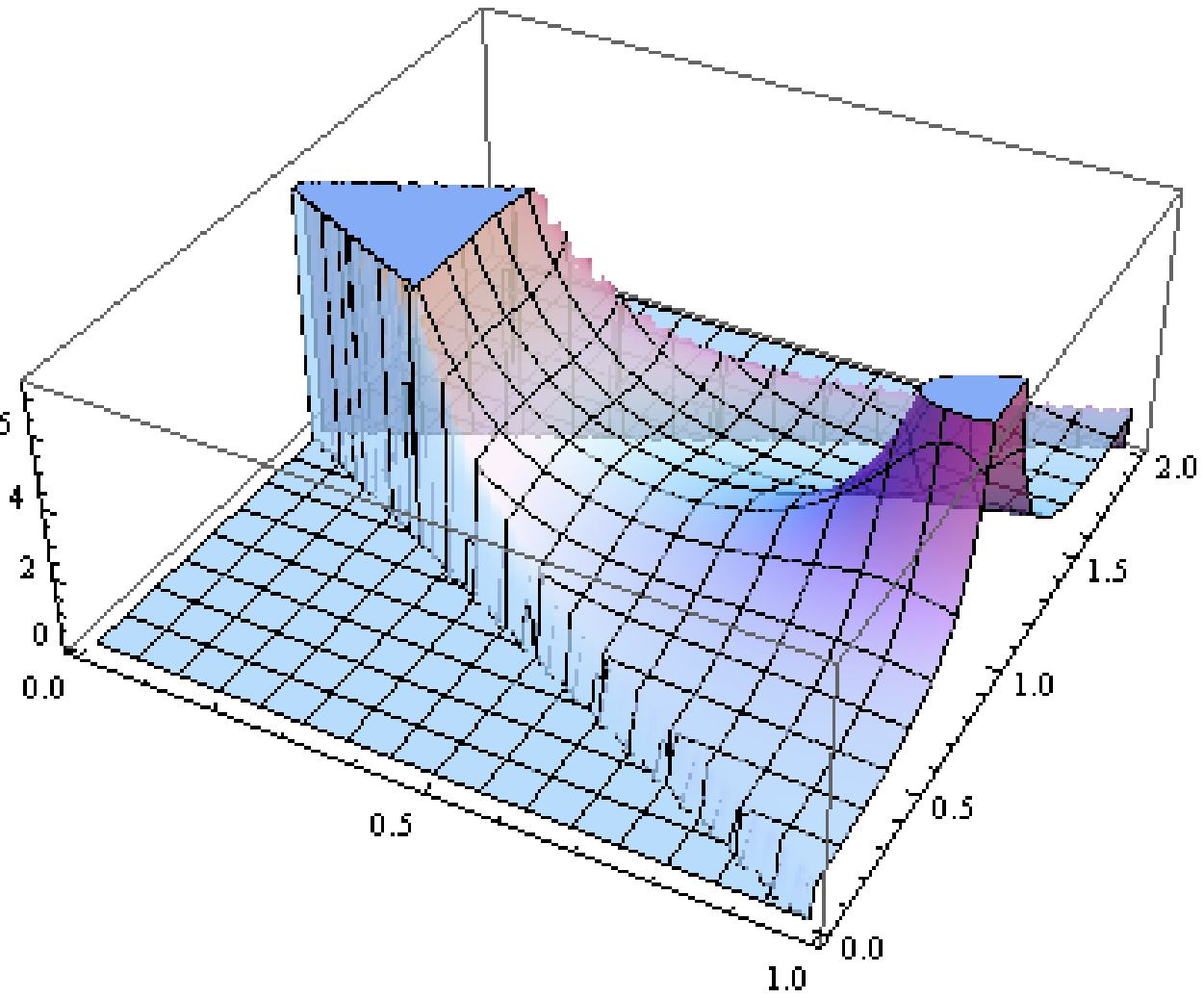}
    \caption{\label{doublesqueeze} In this group of figures, we look
      at the shapes near
      the double squeezed limit: we consider the case where
      ${k}_3={k}_4=k_{12}$ and
      the tetrahedron
  is a planar quadrangle. We plot
      $T_{s1}/(\prod_{i=1}^4 k_i)$,
      $T_{s2}/(\prod k_i)$, $T_{s3}/(\prod k_i)$, $T_{c1}/(\prod k_i)$, $T_{loc1}/(\prod k_i)$ and $T_{loc2}/(\prod k_i)$,
      respectively, as
      functions of $k_{12}/k_1$ and
      $k_{14}/k_1$. Note that, taking the double-squeezed limit $k_4\rightarrow
  0$, the scalar-exchange contributions $T_{s1}/(\prod k_i)$, $T_{s2}/(\prod k_i)$, $T_{s3}/(\prod
  k_i)$ are nonzero and finite, and the contact-interaction
      $T_{c1}/(\prod k_i)$ vanishes.
  As a comparison, the local form terms $T_{loc1}/(\prod k_i)$ and $T_{loc2}/(\prod
  k_i)$ blow up. The different behaviors in the folded and squeezed
      limit can also been seen from this figure (see the main text for details).}
\end{figure}

After checking the consistency relations, now we shall plot the
shape functions. To do so, we shall take various limits to reduce
the number of variables. We set the shape function to zero when the
momenta do not form a tetrahedron. We consider the following
cases:

\begin{enumerate}
  \item Equilateral limit: $k_1=k_2=k_3=k_4$. In Fig. \ref{equilateral}, we
  plot $T_{s1}$, $T_{s2}$, $T_{s3}$, $T_{c1}$, $T_{loc1}$ and $T_{loc2}$ as
  functions of $k_{12}/k_1$ and $k_{14}/k_1$. (We would like to
  remind the reader that unlike the first four
  shape functions, $T_{loc1}$ and $T_{loc2}$ are not obtained in our
  model. We plot them for the purpose of comparison.)
  One observes that
  $T_{loc1}$ blows up at all boundaries. This feature can distinguish
  our shape functions
  from the local shape $T_{loc1}$ originated from the local
  $f_{NL}$.

  \item Folded limit: $k_{12}= 0$. (This limit is also related
  to the parallelogram limit, $\bk_1=\bk_3$, by the symmetry
  (\ref{symeq}).)
  In this limit, $k_1=k_2$
    and $k_3=k_4$. We  plot
  $T_{s1}$, $T_{s2}$, $T_{s3}$, $T_{c1}$ and $T_{loc2}$
    as functions of  $k_{4}/k_1$ and $k_{14}/k_1$ in
  Fig. \ref{folded}.
  (Note that $T_{loc1}$ blows up in this limit). We assumed $k_4<k_1$
  without losing
    generality. Note that $T_{loc2}$ does not vanish in the
    $k_4\rightarrow 0$ limit. This can be used to distinguish our
    shape functions from the local shape originated from $g_{NL}$.

  \item Specialized planar limit: We take $k_1=k_3=k_{14}$, and
    additionally the
    tetrahedron to be a planar quadrangle. In this limit,
    one can solve for $k_{12}$ from \eqref{cosineq}:
\begin{align}
      k_{12}=\left[
k_1^2+\frac{k_2 k_4}{2 k_1^2}\left( k_2 k_4 \pm
\sqrt{(4k_1^2-k_2^2)(4k_1^2-k_4^2)} \right) \right]^{1/2}~.
    \end{align}
The minus sign solution can be related to another plus sign solution
in the $k_1=k_2=k_{14}$ limit through a symmetry discussed in
Appendix~\ref{planarappendix}. We will only consider the plus sign
solution in our following discussion.
     We plot the shape functions as functions
of $k_2/k_1$ and $k_4/k_1$ in Fig. \ref{specplanar}. These figures
illustrate two important distinctions between our shape functions
and the local form shape functions. At the $k_2 \rightarrow k_4$
limit, we have $k_{13} \to 0$, so $T_{loc1}$ blows up, while the
others are all finite. At the $k_2\rightarrow 0$ and $k_4\rightarrow
0$ boundaries, our shapes functions vanish as $\CO(k_2^2)$ and
    $\CO(k_4^2)$ respectively,
while $T_{loc1}$ and $T_{loc2}$ are non-vanishing.

\item Near the double-squeezed limit: we consider the case where
  ${k}_3={k}_4=k_{12}$ and the tetrahedron
  is a planar quadrangle. We are interested in the behavior of the
  shape functions as $k_3=k_4=k_{12} \to 0$, i.e.~as the planar
  quadrangle is doubly squeezed.
  In this case, Eq. \eqref{cosineq} takes the
  equal sign. One can solve for $k_{2}$
    from \eqref{cosineq}. The solution is presented in Eq.
    \eqref{planark2}.
  We plot $T_{s1}/(\prod_{i=1}^4 k_i)$, $T_{s2}/(\prod k_i)$,
  $T_{s3}/(\prod k_i)$, $T_{c1}/(\prod k_i)$, $T_{loc1}/(\prod k_i)$
  and $T_{loc2}/(\prod k_i)$ as
  functions of $k_{12}/k_1$ and $k_{14}/k_1$ in
  Fig. \ref{doublesqueeze}. To reduce the
  range of the plot, we only show the figures partially with
  $k_4<k_1$. Note that in this figure, we divided the
  shape functions by $\prod k_i$ in order to have better distinction
  between contact-interaction and scalar-exchange contributions.
  Fig. \ref{doublesqueeze} shows simultaneously the three
  differences among the four shapes $T_{s1}$ ($\sim T_{s2,3}$),
  $T_{c1}$, $T_{loc1}$ and $T_{loc2}$. 1) In the double-squeezed
  limit, $k_3=k_4\rightarrow
  0$, the scalar-exchange contributions $T_{s1}/(\prod k_i)$,
  $T_{s2}/(\prod k_i)$, $T_{s3}/(\prod
  k_i)$ are nonzero and finite, and the contact-interaction
  $T_{c1}/(\prod k_i)$ vanishes.
  As a comparison, the local form terms $T_{loc1}/(\prod k_i)$ and
  $T_{loc2}/(\prod k_i)$ blow up.
  2) In the folded limits, at the $(k_4/k_1=1,k_{14}/k_1=0)$ corner where
  $k_{14}\to 0$, and close to the $(k_4/k_1=1,k_{14}/k_1=2)$ area where
  $k_{13} \to 0$, $T_{loc1}/(\prod k_i)$ blows up. 3) In the squeezed
  limit, at
  $(k_4/k_1=1,k_{14}/k_1=1)$ where $k_2\to 0$, the $T_{loc1}/(\prod
  k_i)$ and
  $T_{loc2}/(\prod k_i)$ blow up. The last two behaviors have also
  appeared in the previous figures.

\end{enumerate}

In the second, third and fourth limits, the tetrahedron reduce to a
planar quadrangle. We collectively denote this group of limits as the
planar limit.  This planar limit is
  of special importance, because one of the most important ways to
  probe trispectrum is the small (angular) scale CMB experiments.
  These experiments directly measure signals contributed mainly from the planar quadrangles. The
  more general plot for the planar limit is presented in
  Appendix \ref{planarappendix}.
We can see that while very different from the two local shapes, the
three shapes $T_{s1}$, $T_{s2}$ and $T_{s3}$ are overall similar. Of
course like in the bispectrum case, we can tune the parameters to
subtract out the similarities and form new bases for the shapes.

We end this section by emphasizing a couple of important points:
\begin{itemize}
\item {\em The equilateral
  trispectra forms}:
  The scalar-exchange contributions $T_{s1,2,3}$
  and the contact-interaction contribution $T_{c1}$
  are similar at most regions, but
  can be distinguished in the double-squeezed limit (e.g.~$k_3=k_4\to
  0$), where the two
  kinds of forms approach
  zero at different speeds, $T_{s1,2,3} \to \CO(k_3^2)$,
  $T_{c1}\to\CO(k_3^4)$.
  Within the scalar-exchange contributions, the three shapes $T_{s1}$,
  $T_{s2}$, $T_{s3}$ are very similar overall, having only small
  differences.\footnote{For example, in Fig.~\ref{equilateral} or
  \ref{folded}, if
  we look at the double folded limit, $k_{12} \to 0$ and $k_{14} \to
  0$, $T_{s2}$ and $T_{s3}$ go from positive to negative ($T_{s2} \to
  -0.066$ and $T_{s3} \to -0.030$), while $T_{s1}$ remains positive
  ($T_{s1} \to 0.092$).}

  For the purpose of data analyses, one can then use the following two
  representative forms for the ``equilateral
  trispectra''. One is $T_{c1}$, given in (\ref{CT_c1}). This ansatz
  can be used to represent all four leading shapes at most
  regions. For a refined data analysis, for example to distinguish
  shapes in the double-squeezed limit, one can add another form
  $T_{s1}$, given in (\ref{aa1_1}) and (\ref{aa23_1}). This ansatz
  represents very well the three scalar-exchange contributions
  $T_{s1,2,3}$.
  The first ansatz is factorizable (in terms of the six variables
  $k_{1,2,3,4},
  k_{12}, k_{14}$) by introducing an integral $1/K^n=
  (1/\Gamma(n)) \int_0^{\infty} t^{n-1} e^{-Kt}$
  \cite{Smith:2006ud}; while the second ansatz cannot be easily
  factorized due to the presence of $k_{13}$ given by (\ref{k13Ex}).

\item {\em Distinguishing between the equilateral and local forms}:
  In the following limits, the equilateral and local forms behave very
  differently.
  At the folded limit (e.g.~$k_{12}\to 0$), $T_{loc1}$ generically
  blows up, while
  the four equilateral shapes and $T_{loc2}$
  approach constants. At the squeezed limit (e.g.~$k_4 \to 0$), the
  four equilateral shapes all vanish as $\CO(k_4^2)$,
  while the local forms $T_{loc1}$
  and $T_{loc2}$ do not.

\end{itemize}

\section{Examples}
\setcounter{equation}{0}

For DBI inflation
\cite{Silverstein:2003hf,Alishahiha:2004eh,Chen:2004gc,Chen:2005ad,Kecskemeti:2006cg,Shandera:2006ax},
$P = -f(\phi)^{-1} \sqrt{1-2X f(\phi)} + f(\phi)^{-1} -V(\phi)$,
\bea c_s \ll 1 ~, ~~~~ \frac{\lambda}{\Sigma}=\half \left(
\frac{1}{c_s^2}-1 \right) ~, ~~~~ \frac{\mu}{\Sigma}= \frac{1}{4}
\left( \frac{5}{c_2^2} -4 \right) \left( \frac{1}{c_s^2} -1 \right)
~. \eea
The dominant contribution come from the scalar-exchange terms
$\CT_{s1,s2,s3}$ and one contact-interaction term $\CT_{c1}$, which
are of order $1/c_s^4$. The $\CT_{c2}$ and $\CT_{c3}$ are of order
$1/c_s^2$, so belong to subleading contributions.
Therefore the shape function defined in \eqref{shapesum} takes the
form
\begin{align}
  \CT^{\rm DBI}\approx &\left( \frac{T_{s1}}{4}+\frac{T_{s2}}{2}+T_{s3}-T_{c1}
  \right)\frac{1}{c_s^4} ~.
\end{align}
According to the definition (\ref{tNLdef}), $t_{NL}^{\rm DBI} \approx
0.542/c_s^4$.

For k-inflation
\cite{Armendariz-Picon:1999rj,Li:2008qc,Engel:2008fu}, we look at
the example $P \sim (-X+X^2)/\phi^2$, \bea c_s \ll 1 ~, ~~~~
\frac{\lambda}{\Sigma} = \frac{2X}{-1+6X} =\frac{1-c_s^2}{2} ~, ~~~~
\frac{\mu}{\Sigma} = \frac{X}{-1+6X} = \frac{1-c_s^2}{4} ~. \eea
The only dominant term is
one of the scalar-exchanging terms $\CT_{s3}$, the others all belong
to subleading contributions.
So
\begin{align}
  \CT^{\rm K}\approx&\frac{T_{s3}}{c_s^4} ~,
\end{align}
and $t_{NL}^{\rm K} \approx 0.305/c_s^4$.

As mentioned in the previous section, there are some differences
among the shapes $T_{s1}$, $T_{s2}$ and
$T_{s3}$, and especially between them and $T_{c1}$. These differences
may be used to distinguish some special models within this class. But
unfortunately, for the above two examples,
after summing over all contributions for DBI inflation,
the trispectrum difference
between the DBI inflation and this specific k-inflation example
becomes smaller, and we do not find any
features that can very sharply distinguish them. This is because
the four leading shapes $T_{si}$ and $T_{c1}$ are similar in most regions and in the discriminating double-squeezed limit, the trispectra in
both examples take the form $T_{si}$ (which are similar among
themselves) since $T_{c1}$ vanishes faster.

\section{Non-Bunch-Davies vacuum}
\setcounter{equation}{0}

We now study the shape of trispectrum if the initial state of inflation deviates from the standard
Bunch-Davies vacuum of
de Sitter space. This is an interesting question
because
short distance physics may give rise to
such deviation  \cite{EGKS,Danielsson}, and one might argue whether its effects on the power spectrum of the CMB are observable \cite{Kaloper}\footnote{The choice of initial state is often discussed in the context of
trans-Planckian effects \cite{Martin:2000bv} though the issue has more general applicability.
See
e.g.~\cite{Greene:2005aj}
for a review and references, and \cite{BoundaryEFT} for a discussion of how to capture the initial state effects in terms of a boundary effective field theory.}.
The effects of the non-Bunch-Davies vacuum on the bispectrum have been
studied in
\cite{Chen:2006nt,Holman:2007na,Meerburg:2009ys}, where it was found
that the non-Gaussianities are boosted in the folded triangle limit
(e.g.~$k_1+k_2-k_3 \sim 0$).

A general vacuum state for the fluctuation of the inflaton during inflation can be written as
\begin{eqnarray}
u_k = u(\textbf{k},\tau) = \frac{H}{\sqrt{4\epsilon c_s k^3}}
(C_{+}(1+i k c_s\tau )e^{-i k c_s\tau}+ C_{-} (1-i k c_s\tau) e^{i k
  c_s \tau}) ~.
\label{nonBDuk}
\end{eqnarray}
Here a small and non-zero $C_{-}$ parametrizes a deviation from the
standard Bunch-Davies vacuum which has $C_{+}=1, C_{-}=0$. We
consider the corrections of a non-zero $C_{-}$ to the leading shapes
assuming $C_{-}$ is small, we keep only terms up to linear order in $C_{-}$.
Similar to the case of the bispectrum \cite{Chen:2006nt},
the first sub-leading corrections
come from replacing one of the $u(\tau,\textbf{k})$'s with their
$C_{-}$ components, and since the correction from $u(0,\textbf{k})$
only has the same shape as that of the Bunch-Davies vacuum, we only
need to consider the case that $u(\tau,\textbf{k})$ comes from the
interacting Hamiltonian, where $\tau$ is not zero. In the following
we discuss the contact-interaction diagram and the scalar-exchange
diagram respectively.

For the contact-interaction diagram, the corrections consist of four
terms from replacing $k_i$ with $-k_i$ in the shape for Bunch-Davies
vacuum. For the leading shape $T_{c1}$ we denote the
correction as $\tilde{T}_{c1}$ and we find
\begin{eqnarray} \label{nonBDc1}
\tilde{T}_{c1} &=& 36 ~{\rm Re}(C_{-}) \prod_{i=1}^4 k_i^2 \left[\frac{1}{(k_1+k_2+k_3-k_4)^5}+\frac{1}{(k_1+k_2-k_3+k_4)^5} \right.\nonumber \\
&& +\left.
\frac{1}{(k_1-k_2+k_3+k_4)^5}+\frac{1}{(-k_1+k_2+k_3+k_4)^5}\right]~.
\end{eqnarray}


Then let us consider scalar-exchange diagram. For illustration we
only consider the $T_{s1}$ term. The calculations for $T_{s2}$ and
$T_{s3}$ are similar but more complicated. Now we have six $u_k$
modes from the interaction Hamiltonian. Replacing each mode with
its $C_{-}$ component gives rise to six terms in the corrections.
They correspond to replacing $k_i$ with $-k_i$, or one of the two
$k_{12}$'s with $-k_{12}$ in the calculations for the Bunch-Davies
vacuum. The corrections to Eq.~(\ref{aa1_1}) are
\begin{eqnarray} \label{nonBDs11}
&& \frac{9~{\rm Re}(C_{-})}{8}
k_1^2k_2^2k_3^2k_4^2k_{12}\left\{\left[\sum _{k_i\rightarrow -k_i,
i=1}^4
\frac{1}{(k_3+k_4+k_{12})^3}\frac{1}{(k_1+k_2+k_{12})^3}\right]\right.
\nonumber \\
&&\left. + \frac{1}{(k_1+k_2-k_{12})^3}\frac{1}{(k_3+k_4+k_{12})^3}
+\frac{1}{(k_1+k_2+k_{12})^3}\frac{1}{(k_3+k_4-k_{12})^3} \right\}
\nonumber \\&& + \textrm{23 perm.}
\end{eqnarray}

In Eq.~(\ref{aa23_1}), the two $k_{12}$
cancelled in the final expression for the Bunch-Davies vacuum, so we
have to recover them in the calculations. Denoting
$M=k_3+k_4+k_{12}$ and $K=k_1+k_2+k_3+k_4$, we find the corrections
\begin{eqnarray} \label{nonBDs12}
&& \frac{9~{\rm Re}(C_{-})}{4} k_1^2k_2^2k_3^2k_4^2 k_{12}
\left\{\left[\sum _{k_i\rightarrow -k_i, i=1}^4
\frac{1}{M^3}\left(\frac{6M^2}{K^5}+\frac{3M}{K^4}+\frac{1}{K^3}\right)\right]\right.
\nonumber \\
&& +\frac{1}{(k_3+k_4-k_{12})^3}\left(\frac{6(k_3+k_4-k_{12})^2}{(K-2k_{12})^5}+\frac{3(k_3+k_4-k_{12})}{(K-2k_{12})^4}+\frac{1}{(K-2k_{12})^3}\right) \nonumber \\
&& \left.
+\frac{1}{M^3}\left(\frac{6M^2}{(K+2k_{12})^5}+\frac{3M}{(K+2k_{12})^4}+\frac{1}{(K+2k_{12})^3}\right)\right\}
+ \textrm{23 perm.}
\end{eqnarray}
To summarize, the correction $\tilde T_{s1}$ to
$T_{s1}$ is the summation of Eqs. \eqref{nonBDs11}
and \eqref{nonBDs12}.


\begin{figure}
  \center
  \includegraphics[width=0.4\textwidth]{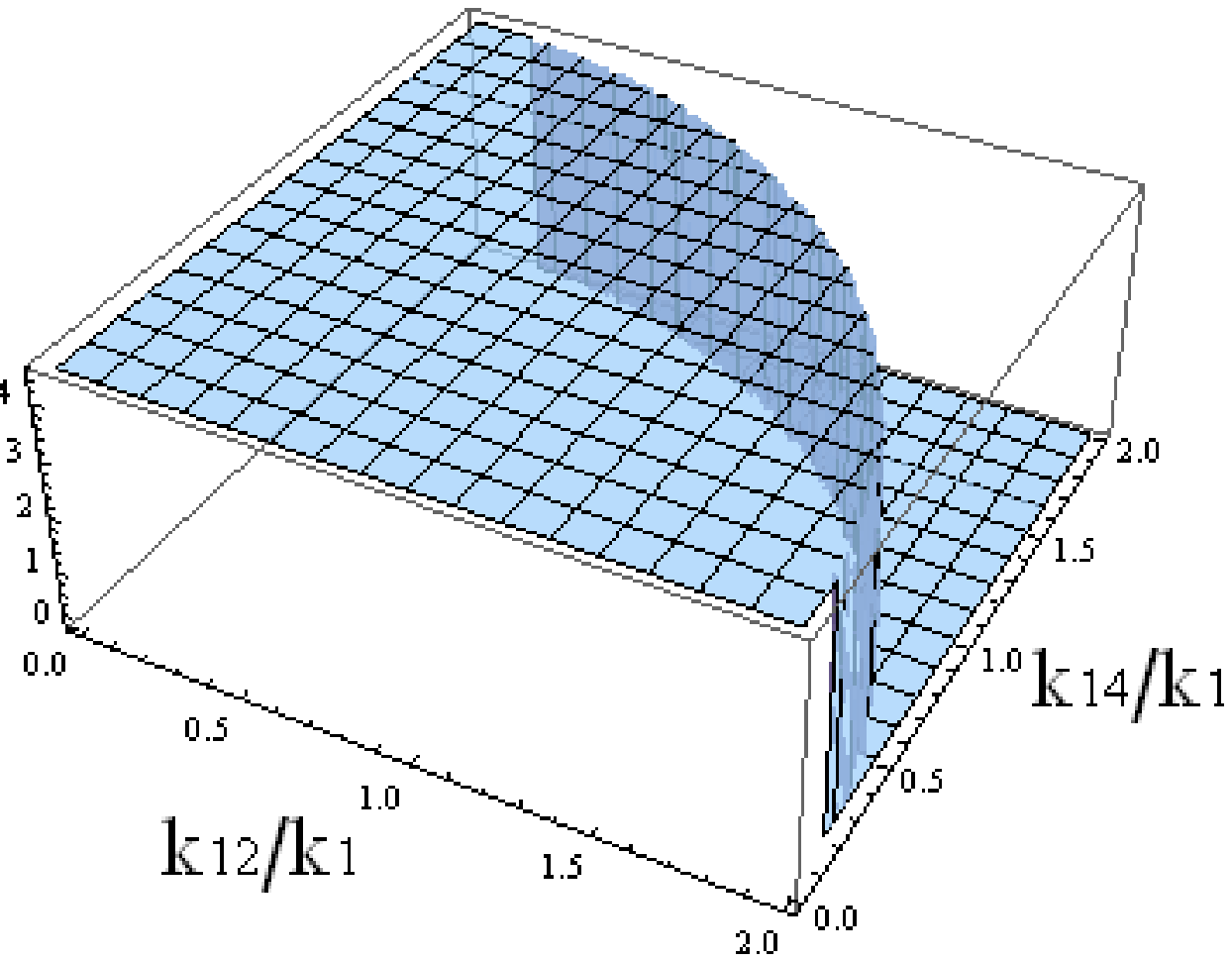}
\hspace{0.05\textwidth}
  \includegraphics[width=0.4\textwidth]{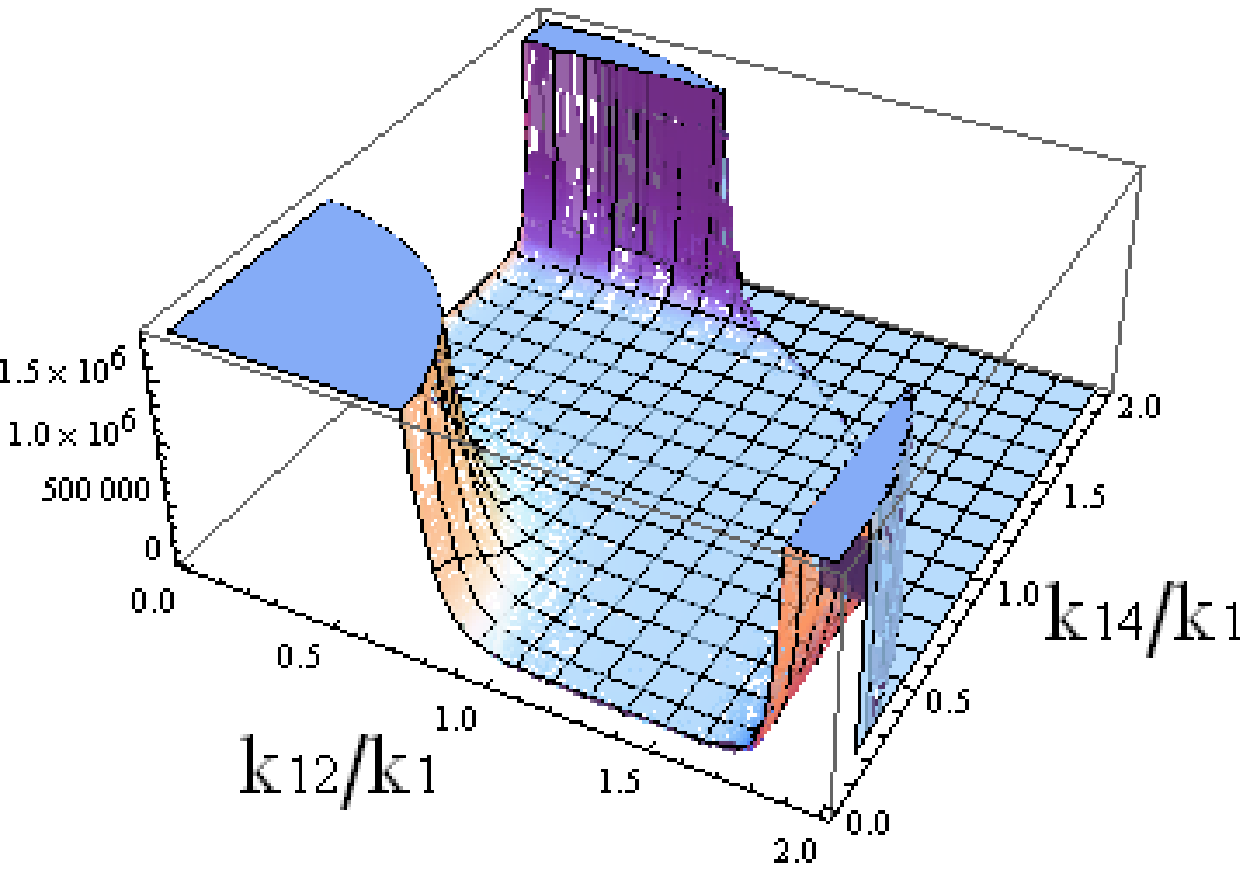}

  \includegraphics[width=0.4\textwidth]{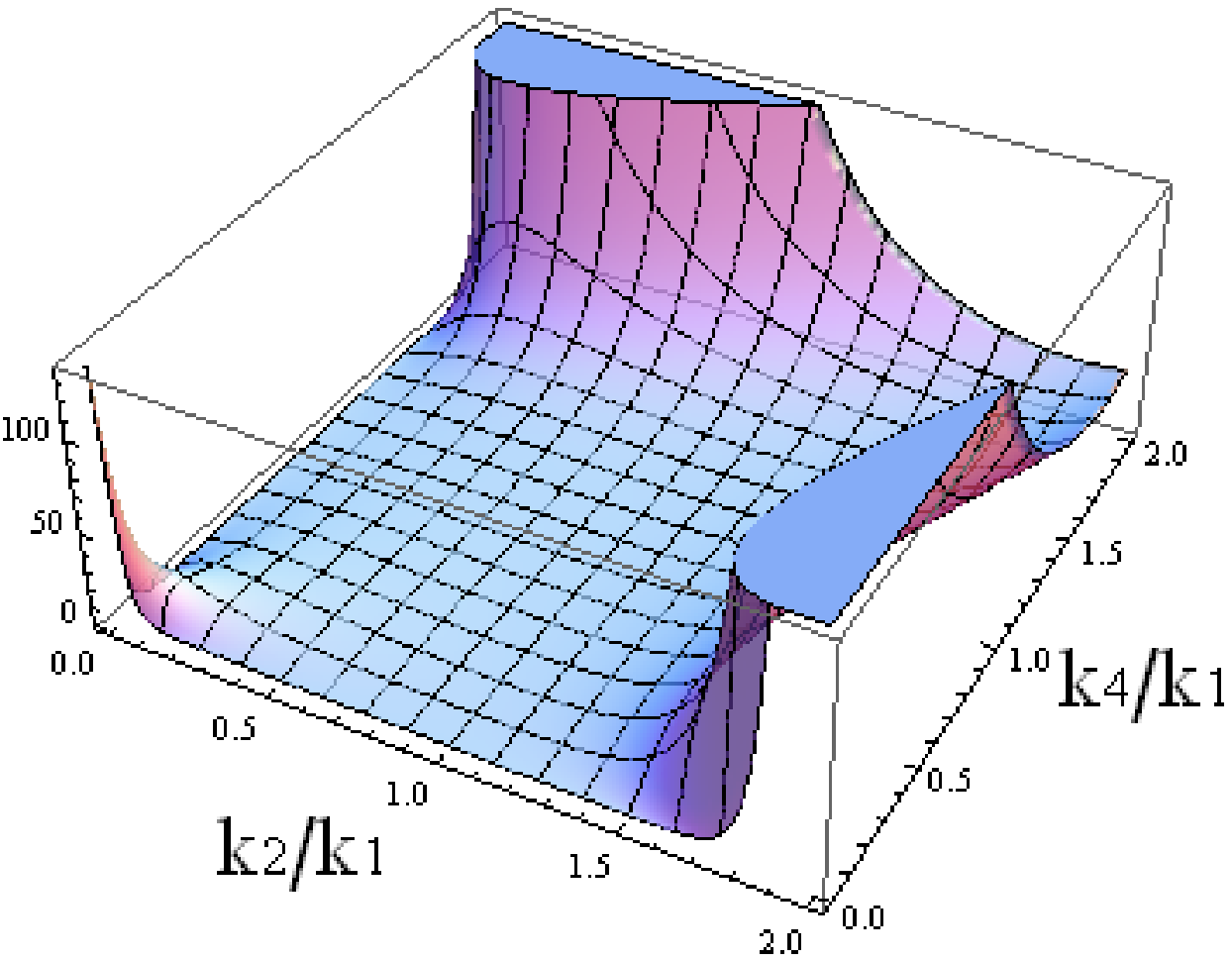}
\hspace{0.05\textwidth}
  \includegraphics[width=0.4\textwidth]{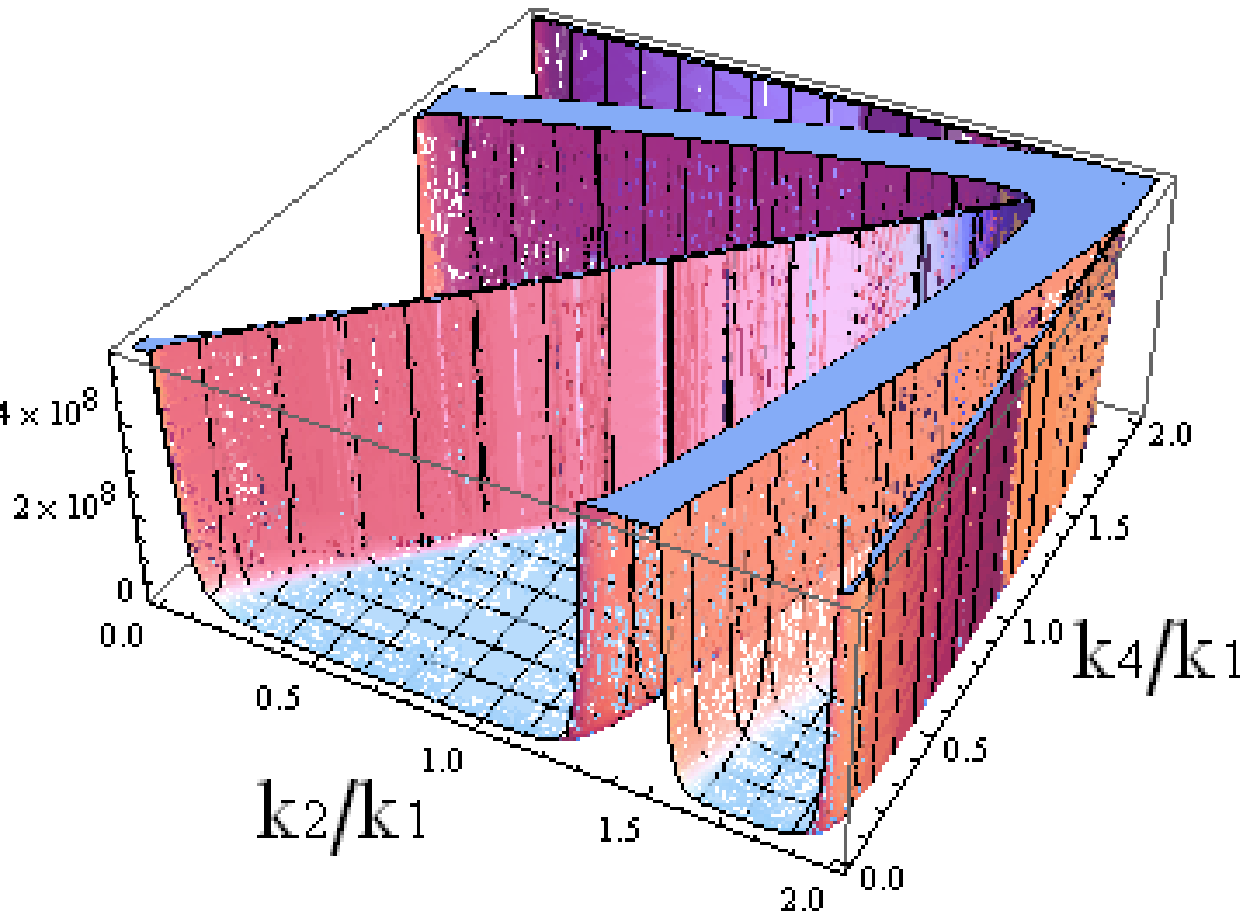}

  \includegraphics[width=0.4\textwidth]{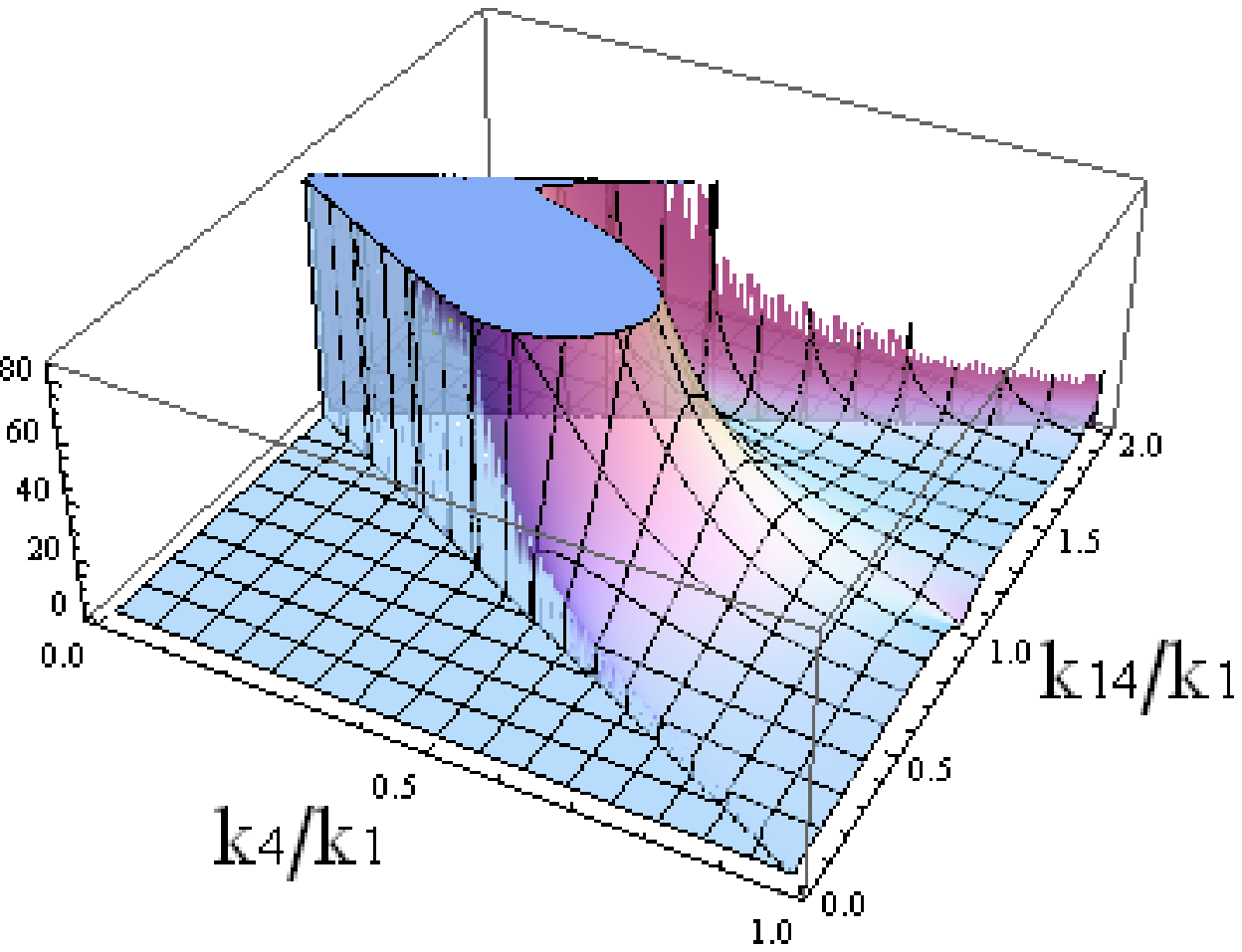}
\hspace{0.05\textwidth}
  \includegraphics[width=0.4\textwidth]{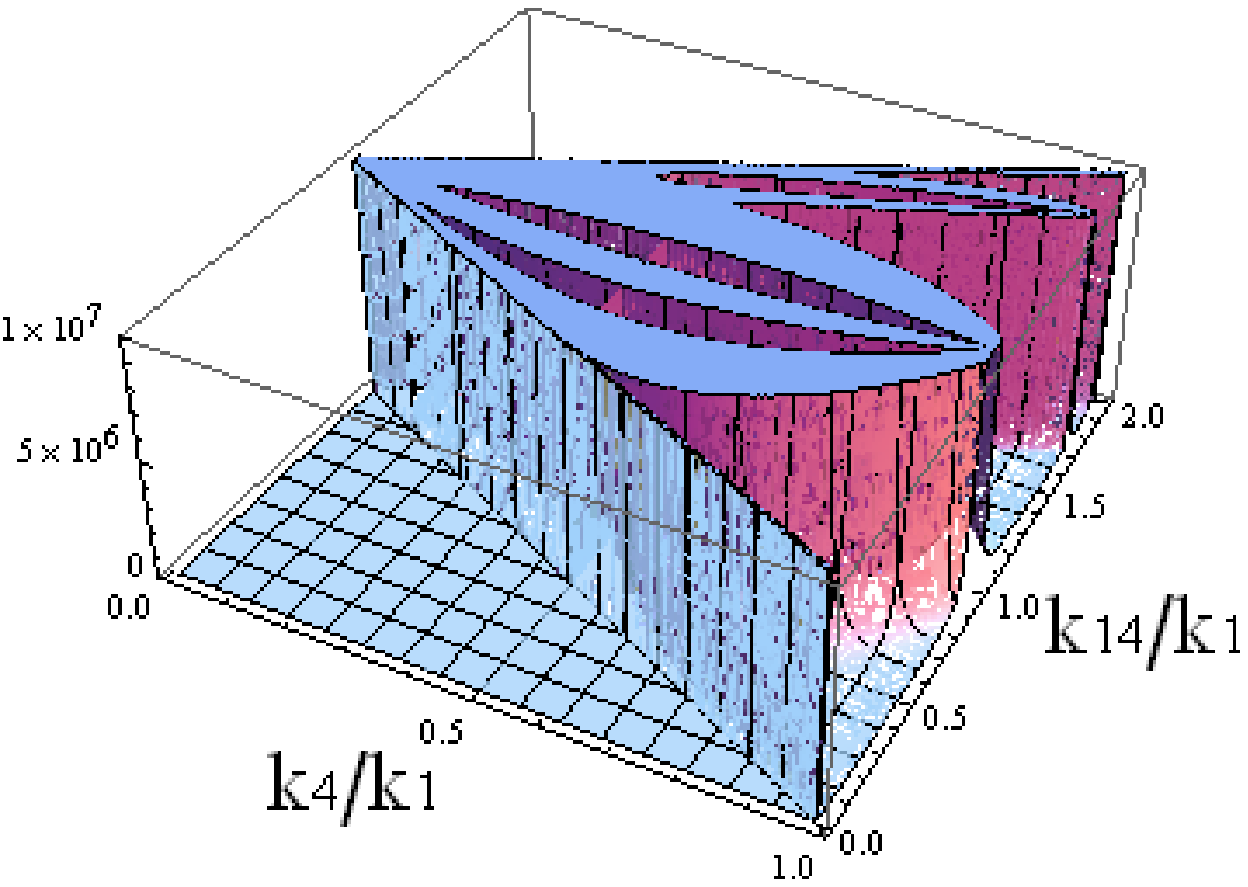}
    \caption{\label{NBDplot} In this group of figures, we plot
      the $\tilde T_{c1}/{\rm
Re}(C_-)$ (the left column) and $\tilde T_{s1}/{\rm Re}(C_-)$ (the
right column) in the equilateral limit, specialized planar limit,
and near double squeezed limit (we plot $\tilde T_{c1}/[{\rm
Re}(C_-)\Pi_i k_i]$ and $\tilde T_{s1}/[{\rm Re}(C_-)\Pi_i k_i]$ in
near double squeezed limit) respectively. Note that, in order to show
      clearly the locations of the divergence, in some figures we have taken the
      cutoffs of the z-axes to be extremely large.}
\end{figure}

We can look for the analogue of the folded triangle limit discovered
in the study of bispectrum where the corrections due to
deviation from the Bunch-Davies vacuum diverge. Here we see when any one
triangle, e.g., $(k_1,k_2,k_{12})$, in the momentum tetrahedron
becomes folded, the corrections to $T_{s1}$ (see (\ref{nonBDs11}) and (\ref{nonBDs12}))
become divergent. Furthermore, when $k_1+k_2+k_3-k_4=0$, the two
triangles $(k_1,k_2,k_{12})$ and $(k_3,k_4,k_{12})$ become folded
simultaneously,  and the correction (\ref{nonBDc1}) to $T_{c1}$
also diverges. We will refer to such configurations as the folded
sub-triangle configurations.\footnote{Due to permutations, the three
  momenta do not have to be next to each other in terms of
  Fig.~\ref{tFig}, for example, $k_1,k_3,k_{13}$.}

As discussed in \cite{Chen:2006nt}, these divergences are
artificial, and do not correspond to real infinities in observables.
Rather, the divergences appear because it is not realistic to assume
a non-standard vacuum to exist in the infinite past. A cutoff on
momenta should be imposed at the same time when a non-Bunch-Davies
vacuum is considered.

We would like to point out two interesting aspects of the effects of
the non-Bunch-Davies vacuum
on trispectra, and more generally on higher point
functions.

Let us first look at the regions away from the folded sub-triangle
configurations.
In the regular tetrahedron limit, in terms of \eqref{tNLdef}, the
corresponding $t_{NL}$ for the non-Bunch-Davies contribution are
\begin{equation}
\tilde t_{NL}^{c1}=4.50 ~ {\rm
  Re}(C_-)\left(\frac{\mu}{\Sigma}-\frac{9\lambda^2}{\Sigma^2}\right)~,\qquad
  \tilde t_{NL}^{s1}=401 ~ {\rm
  Re}(C_-)\left(\frac{\lambda}{\Sigma}\right)^2~.
\end{equation}
Note that $\tilde t_{NL}^{c1}/{\rm Re}(C_-)$ is 128 times larger
than $t_{NL}^{c1}$; $\tilde t_{NL}^{s1}/{\rm Re}(C_-)$ is about 1600
times larger than $t_{NL}^{s1}$. These large numbers arise because
some plus signs become minus signs in the denominators, and there are
also more terms to consider in the non-Bunch-Davies case. So in the
context of general single field inflation, even if ${\rm Re}(C_-)$
is as small as one part in one thousand, it becomes important
phenomenologically. Generalize this to higher point functions, we see
that, no matter how small the $C_-$ is,
there always exists a high point function beyond
which the contribution from the $C_-$ component becomes comparable to
that from the $C_+$ component. For such functions,
we should use the whole
wavefunction (\ref{nonBDuk}) to get the correct shapes instead of
treating the $C_-$ component as a correction, even away from the
folded sub-triangle limit.
Therefore generally speaking, higher
point function is a more sensitive probe to the
non-Bunch-Davies component. However, on the other hand,
higher point functions
contribute less to the total non-Gaussianities and will be more
difficult to measure experimentally.

We next look at the region near the folded sub-triangle limits.
Because the denominators here have larger
powers than those in 3pt, the corrections grow faster as we
approach the folded sub-triangle limit. This also indicates that
the trispectrum, and more generally higher point functions, is a
nice probe of the non-Bunch-Davies vacuum.
But, on the other hand, because the higher point function has a
larger momentum phase space (three more dimensions here in trispectra)
than the
3pt, the phase space for the folded limits becomes relatively
smaller.

It will be interesting to see how the two factors in each of the
above two
aspects play out in the data analyses.

In Fig. \ref{NBDplot}, we plot $\tilde T_{c1}/{\rm Re}(C_-)$ and
$\tilde T_{s1}/{\rm Re}(C_-)$ in the equilateral limit, specialized
planar limit, and near double squeezed limit ($\tilde T_{c1}/[{\rm
Re}(C_-)\Pi_i k_i]$ and $\tilde T_{s1}/[{\rm Re}(C_-)\Pi_i k_i]$ in
this case) respectively.

\section{Conclusion}
\setcounter{equation}{0}

To conclude, we have calculated the leading order trispectra for
general single field inflation. As in the case of bispectra, the
trispectra turns out to be of ``equilateral shape'' in general
single field inflation. Compared with the local shape trispectra,
the equilateral shape trispectra has not been extensively
investigated in the literature. It is clear that there are a lot of
work worthy to be done on this topic in the future. Directions for
future work include:

\begin{itemize}
  \item It is useful to extend our calculation to more
  general cases. We have focused here on the leading order
  contribution and single field inflation. It is interesting to
  generalize this calculation to next-to-leading order which may be
  also potentially observable, or
  multifield inflation \cite{XGao}. It is also useful to perform a unified
  analysis for general single field inflation and slow roll
  inflation.

  \item The shape of the trispectra is much more complicated
  compared with that of the bispectra. In our paper, we have
  obtained a lot of features of the shape functions by taking various
  limits. However, it is still a challenge to find improved
  representations to understand the shape functions. For example,
  one can find new bases for the shapes by tuning parameters to subtract out the
  similarities. Also, we focus on the planar limit in plotting
  the figures (except for Fig. \ref{equilateral}), because the planar
  limit is of special importance for the CMB data analysis. It is
  interesting to investigate the non-planar parameter region in more
  details for the large scale structure and 21-cm line surveys.

  \item In the discussion of non-Bunch-Davies vacuum, we have
  calculated contributions from the two representative shapes.
  However, as we have seen in Section 6, trispectra is a powerful
  probe for non-Bunch-Davies vacuum. Even one part in $10^3$
  deviation from the Bunch-Davies vacuum could lead to an order
  one correction to the trispectra. So it is valuable to perform the
  full calculation for the non-Bunch-Davies vacuum, and to study the
  effects of the cutoff.

  \item Most importantly, one would like to apply these shape
  functions to data analyses and see how they are constrained.
\end{itemize}

\noindent {\it Note added:} On the day this work appeared on the
arXiv, the paper \cite{Arroja:2009pd} was also submitted to the
arXiv, which overlaps with our Sec.3 and 5.

\medskip
\section*{Acknowledgments}

We thank Bin Chen for his participation in the early stage of this
work. We thank Eiichiro Komatsu, Miao Li, Eugene Lim for helpful
discussions. XC and MH would like to thank the hospitality of the
organizers of the program ``Connecting fundamental physics with
observations'' and the KITPC, where this work was initiated.
GS would like to thank Kazuya Koyama for independently
pointing out to him the
importance of the scalar-exchange term.
XC was
supported by the US Department of Energy under cooperative research
agreement DEFG02-05ER41360. BH was supported in part by the Chinese
Academy of Sciences with Grant No. KJCX3-SYW-N2 and the NSFC with
Grant No. 10821504 and No. 10525060.
GS  was supported in part by NSF CAREER Award No. PHY-0348093, DOE grant DE-FG-02-95ER40896, a Research Innovation Award and a Cottrell Scholar Award from Research
Corporation, a Vilas Associate Award from the University of Wisconsin, and a John Simon Guggenheim Memorial Foundation Fellowship. GS would also like to acknowledge support from the Ambrose Monell Foundation during his stay at the Institute for Advanced Study.
YW was supported in part by a NSFC grant No. 10535060/A050207, a NSFC group grant No. 10821504, and a 973 project grant No. 2007CB815401.

\appendix

\section{Commutator form of the in-in formalism}
\label{AppCom}
\setcounter{equation}{0}

In Sec.~\ref{SecSE}, we have used the original definition of the
in-in formalism (\ref{Def}) and (\ref{4pt3terms}) to calculate the
correlation function due to the scalar-exchange diagram. Another
equivalent and commonly-used form is written in terms of the nested
commutators. For the diagrams that we considered in
Sec.~\ref{SecSE}, it takes the following form, \bea\ \label{in in}
\langle \zeta^4 \rangle \supset - \int_{t_0}^t dt' \int_{t_0}^{t'}
dt'' \langle 0| \left[\left[\zeta_I^4(t), H_I(t') \right],
H_I(t'')\right] |0 \rangle ~. \eea The main difference is that now
the first term in (\ref{4pt3terms}) is separated into two integrals,
each has a time- or anti-time-ordered double integration, \bea
\int_{t_0}^t dt' \int_{t_0}^{t'} dt'' ~\langle 0| ~H_I(t'')
~\zeta^4_I(t) ~H_I(t') ~|0\rangle + \int_{t_0}^t dt' \int_{t_0}^{t'}
dt'' ~\langle 0| ~H_I(t') ~\zeta^4_I(t) ~H_I(t'') ~|0\rangle ~.
\label{ComForm} \eea If one uses this form, besides the fact that
the algebra becomes much more complicated due to the time-ordered
double integrations, one also encounters spurious divergences at
special momentum configurations, such as $k_1 +k_2 - k_3 -k_4 =0$, for each
of the two terms in (\ref{ComForm}). These divergences can be seen,
after some complicated algebra, to cancel each other once the two
terms are summed up \cite{Seery:2008ax,Gao:2009gd,Adshead:2009cb}.
Therefore using the form of the first term in (\ref{4pt3terms}) is
both algebraically simpler and free of spurious divergences. This
conclusion can be generalized to the more nested terms.

For example, using formula (\ref{in in}) we can get
\bea\label{aa
appen}
\langle\zeta^4\rangle_{aa}&\propto&\frac{9}{8}\frac{\lambda^2}{\Sigma^2}
\frac{8k_1^2k_2^2k_3^2k_4^2k_{12}}{K^5_{(-)}K^5_{(+)}\left(2k_{12}-K_{(-)}+K_{(+)}\right)^3}
\left\{3\left(K^7_{(-)}+K^7_{(+)}\right)+\left(12k^2_{12}-9K_{(-)}K_{(+)}\right)\right.\nonumber\\
&&\times\left(K^5_{(-)}+K^5_{(+)}\right)+8K^2_{(-)}K^2_{(+)}\left(K^3_{(-)}+K^3_{(+)}\right)
-6k_{12}\left[2\left(K^6_{(-)}-K^6_{(+)}\right)\right.\nonumber\\
&&\left.\left.-3K_{(-)}K_{(+)}\left(K^4_{(-)}-K^4_{(+)}\right)\right]\right\}+
{\rm 23~perm.}\;,
\eea
with $K_{(-)}\equiv k_1+k_2-k_3-k_4$ and
$K_{(+)}\equiv k_1+k_2+k_3+k_4$. Although the terms before ${\rm
23~perm.}$ are not equivalent to the one in (\ref{aa1_1}) plus
(\ref{aa23_1}), however, after including the permutation terms and
performing lots of lengthy but straightforward calculations, one can
find that the two expressions are the same.

\section{Details on the scalar-exchange diagram}
\label{AppSEDetail}
\setcounter{equation}{0}

In this Appendix, we give the details of
the scalar-exchange diagram.
We denote
\bea
\bk_{12}=\bk_1+\bk_2 ~, ~~~
M=k_3+k_4+k_{12}~, ~~~ K=k_1+k_2+k_3+k_4~.
\eea
The following are the various
contributions to the trispectrum form factor $\CT$ defined
as
\bea
\langle \zeta^4 \rangle \equiv (2\pi)^9
P_\zeta^3 \delta^3(\sum_{i=1}^4 \bk_i)
\prod_{i=1}^4 \frac{1}{k_i^3} ~ \CT(k_1,k_2,k_3,k_4,k_{12},k_{14}) ~.
\eea
The interaction Hamiltonian has two components, (\ref{H_a}) and
(\ref{H_b}). So there are four different combinations.

\subsection{The component $\langle \zeta^4 \rangle_{aa}$}

This component is given in (\ref{aa1}) and (\ref{aa2}). The
contribution from the first term of Eq.~(\ref{4pt3terms}):

\bea
\frac{9}{8} \left( \frac{\lambda}{\Sigma} \right)^2
k_1^2 k_2^2 k_3^2 k_4^2 k_{12}
\frac{1}{(k_1+k_2+k_{12})^3 M^3} + {\rm 23~ perm.} ~.
\label{aa1_1}
\eea

\medskip

\noindent The contribution from the second and third terms of
Eq.~(\ref{4pt3terms}):

\bea
\frac{9}{4} \left( \frac{\lambda}{\Sigma} \right)^2
k_1^2 k_2^2 k_3^2 k_4^2 k_{12} \frac{1}{M^3}
\left( \frac{6M^2}{K^5} + \frac{3M}{K^4} + \frac{1}{K^3} \right)
+ {\rm 23~ perm.} ~.
\label{aa23_1}
\eea

\subsection{The component $\langle \zeta^4 \rangle_{ab}$}

There are two sub-diagrams contributing to this case. In the first
sub-diagram, the exchanged scalar propagator is due to
the contraction between two $\dot
\zeta$'s; in the second, between $\dot \zeta$ and $\zeta$.
The result of these
two sub-diagrams for the first term of Eq. \eqref{4pt3terms}, and
the second and third term of Eq. \eqref{4pt3terms} are as follows.
\medskip

\noindent The contribution from the first term:

$\bullet$ The first sub-diagram: \bea -\frac{3}{32}
\frac{\lambda}{\Sigma} \left( \frac{1}{c_s^2} - 1 \right) (\bk_3
\cdot \bk_4) k_{12} k_1^2 k_2^2 \frac{1}{(k_1+k_2+k_{12})^3}
F(k_3,k_4,M) + 23~ {\rm perm.} ~. \label{ab1_1} \eea

$\bullet$ The second sub-diagram: \bea -\frac{3}{16}
\frac{\lambda}{\Sigma} \left( \frac{1}{c_s^2} -1 \right) (\bk_{12}
\cdot \bk_4) \frac{k_1^2 k_2^2 k_3^2}{k_{12}}
\frac{1}{(k_1+k_2+k_{12})^3} F(k_{12},k_4,M) + 23~ {\rm perm.} ~.
\label{ab1_2} \eea

\medskip

\noindent The contribution from the second and third terms:

$\bullet$ The first sub-diagram: \bea - \frac{3}{16}
\frac{\lambda}{\Sigma} \left( \frac{1}{c_s^2} -1 \right) (\bk_3
\cdot \bk_4) k_1^2 k_2^2 k_{12} ~ G_{ab}(k_3,k_4) +23~{\rm perm.} ~.
\label{ab23_1} \eea

$\bullet$ The second sub-diagram: \bea - \frac{3}{8}
\frac{\lambda}{\Sigma} \left( \frac{1}{c_s^2} -1 \right) (\bk_{12}
\cdot \bk_4) \frac{k_1^2 k_2^2 k_3^2}{k_{12}} ~ G_{ab}(k_{12},k_4)
+23~{\rm perm.} ~. \label{ab23_2} \eea

\subsection{The component $\langle \zeta^4 \rangle_{ba}$}

Similar to the $\langle \zeta^4 \rangle_{ab}$ case, here we also
have two terms, each term includes two sub-diagrams:

\medskip

\noindent The contribution from the first term:

Same as (\ref{ab1_1}) and (\ref{ab1_2}).

\medskip

\noindent The contribution from the second and third terms:

$\bullet$ The first sub-diagram: \bea - \frac{3}{16}
\frac{\lambda}{\Sigma} \left( \frac{1}{c_s^2} -1 \right) (\bk_1
\cdot \bk_2) k_3^2 k_4^2 k_{12} ~ G_{ba}(k_1,k_2) +23~{\rm perm.} ~.
\label{ba23_1} \eea

$\bullet$ The second sub-diagram: \bea \frac{3}{8}
\frac{\lambda}{\Sigma} \left( \frac{1}{c_s^2} -1 \right) (\bk_2
\cdot \bk_{12}) \frac{k_1^2 k_3^2 k_4^2}{k_{12}} ~
G_{ba}(-k_{12},k_2) +23~{\rm perm.} ~. \label{ba23_2} \eea

\subsection{The component $\langle \zeta^4 \rangle_{bb}$}

In this case, we have four sub-diagrams for each term. In the first
sub-diagram, the exchanged scalar propagator
is due to the contraction between
two $\dot \zeta$'s; in the second, between $\dot \zeta$ and
$\zeta$; in the third, between $\zeta$ and $\dot \zeta$; in the
fourth, between two $\zeta$'s.
The result of these
four sub-diagrams for the first term of Eq. \eqref{4pt3terms}, and
the second and third term of Eq. \eqref{4pt3terms} are as follows.

\medskip

\noindent The contribution from the first term:

$\bullet$ The first sub-diagram: \bea \frac{1}{2^7} \left(
\frac{1}{c_s^2} -1 \right)^2 (\bk_1 \cdot \bk_2)(\bk_3 \cdot \bk_4)
k_{12} ~ F(k_1,k_2,k_1+k_2+k_{12}) F(k_3,k_4,M) + {\rm 23~perm.} ~.
\label{bb1_1} \eea

$\bullet$ The second and third sub-diagrams: \bea \frac{1}{2^5}
\left( \frac{1}{c_s^2} -1 \right)^2 (\bk_1 \cdot \bk_2)(\bk_{12}
\cdot \bk_4) \frac{k_3^2}{k_{12}} ~ F(k_1,k_2,k_1+k_2+k_{12})
F(k_{12},k_4,M) + {\rm 23~perm.} ~. \eea

$\bullet$ The fourth sub-diagram: \bea -\frac{1}{2^5} \left(
\frac{1}{c_s^2} -1 \right)^2 (\bk_{12} \cdot \bk_2)(\bk_{12} \cdot
\bk_4) \frac{k_1^2 k_3^2}{k_{12}^3} ~ F(k_{12},k_2,k_1+k_2+k_{12})
F(k_{12},k_4,M) + {\rm 23~perm.} ~. \eea

\medskip

\noindent The contribution from the second and third terms:

$\bullet$ The first sub-diagram: \bea \frac{1}{2^6} \left(
\frac{1}{c_s^2} -1 \right)^2 (\bk_1 \cdot \bk_2)(\bk_3 \cdot \bk_4)
k_{12} ~ G_{bb}(k_1,k_2,k_3,k_4) + {\rm 23~perm.} ~. \eea

$\bullet$ The second sub-diagram: \bea \frac{1}{2^5} \left(
\frac{1}{c_s^2} -1 \right)^2 (\bk_1 \cdot \bk_2)(\bk_{12} \cdot
\bk_4) \frac{k_3^2}{k_{12}} ~ G_{bb}(k_1,k_2,k_{12},k_4) + {\rm
23~perm.} ~. \eea

$\bullet$ The third sub-diagram: \bea -\frac{1}{2^5} \left(
\frac{1}{c_s^2} -1 \right)^2 (\bk_{12} \cdot \bk_2)(\bk_3 \cdot
\bk_4) \frac{k_1^2}{k_{12}} ~ G_{bb}(-k_{12},k_2,k_3,k_4) + {\rm
23~perm.} ~. \eea

$\bullet$ The fourth sub-diagram: \bea -\frac{1}{2^4} \left(
\frac{1}{c_s^2} -1 \right)^2 (\bk_{12} \cdot \bk_2)(\bk_{12} \cdot
\bk_4) \frac{k_1^2k_3^2}{k_{12}^3} ~ G_{bb}(-k_{12},k_2,k_{12},k_4)
+ {\rm 23~perm.} ~. \label{bb23_4} \eea

\medskip
\noindent
The function $F$, $G_{ab}$, $G_{ba}$, $G_{bb}$ are defined as follows:
\bea
&&F(\alpha_1,\alpha_2,m)
\cr
&\equiv& \frac{1}{m^3}
\left[ 2\alpha_1\alpha_2 + (\alpha_1+\alpha_2)m +m^2 \right] ~,
\label{Fdef}
\\
\nonumber \\
&&G_{ab}(\alpha_1,\alpha_2)
\cr
&\equiv&
\frac{1}{M^3K^3}
\left[ 2\alpha_1 \alpha_2 +(\alpha_1+\alpha_2)M +M^2 \right]
\cr
&+&
\frac{3}{M^2K^4}
\left[ 2\alpha_1 \alpha_2 +(\alpha_1+\alpha_2)M \right]
+ \frac{12}{MK^5} \alpha_1\alpha_2 ~,
\\
\nonumber \\
&&G_{ba}(\alpha_1,\alpha_2)
\cr
&\equiv&
\frac{1}{M^3K} + \frac{1}{M^3K^2}(\alpha_1+\alpha_2+M)
+\frac{1}{M^3K^3}
\left[ 2\alpha_1\alpha_2 + 2(\alpha_1+\alpha_2)M +M^2 \right]
\cr
&+&
\frac{3}{M^2K^4}
\left[ 2\alpha_1\alpha_2 + (\alpha_1+\alpha_2)M \right]
+ \frac{12}{MK^5}\alpha_1\alpha_2 ~,
\label{Gba}
\\
\nonumber \\
&&G_{bb}(\alpha_1,\alpha_2,\alpha_3,\alpha_4)
\cr
&\equiv&
\frac{1}{M^3 K} \left[ 2 \alpha_3 \alpha_4 + (\alpha_3+\alpha_4)M +
M^2 \right]
\cr
&+& \frac{1}{M^3 K^2}
\left[ 2 \alpha_3\alpha_4(\alpha_1 + \alpha_2)
+ \left( 2 \alpha_3 \alpha_4 +
(\alpha_1 + \alpha_2)(\alpha_3+\alpha_4) \right) M
+ \sum_{i=1}^4 \alpha_i M^2 \right]
\cr
&+& \frac{2}{M^3 K^3}
\left[ 2 \prod_{i=1}^4 \alpha_i +
\left(2\alpha_3\alpha_4(\alpha_1+\alpha_2) +
\alpha_1\alpha_2(\alpha_3+\alpha_4) \right) M
+ \sum_{i<j} \alpha_i \alpha_j M^2 \right]
\cr
&+& \frac{6}{M^2 K^4} \left(\prod_{i=1}^4 \alpha_i \right)
\left( 2+M \sum_{i=1}^4 \frac{1}{\alpha_i} \right)
+ \frac{24}{MK^5} \prod_{i=1}^4 \alpha_i
~.
\label{Gbb}
\eea
Note that in $G_{ab}$, $G_{ba}$ and $G_{bb}$, the $K$ and $M$ are
defined as $K=k_1+k_2+k_3+k_4$ and $M=k_3+k_4+k_{12}$, but not in terms
of $\alpha_i$'s.

To summarize we denote the overall contribution from the
scalar-exchange diagrams as
\bea
\CT_s = \left( \frac{\lambda}{\Sigma} \right)^2 T_{s1}
+ \frac{\lambda}{\Sigma} \left( \frac{1}{c_s^2}-1 \right) T_{s2}
+ \left( \frac{1}{c_s^2}-1 \right)^2 T_{s3} ~,
\eea
where $T_{s1}$ is given by (\ref{aa1_1}) and (\ref{aa23_1}), $T_{s2}$ is
given by (\ref{ab1_1})-(\ref{ba23_2}), $T_{s3}$ is given by
(\ref{bb1_1})-(\ref{bb23_4}).

\section{The planar limit of the trispectra}
\label{planarappendix}
\setcounter{equation}{0}

In Sec. \ref{shapesection}, we discussed various properties of the
shape functions. As we have stated, the planar limit has special
importance for CMB experiments. So in this appendix, we investigate
the trispectra in more detail in the planar limit and perform a survey
of parameters for the shape functions.

\begin{figure}
\center
\includegraphics[width=0.8\textwidth]{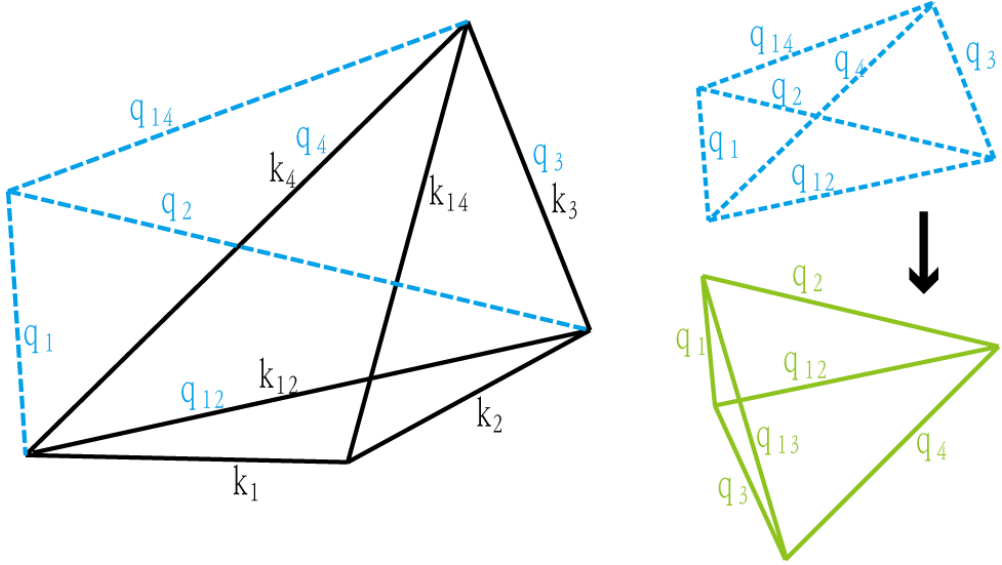}
\caption{\label{symquadrangles} The quadrangles in black and blue
represents the solution (\ref{planark2}) with minus and plus sign
respectively. The term with plus sign corresponds to the blue
quadrangle, and can be transformed into another quadrangle
corresponding to the minus solution by a symmetry discussed in Eq.
\eqref{symeq}.}
\end{figure}

In the planar limit, Eq. \eqref{cosineq} takes the equal sign. One can
solve $k_2$ from Eq. \eqref{cosineq},
  \begin{align}\label{planark2}
   k_2= \frac{\sqrt{k_1^2 \left(-k_{12}^2+k_3^2+k_4^2\right)\pm k_{s1}^2 k_{s2}^2+k_{12}^2 k_{14}^2+k_{12}^2 k_4^2+k_{14}^2
   k_4^2-k_{14}^2 k_3^2-k_4^4+k_3^2 k_4^2}}{\sqrt{2} k_4}~,
  \end{align}
where $k_{s1}$ and $k_{s2}$ are defined as
\begin{align}
&  k_{s1}^2\equiv 2\sqrt{(k_1 k_4+{\bf k}_1 \cdot {\bf k}_4)(k_1
k_4-{\bf k}_1 \cdot {\bf
  k}_4)}~,\nonumber\\ &
k_{s2}^2\equiv 2\sqrt{(k_3 k_4+{\bf k}_3 \cdot {\bf k}_4)(k_3
k_4-{\bf k}_3 \cdot {\bf
  k}_4)}~.
\end{align}
We first take a close look at the $\pm$ sign in Eq.
\eqref{planark2}. The $-$ sign and the $+$ sign correspond to two
different quadrangles. In Fig. \ref{symquadrangles}, the $-$ and $+$
solutions correspond to the black (with edge $k_i$) and blue (with
edge $q_i$) quadrangles respectively.
The former has all internal angles $\le \pi$, while the latter has one
$>\pi$.
The blue quadrangle can be
transformed into another quadrangle belonging to the same class as
the black one by the symmetry discussed in Eq. \eqref{symeq}. So
without losing generality, we will only consider the $-$ solution in
the following discussion.

\begin{figure}[p]
  \center
  \includegraphics[width=0.22\textwidth]{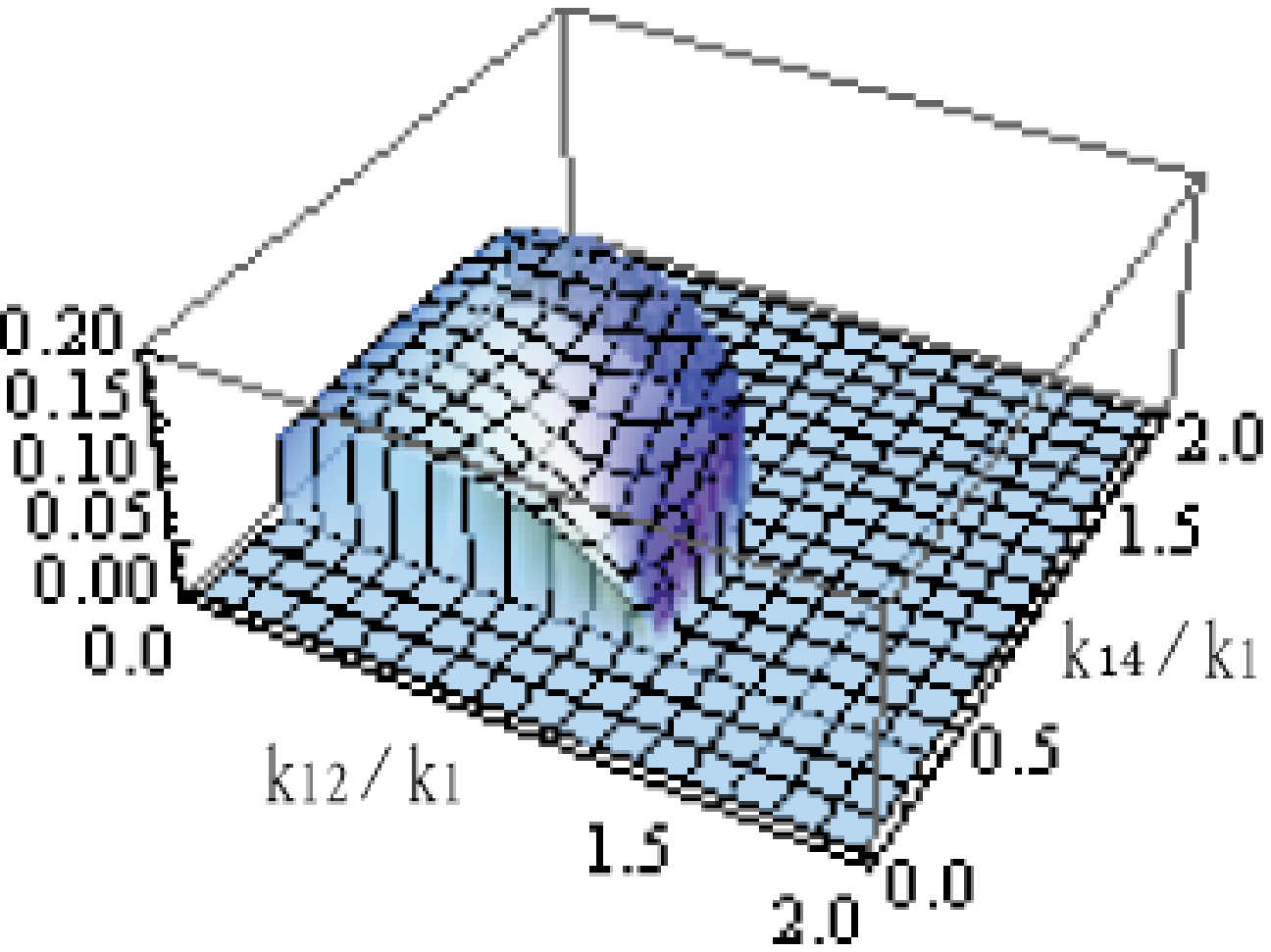}
  \includegraphics[width=0.22\textwidth]{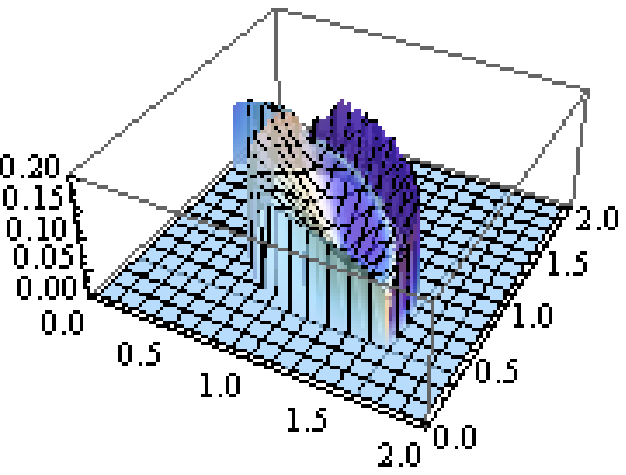}
  \includegraphics[width=0.22\textwidth]{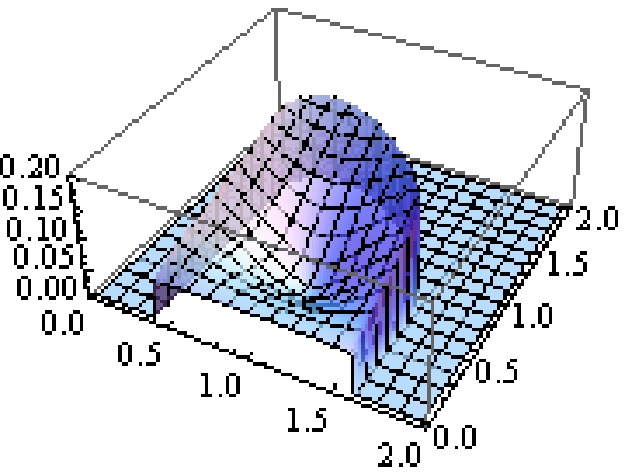}
  \includegraphics[width=0.22\textwidth]{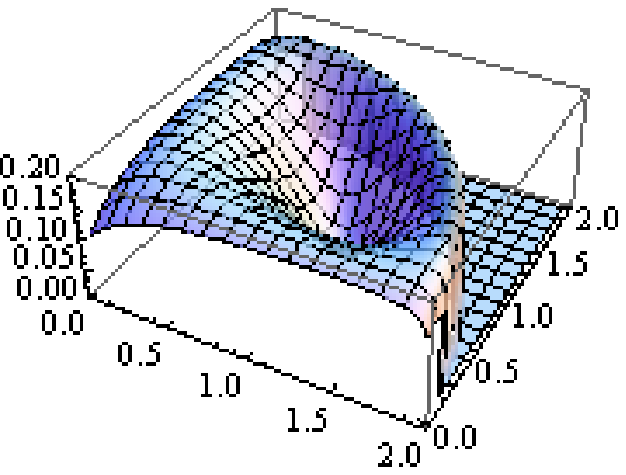}

  \includegraphics[width=0.22\textwidth]{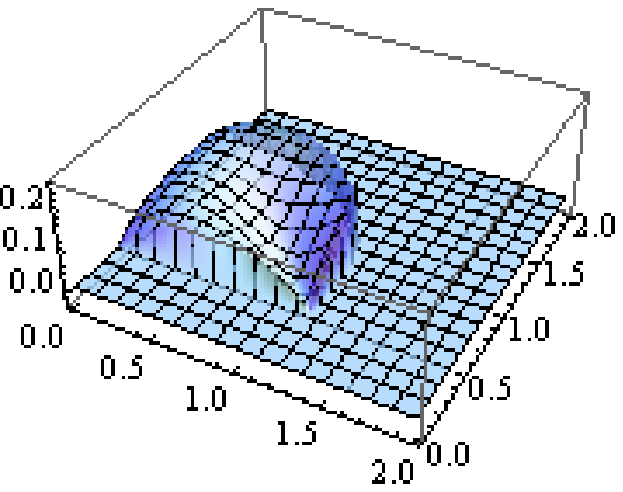}
  \includegraphics[width=0.22\textwidth]{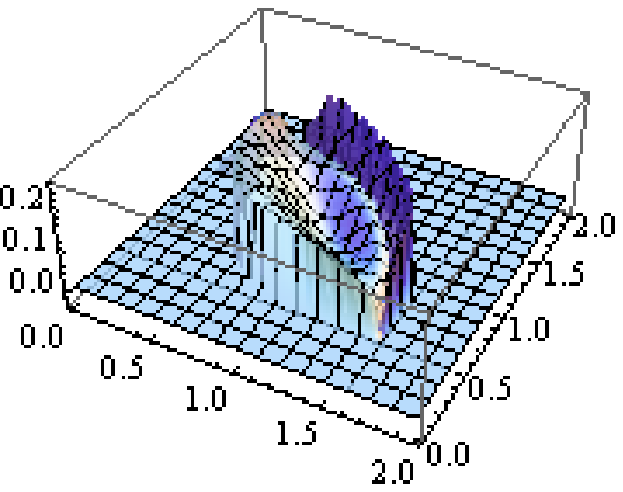}
  \includegraphics[width=0.22\textwidth]{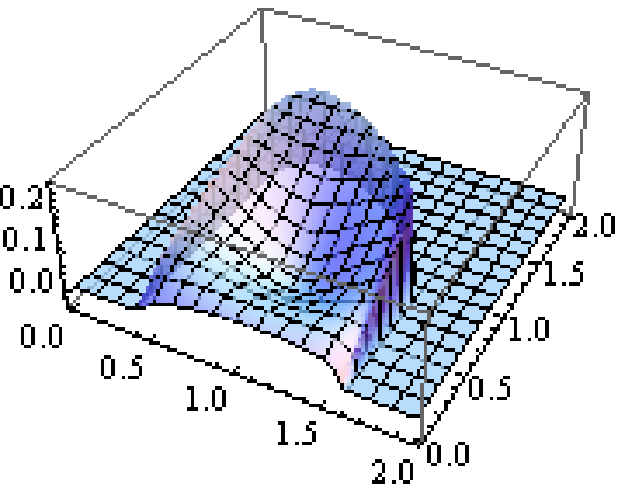}
  \includegraphics[width=0.22\textwidth]{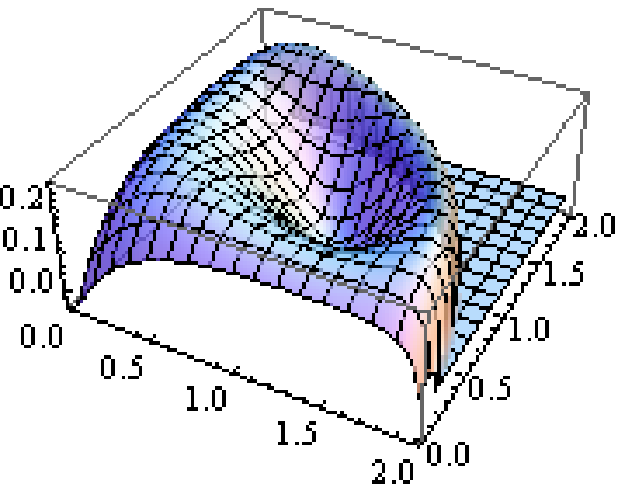}

  \includegraphics[width=0.22\textwidth]{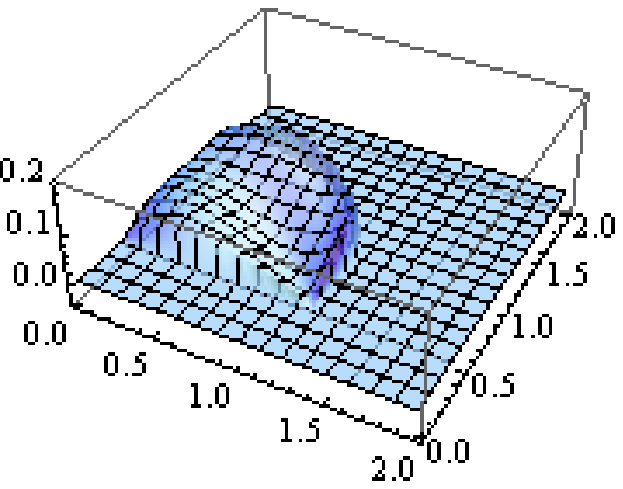}
  \includegraphics[width=0.22\textwidth]{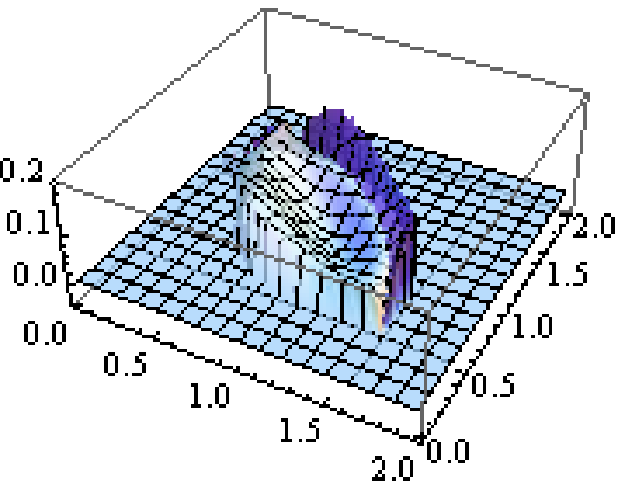}
  \includegraphics[width=0.22\textwidth]{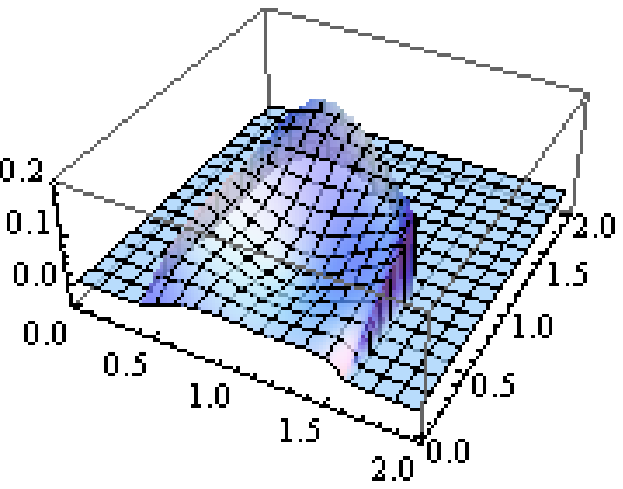}
  \includegraphics[width=0.22\textwidth]{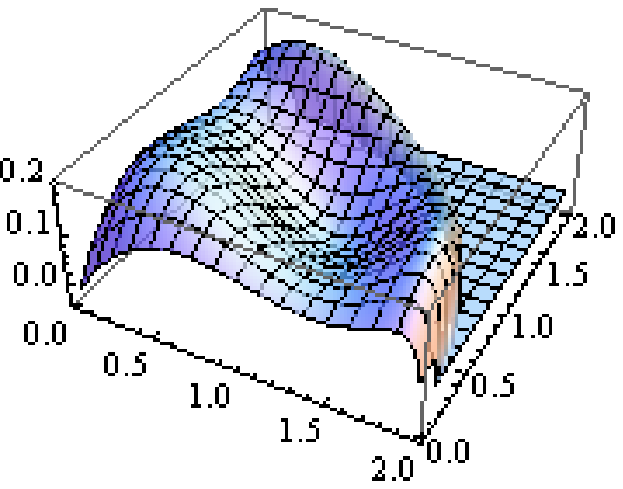}

  \includegraphics[width=0.22\textwidth]{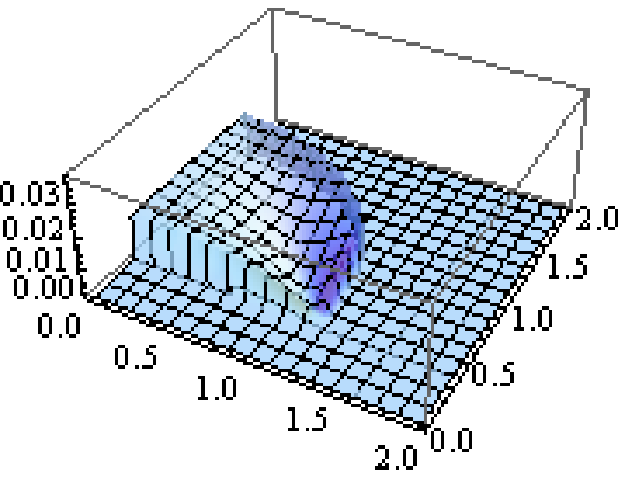}
  \includegraphics[width=0.22\textwidth]{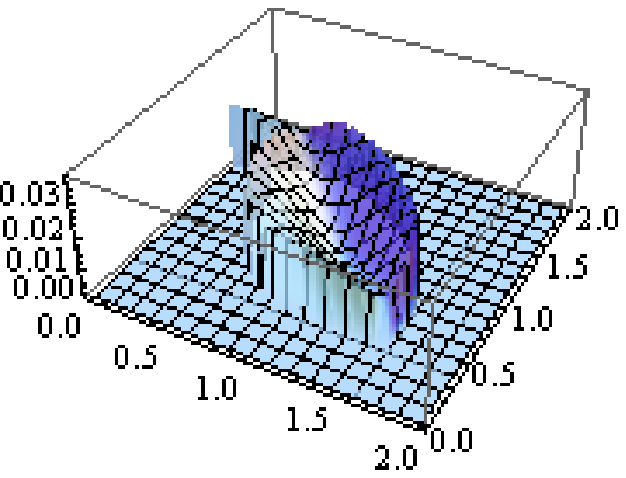}
  \includegraphics[width=0.22\textwidth]{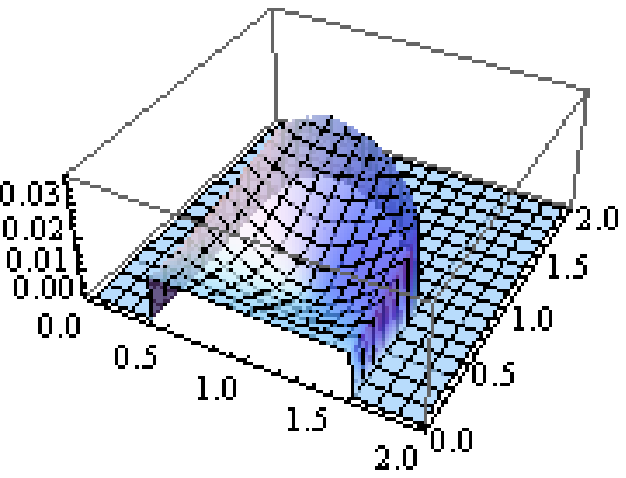}
  \includegraphics[width=0.22\textwidth]{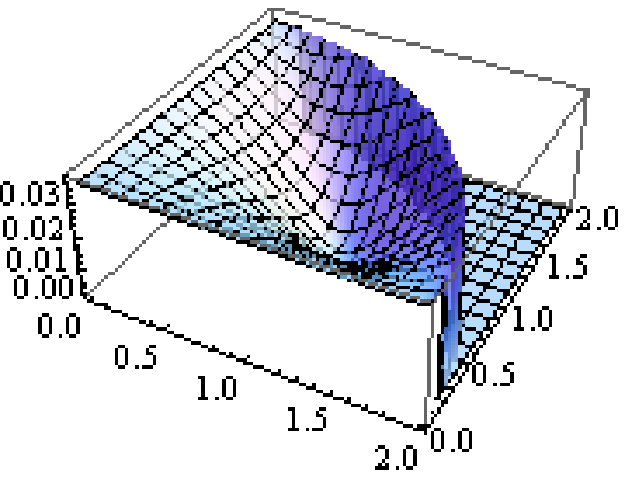}

  \includegraphics[width=0.22\textwidth]{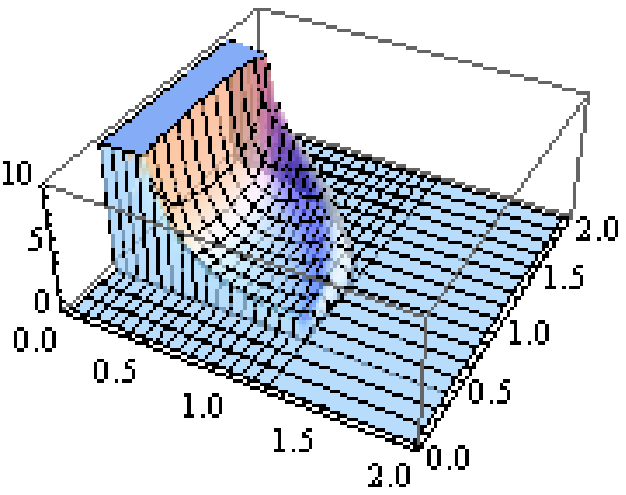}
  \includegraphics[width=0.22\textwidth]{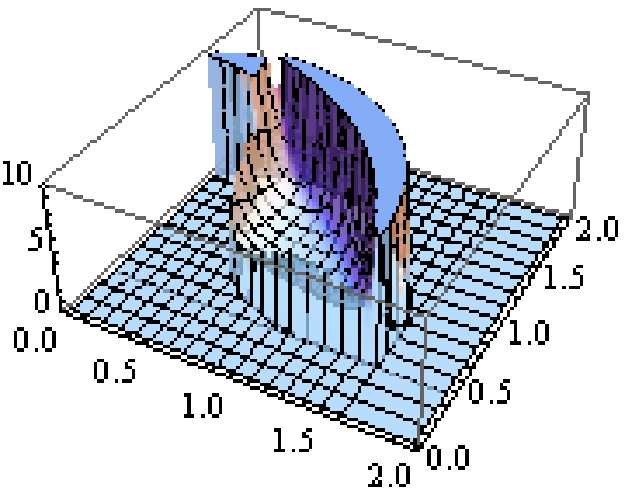}
  \includegraphics[width=0.22\textwidth]{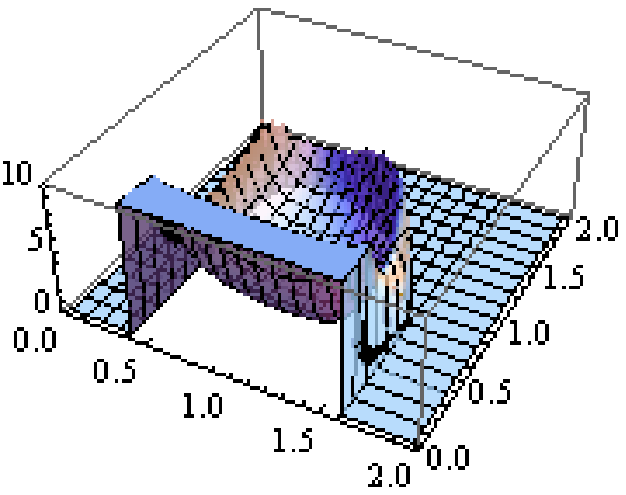}
  \includegraphics[width=0.22\textwidth]{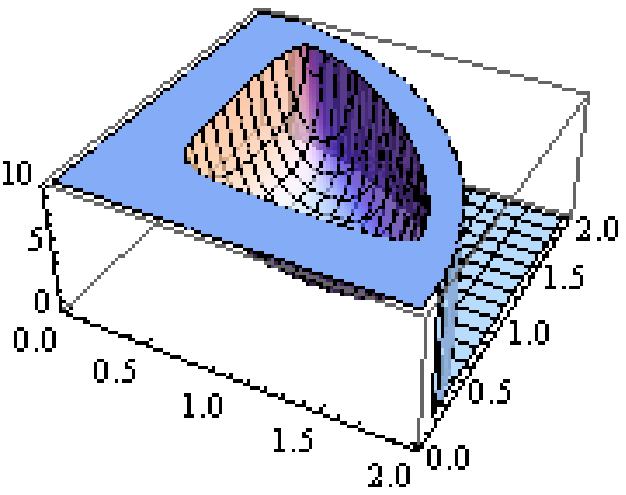}

  \includegraphics[width=0.22\textwidth]{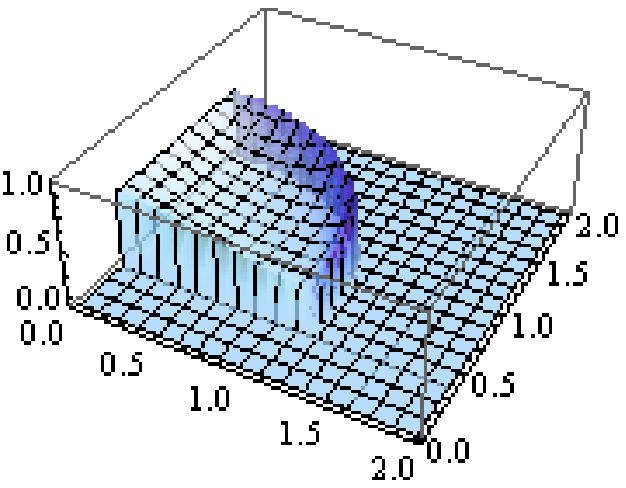}
  \includegraphics[width=0.22\textwidth]{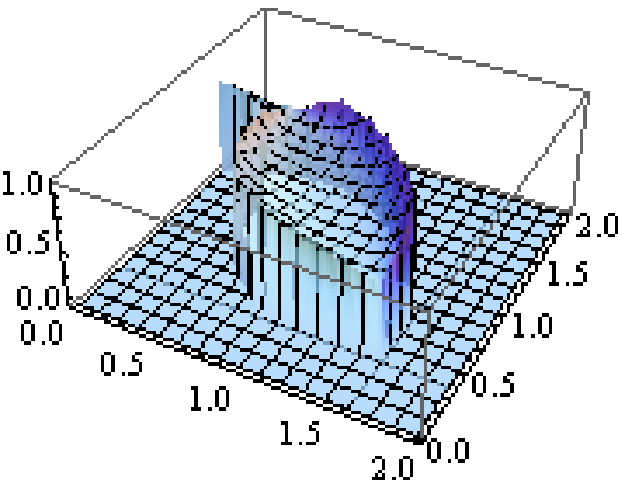}
  \includegraphics[width=0.22\textwidth]{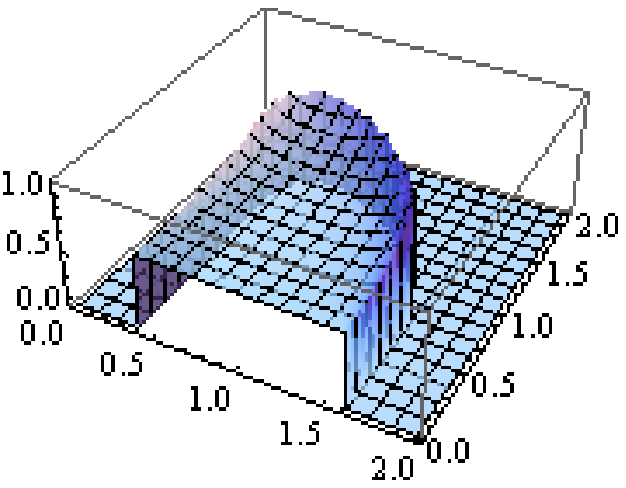}
  \includegraphics[width=0.22\textwidth]{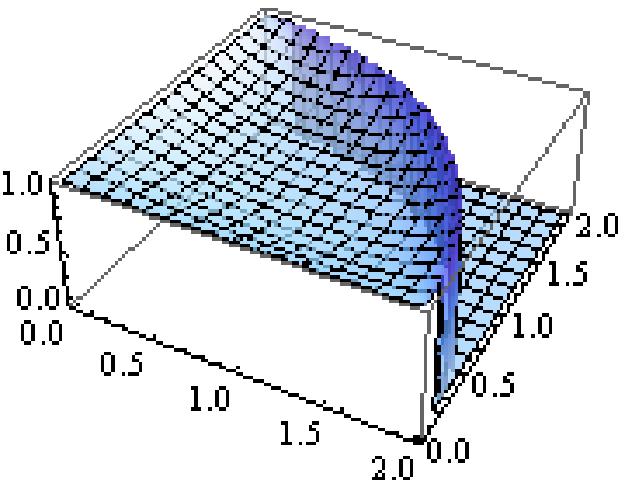}

    \caption{\label{planar1} In the six rows, we plot $T_{s1}$, $T_{s2}$,
$T_{s3}$, $T_{c1}$, $T_{loc1}$ and $T_{loc2}$ respectively
      as functions of $k_{12}/k_1$ and
      $k_{14}/k_1$ in the planar limit.
Within each row, the momenta configuration is $(k_3/k_1,k_4/k_1)=\{(0.6,0.6),(0.6,1.0),(1.0,0.6),(1.0,1.0)\}$
respectively in the four columns, as given in \eqref{groupfigk}.
}
\end{figure}

To scan the parameter space, We plot  $T_{s1}$, $T_{s2}$,
$T_{s3}$, $T_{c1}$, $T_{loc1}$ and $T_{loc2}$ as
  functions of $k_{12}/k_1$ and $k_{14}/k_1$ for different values of
  $k_3/k_1$ and $k_4/k_1$ in Fig.\ref{planar1}. The
  momenta $(k_3/k_1,k_4/k_1)$ take the following values in the 4
  figures in each group.
\begin{equation}\label{groupfigk}
(k_3/k_1,k_4/k_1)=\{(0.6,0.6),(0.6,1.0),(1.0,0.6),(1.0,1.0)\}~.
\end{equation}
We assume $k_1>k_2,k_3,k_4$ in the plot without losing generality.
Note that when $k_3=k_4=k_{12}=k_{14}=k_1$ (the center of the last
figure in each group),  $T_{s1}$, $T_{s2}$, $T_{s3}$ and $T_{c1}$
vanishes because in this case $k_2=0$. We can see from these graphs
that the shapes of $T_{s1}$, $T_{s2}$ and $T_{s3}$ are overall very
similar.

\end{document}